\documentclass[11pt,a4paper]{article}
\usepackage{amsmath,amsthm,amsfonts,amssymb,mathrsfs}
\usepackage{graphicx}
\usepackage{subfigure}
\usepackage[latin1]{inputenc}
\usepackage[swedish,english]{babel}

\begin{document}

\title{Solving integral equations on piecewise smooth boundaries using
  the RCIP method: a tutorial} 
\author{Johan Helsing\\
  {\it Centre for Mathematical Sciences}\\
  {\it Lund University, Box 118, SE-221 00 Lund, Sweden}}
\date{(revised January 7, 2022)}
\maketitle

\begin{abstract}
  Recursively compressed inverse preconditioning (RCIP) is a
  kernel-independent and purely numerical method for solving Fredholm
  second kind boundary integral equations in situations where the
  boundary shape induces a non-smooth behavior in the solution. The
  method originated in 2008 within a scheme for Laplace's equation in
  two-dimensional domains with corners. In a series of subsequent
  papers the method was then refined and extended as to apply to
  integral equation formulations of a broad range of boundary value
  problems in physics and engineering. The purpose of the present
  tutorial is threefold: First, to review the RCIP method in a simple
  setting. Second, to show how easily the method can be implemented in
  {\sc Matlab}. Third, to present new applications.
\end{abstract}

\section{Introduction} 

This tutorial is about an efficient numerical solver for elliptic
boundary value problems in domains whose boundaries contain some sort
of singular points. Such a solver is useful for applications in
physics and engineering, where computational domains of interest often
have corners, triple junctions, and close-to-touching boundary parts.
Furthermore, these problems are difficult to solve irrespective of
what numerical method is used. The reason being that the solution, or
the quantity representing the solution, often exhibits a non-smooth
behavior close to boundary singularities. That behavior is hard to
resolve by polynomials, which underlie most approximation schemes.
Mesh refinement is needed. This is costly and may lead to artificial
ill-conditioning and the loss of accuracy.

The numerical solver we propose takes its starting point in an
integral equation reformulation of the boundary value problem at hand.
We assume that the problem can be modeled as a Fredholm second kind
integral equation with compact integral operators away from singular
boundary points and whose solution is a layer density representing the
solution to the original problem. We seek a discrete approximation to
the layer density using Nyström discretization~\cite[Chapter
4]{Atki97}. At the heart of the solver lies an integral transform
whose inverse modifies the kernels of the integral operators in such a
way that the layer density becomes piecewise smooth and simple to
resolve by polynomials. The inverse is constructed recursively on
small, locally refined, temporary meshes. Conceptually, this
corresponds to applying a fast direct solver~\cite{Kong11} locally to
regions with troublesome geometry. A global iterative method is then
applied. Finally, the original layer density is reconstructed by
running the recursion backwards, should it be explicitly needed. This
gives us many of the advantages of fast direct methods, for example
the ability to deal with certain classes of operators whose spectra
make them unsuitable for iterative methods. In addition, the approach
is typically much faster than using only a fast direct solver.

Our method, or scheme, has been referred to as {\it recursive
  compressed inverse
  preconditioning}~\cite{Hels09JCP,Hels11SISC,Hels08b,Hels09IJSS} and
there is a good reason for that name: the scheme relies on applying a
relieving right inverse to the integral equation; on compressing this
inverse to a low-dimensional subspace; and on carrying out the
compression in a recursive manner. Still, the name {\it recursive(ly)
  compressed inverse preconditioning} is a bit awkward and we will
here simply use the acronym RCIP.

A strong motivation for writing the present tutorial is that the
original references~\cite{Hels09JCP,Hels11SISC,Hels08b,Hels09IJSS} are
hard to read. Certain derivations
in~\cite{Hels09JCP,Hels11SISC,Hels08b,Hels09IJSS} use complicated
intermediary constructions, application specific issues obscure the
general picture, and the notation has evolved from paper to paper.
Here we focus on the method itself, on how it works and how it can be
implemented, and refer to the original research papers for details.
Demo programs in {\sc Matlab}, updated as of December 2021, are a part
of the exposition and can be downloaded from the web page:

\noindent
\centerline{\tt http://www.maths.lth.se/na/staff/helsing/Tutor/}

Section~\ref{sec:back} provides a historical background.
Section~\ref{sec:summary} is a summary of the main features of RCIP.
The basics of the method are then explained by solving a simple model
problem in Sections~\ref{sec:model}--\ref{sec:rec}.
Sections~\ref{sec:SB}--\ref{sec:sing} review general algorithmic
improvements. Sections~\ref{sec:HelmDiri}--\ref{sec:HelmTrans} contain
applications to scattering problems.
Sections~\ref{sec:ctt}--\ref{sec:Vshape} deal with close-to-touching
objects, mixed (Zaremba) boundary conditions, Steklov eigenvalue
problems, limit polarizability, vertex singularity exponents, and
planar crack problems. Some of this material is new and has not been
published elsewhere.

\section{Background}
\label{sec:back}

The line of research on fast solvers for elliptic boundary value
problems in piecewise smooth domains, leading up to the RCIP method,
grew out of work in computational fracture mechanics. Early efforts
concerned finding efficient integral equation formulations. Corner
singularities were either resolved by brute force or by using special
basis functions~\cite{Hels00,IJNME02,SIAP99}. Such strategies, in
combination with fast multipole~\cite{Gree87} accelerated iterative
solvers, work well for simple small-scale problems.

Real world physics is more complicated and, for example, the
study~\cite{Englund07} on a high-order time-stepping scheme for crack
propagation (a series of biharmonic problems for an evolving piecewise
smooth surface) shows that radically better methods are needed.
Special basis functions are too complicated to construct and brute
force is not economical -- merely storing the discretized solution
becomes too costly in a large-scale simulation.

A breakthrough came in 2007, when a scheme was created that resolves
virtually any problem for Laplace's equation in piecewise smooth
two-dimensional domains in a way that is fully automatic, fast,
stable, memory efficient, and whose computational cost scales linearly
with the number of corners in the computational domain. The resulting
paper~\cite{Hels08b} constitutes the origin of the RCIP method.
Unfortunately, however, there are some flaws in~\cite{Hels08b}. For
example, the expressions in~\cite[Section~9]{Hels08b} are not
generally valid and the paper fails to apply RCIP in its entirety to
the biharmonic problem of~\cite[Section~3]{Hels08b}, which was the
ultimate goal.

The second paper on RCIP~\cite{Hels09IJSS} deals with elastic grains.
The part~\cite[Appendix B]{Hels09IJSS}, on speedup and enhanced
stability, is particularly useful.

The third paper on RCIP~\cite{Hels09JCP} contains improvement relative
to the earlier papers, both in the notation and in the discretization
of singular operators. The overall theme is mixed boundary conditions,
which pose similar difficulties as do piecewise smooth boundaries.

The fourth paper on RCIP~\cite{Hels11SISC}, finally, solves the
problem of~\cite[Section~3]{Hels08b} in a broad setting, involving
dominant integral operators with non-zero Fredholm indices and
compositions of integral operators. In this context, too, some
subsequent improvements have been made. See~\cite{HelsKarl13} and
Sections~\ref{sec:compose}, \ref{sec:regul}, and \ref{sec:Vshape},
below.

Further work on developing RCIP deal with more general boundary
conditions~\cite{Ojala12}, with problem-specific stabilization
techniques~\cite{Hels18SISC}, with singular right-hand
sides~\cite{Hels22JCP}, with problems in three
dimensions~\cite{HelsKarl16,HelsP12,HelsP17}, and with large-scale
applications to aggregates of millions of
grains~\cite{Hels11JCPa,Hels11JCPb}.

We end this retrospection by noting that several research groups in
recent years have proposed numerical schemes for integral equations
stemming from elliptic partial differential equations (PDEs) in
domains with boundary singularities. See, for
example,~\cite{Akhm15,Brem12a,Brem12b,Brem10a,Brem10b,Brem10c,Brun09,Hosk19,Hosk20,Rach20,Serkh16}.
There is also a widespread notion that a slight rounding of corners is
a good idea for numerics. While rounding may work in particular
situations, we do not believe it is a generally viable method. For one
thing, how does one round a triple junction?

\section{Summary of RCIP}
\label{sec:summary}

This section summarizes Sections~\ref{sec:model}--\ref{sec:asymp},
below, and reviews the most important features of the RCIP method.

The starting point is an integral equation on a boundary $\Gamma$
containing a corner
\begin{equation}
\left(I+\lambda K\right)\rho(r)=h(r)\,,\quad r\in \Gamma\,.
\label{eq:1}
\end{equation}
Here $I$ is the identity, $\lambda$ is a parameter, $K$ is an integral
operator that is compact away from the corner, $h(r)$ is a piecewise
smooth right-hand side, and $\rho(r)$ is an unknown layer density to
be solved for.

Let the operator $K$ be split into two parts
\begin{equation}
K=K^\star+K^\circ\,,
\label{eq:2}
\end{equation}
where $K^\star$ describes the kernel interaction close to the corner
and $K^\circ$ is a compact operator. Now introduce the {\it
  transformed density}
\begin{equation}
\tilde{\rho}(r)=\left(I+\lambda K^\star\right)\rho(r)\,.
\label{eq:3}
\end{equation}
Then use~(\ref{eq:2}) and~(\ref{eq:3}) to rewrite~(\ref{eq:1}) as
\begin{equation}
\left(I+\lambda K^\circ(I+\lambda K^\star)^{-1}\right)\tilde{\rho}(r)
 =h(r)\,,\quad r\in \Gamma\,.
\label{eq:4}
\end{equation}
Although~(\ref{eq:4}) looks similar to~(\ref{eq:1}), there are
advantages with using~(\ref{eq:4}) from a numerical point of view.

The RCIP method discretizes~(\ref{eq:4}) chiefly on a grid on a {\it
  coarse mesh} on $\Gamma$ that is sufficient to resolve $K^\circ$ and
$h(r)$. Only $(I+\lambda K^\star)^{-1}$ needs a grid on a locally
refined {\it fine mesh}. Nyström discretization is used. The
discretization of~(\ref{eq:4}) assumes the form
\begin{equation}
\left({\bf I}_{\rm coa}+\lambda{\bf K}_{\rm coa}^\circ{\bf R}\right)
\tilde{\boldsymbol{\rho}}_{\rm coa}={\bf h}_{\rm coa}\,,
\label{eq:5}
\end{equation}
where ${\bf R}$ is a sparse block matrix called the {\it compressed
  inverse}. Note that~(\ref{eq:5}) is a discrete system on the coarse
grid only.

The power of RCIP lies in the construction of ${\bf R}$. In theory,
${\bf R}$ corresponds to a discretization of $(I+\lambda
K^\star)^{-1}$ on the fine grid, followed by a lossless compression to
the coarse grid. In practice, ${\bf R}$ is constructed via a {\it
  forward recursion}~(\ref{eq:recur}) where refinement and compression
occur in tandem. The recursion starts on the smallest panels in a
hierarchy of nested meshes around the corner, gradually moves up the
hierarchy, and finally reaches the coarse mesh. At each refinement
level a small matrix ${\bf K}^\circ_{i{\rm b}}$ is needed as input and
a small matrix ${\bf R}_i$ is generated as output. The computational
cost grows, at most, linearly with the number of refinement levels.

Now, with access to the coarse-grid quantities ${\bf R}$ and
$\tilde{\boldsymbol{\rho}}_{\rm coa}$ only, surprisingly much is known
about the solution $\rho(r)$ on the fine grid. For example, the {\it
  weight-corrected density}
\begin{equation}
\hat{\boldsymbol{\rho}}_{\rm coa}={\bf R}\tilde{\boldsymbol{\rho}}_{\rm coa}
\end{equation}
can be used to compute numerical approximations of integrals of
$\rho(r)$ against smooth functions $f(r)$ as if they were carried out
on the fine mesh
\begin{equation}
\int_\Gamma f(r)\rho(r)\,{\rm d}\ell
\approx\sum_j f_{{\rm fin}_j}\rho_{{\rm fin}_j}w_{\Gamma{\rm fin}_j}
    =  \sum_j f_{{\rm coa}_j}\hat{\rho}_{{\rm coa}_j}w_{\Gamma{\rm coa}_j}\,.
\end{equation}
Here $w_{\Gamma{\rm fin}_j}$ and $w_{\Gamma{\rm coa}_j}$ are
quadrature weights suitable for integrating polynomials on the fine
grid and the coarse grid, respectively.

With access also to the matrices ${\bf K}^\circ_{i{\rm b}}$ and ${\bf
  R}_i$, everything is known about $\rho(r)$ on the fine grid: The
discrete density $\boldsymbol{\rho}_{\rm fin}$ can be reconstructed
via a {\it backward recursion}~(\ref{eq:back}); The eigenvalues of a
certain backward recursion submatrix ${\bf C}^\star$ contains
information about the asymptotics of $\rho(r)$ close to the corner
vertex; The backward recursion acting on smooth basis functions
automatically generates a tailor-made (singular) basis for $\rho(r)$.
Such a basis is helpful when RCIP is used for integral equations on
non-smooth domains with edges in three dimensions.

It is important to observe that the forward recursion~(\ref{eq:recur})
is fast. It can be executed on the fly, even when the layer density
$\rho(r)$ is strongly singular and does not lie in any usual $L^p$
space. Deep inte the corner, the sequence of matrices ${\bf
  K}^\circ_{i{\rm b}}$ have often converged to a beforehand given
precision and~(\ref{eq:recur}) assumes the form of a fixed-point
iteration. This opens up for the use of Newton's method. The
computational cost for obtaining ${\bf R}$ can then be said to grow
sub-linearly with respect to the number of refinement levels.

\section{A one-corner model problem}
\label{sec:model}

\begin{figure}
\centering 
\includegraphics[height=73mm]{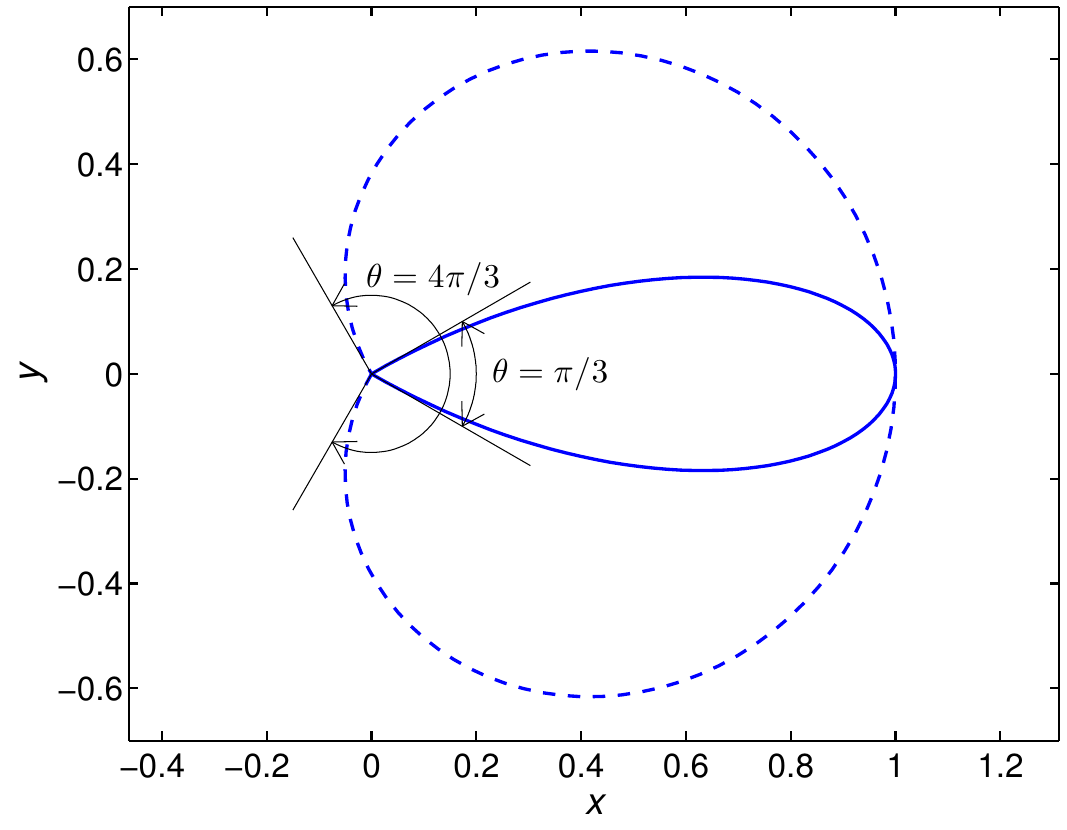}
\caption{\sf The contour $\Gamma$ of~(\ref{eq:gamma}) with a corner at
  the origin. The solid curve corresponds to opening angle
  $\theta=\pi/3$. The dashed curve has $\theta=4\pi/3$.}
\label{fig:sno1}
\end{figure}

Let $\Gamma$ be the closed contour of Figure~\ref{fig:sno1} with the
parameterization
\begin{equation}
r(s)=\sin(\pi s)\left(\cos((s-0.5)\theta),\sin((s-0.5)\theta)\right)\,,
\quad s\in[0,1]\,.
\label{eq:gamma}
\end{equation}
Let $G(r,r')$ be the fundamental solution to Laplace's equation
in the plane:
\begin{equation}
G(r,r')=-\frac{1}{2\pi}\log|r-r'|\,.
\end{equation}
We shall solve the integral equation
\begin{equation}
\rho(r)+2\lambda\int_{\Gamma}\frac{\partial G}{\partial\nu}(r,r')\rho(r')
\,{\rm d}\ell'=2\lambda\left(e\cdot \nu\right)\,,
\quad r\in \Gamma\,,
\label{eq:inteq1a}
\end{equation}
numerically for the unknown layer density $\rho(r)$. Here $\nu$ is the
exterior unit normal at $r\in\Gamma$, ${\rm d}\ell$ is an element of
arc length, $\lambda$ is a parameter, $e$ is a unit vector, and
\begin{equation}
\frac{\partial G}{\partial\nu}(r,r')=\frac{\nu\cdot(r'-r)}
{2\pi|r'-r|^2}\,.
\end{equation}
The equation~(\ref{eq:inteq1a}) models an electrostatic transmission
problem~\cite{Hels08b} where $e$ is an applied electric field.

Using complex notation, where vectors $r$, $r'$, $\nu$, and $e$ in the
real plane~$\mathbb{R}^2$ correspond to points $z$, $\tau$, $n$, and
$e$ in the complex plane $\mathbb{C}$, one can
write~(\ref{eq:inteq1a}) as
\begin{equation}
\rho(z)+\frac{\lambda}{\pi}\int_{\Gamma}\rho(\tau)
\Im\left\{\frac{n_z\bar{n}_{\tau}\,{\rm d}\tau}{\tau-z}\right\}=
2\lambda\Re\left\{\bar{e}n_z\right\}\,,\quad z\in\Gamma\,,
\label{eq:inteq2}
\end{equation}
where the overbar symbol denotes the complex conjugate.
Equation~(\ref{eq:inteq2}) is a simplification over~(\ref{eq:inteq1a})
from a programming point of view.

In many contexts it is advantageous to abbreviate~(\ref{eq:inteq1a})
as
\begin{equation}
\left(I+\lambda K\right)\rho(r)=\lambda g(r)\,,\quad r\in \Gamma\,,
\label{eq:inteq3}
\end{equation}
where $I$ is the identity. If $\Gamma$ is smooth,
then~(\ref{eq:inteq3}) is a Fredholm second kind integral equation
with a compact, non-self-adjoint, integral operator $K$ whose spectrum
is discrete, bounded by one in modulus, and accumulates at zero.

We also need a way to monitor the convergence of solutions $\rho(r)$
to~(\ref{eq:inteq3}). For this purpose we introduce a quantity $q$,
which corresponds to dipole moment or (un-normalized)
polarizability~\cite{HelsP12}
\begin{equation}
q\equiv\int_{\Gamma}\rho(r)(e\cdot r)\,{\rm d}\ell
      =\int_{\Gamma}\rho(z)\Re\left\{\bar{e}z\right\}{\rm d}|z|\,.
\label{eq:q}
\end{equation}

\medskip\noindent
{\bf Remark:} Existence issues are important. Loosely speaking, the
boundary value problem modeled by~(\ref{eq:inteq1a}) has a unique
finite-energy solution for a large class of non-smooth $\Gamma$ when
$\lambda$ is either off the real axis or when $\lambda$ is real and
$\lambda\in[-1,1)$. See~\cite{HelsP12} for sharper statements. The
precise meaning of a {\it numerical solution} to an integral equation
such as~(\ref{eq:inteq1a}) also deserves comment. In this paper, a
numerical solution refers to approximate values of $\rho(r)$ at a
discrete set of points $r_i\in\Gamma$. The values $\rho(r_i)$ should,
in a post-processor, enable the extraction of quantities of interest
including values of $\rho(r)$ at arbitrary points $r\in\Gamma$,
functionals of $\rho(r)$ such as $q$ of~(\ref{eq:q}), and the solution
to the underlying boundary value problem at points in the domain where
that problem was set.

\begin{figure}
  \centering \includegraphics[height=30mm]{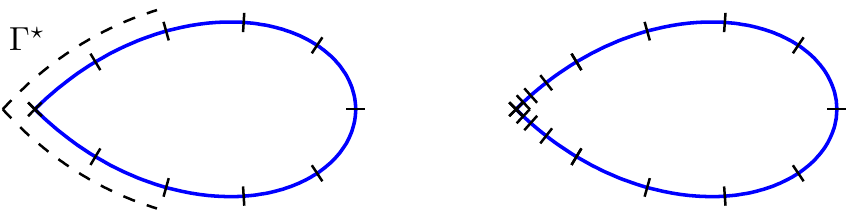}
\caption{\sf Left: A coarse mesh with ten panels on the contour
  $\Gamma$ of~(\ref{eq:gamma}) with opening angle $\theta=\pi/2$. A
  subset of $\Gamma$, called $\Gamma^\star$, covers four coarse
  panels as indicated by the dashed curve. Right: A fine mesh created
  from the coarse mesh by subdividing the panels closest to the corner
  $n_{\rm sub}=3$ times.}
\label{fig:sno2}
\end{figure}

\section{Discretization on two meshes}
\label{sec:disc}

We discretize~(\ref{eq:inteq3}) using standard Nyström discretization
based on composite 16-point Gauss--Legendre quadrature on two
different meshes: a {\it coarse mesh} with $n_{\rm pan}$ quadrature
panels and a {\it fine mesh} which is constructed from the coarse mesh
by $n_{\rm sub}$ times dyadically subdividing the panels closest to
the corner in a direction toward the corner. The discretization is in
parameter. The four panels on the coarse mesh that are closest to the
corner should be equi-sized in parameter. These {\it innermost} four
panels form a subset of $\Gamma$ called $\Gamma^\star$. See
Figure~\ref{fig:sno2}.

The linear systems resulting from the discretization on the coarse
mesh and on the fine mesh can be written formally as
\begin{align}
\left({\bf I}_{\rm coa}+\lambda{\bf K}_{\rm coa}\right)
\boldsymbol{\rho}_{\rm coa}&=\lambda{\bf g}_{\rm coa}\,,
\label{eq:inteq3a}\\
\left({\bf I}_{\rm fin}+\lambda{\bf K}_{\rm fin}\right)
\boldsymbol{\rho}_{\rm fin}&=\lambda{\bf g}_{\rm fin}\,,
\label{eq:inteq3b}
\end{align}
where ${\bf I}$ and ${\bf K}$ are square matrices and
$\boldsymbol{\rho}$ and ${\bf g}$ are column vectors. The subscripts
{\footnotesize fin} and {\footnotesize coa} indicate what type of mesh
is used. Discretization points on a mesh are said to constitute a {\it
  grid}. The coarse grid has $n_{\rm p}=16n_{\rm pan}$ points. The
fine grid has $n_{\rm p}=16(n_{\rm pan}+2n_{\rm sub})$ points.

The discretization of~(\ref{eq:inteq3}) is carried out by first
rewriting~(\ref{eq:inteq2}) as
\begin{equation}
\rho(z(s))+\frac{\lambda}{\pi}\int_0^1\rho(\tau(t))
\Re\left\{\frac{n_{z(s)}|\dot{\tau}(t)|\,{\rm d}t}
{\tau(t)-z(s)}\right\}=
2\lambda\Re\left\{\bar{e}n_{z(s)}\right\}\,,\quad s\in[0,1]\,,
\label{eq:inteq2b}
\end{equation}
where $\dot{\tau}(t)={\rm d}\tau(t)/{\rm d}t$. Then Nyström
discretization with $n_{\rm p}$ points $z_i$ and weights $w_i$ on
$\Gamma$ gives
\begin{equation}
\rho_i+\frac{\lambda}{\pi}\sum_{j=1}^{n_{\rm p}}\rho_j
\Re\left\{\frac{n_i|\dot{z}_j|w_j}
{z_j-z_i}\right\}=
2\lambda\Re\left\{\bar{e}n_i\right\}\,,\quad i=1,2,\ldots,n_{\rm p}\,.
\label{eq:inteq2c}
\end{equation}

\begin{figure}
\centering 
\noindent\makebox[\textwidth]{
\begin{minipage}{1.1\textwidth}
\includegraphics[height=66mm]{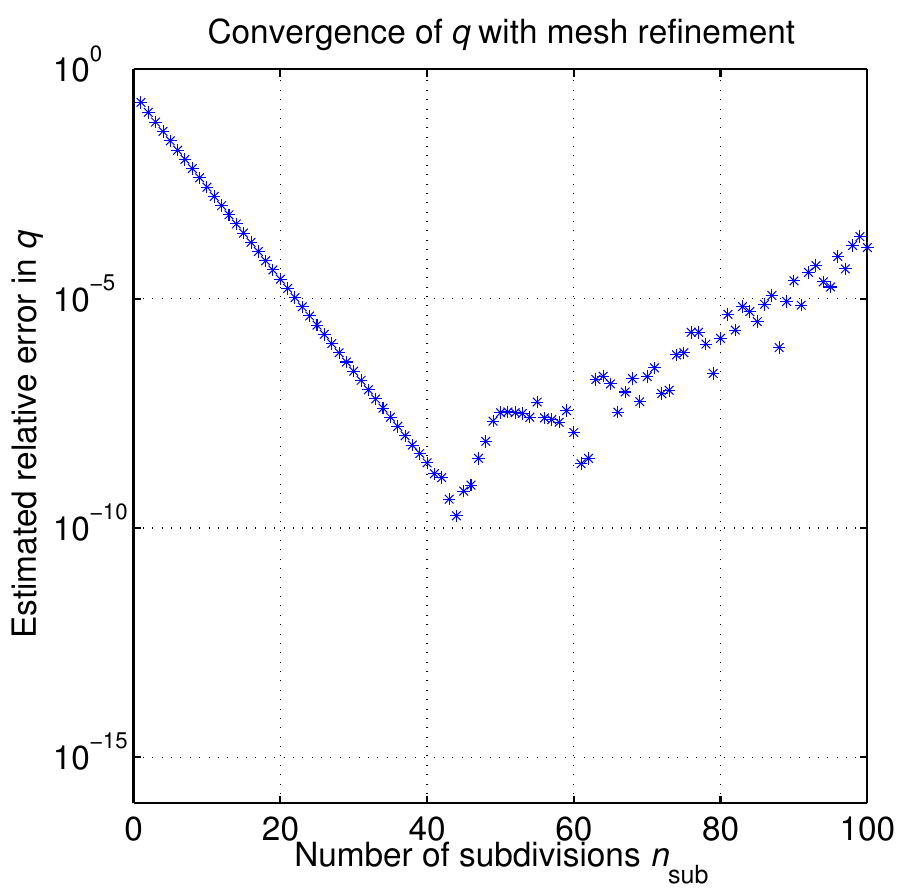}
\includegraphics[height=66mm]{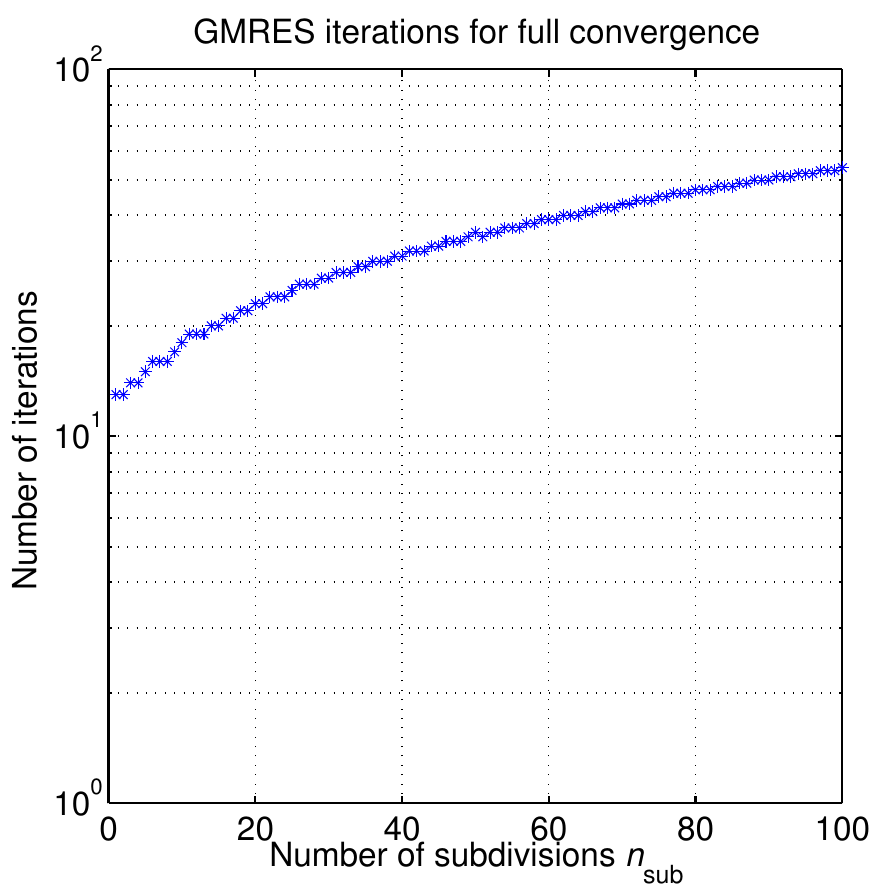}
\end{minipage}}
\caption{\sf Convergence for $q$ of~(\ref{eq:q})
  using~(\ref{eq:inteq3b}) and the program {\tt demo1b.m} (a loop of
  {\tt demo1.m}) with {\tt lambda=0.999}, {\tt theta=pi/2}, {\tt
    npan=10}, and {\tt evec=1}. The reference value is
  $q=1.1300163213105365$. There are $n_{\rm p}=160+32n_{\rm sub}$
  unknowns in the main linear system. Left: Convergence with $n_{\rm
    sub}$. Right: The number of iterations needed to meet an estimated
  relative residual of $\epsilon_{\rm mach}$.}
\label{fig:conv1}
\end{figure}

The program {\tt demo1.m} sets up the system~(\ref{eq:inteq3b}),
solves it using the GMRES iterative solver~\cite{Saad86} incorporating
a low-threshold stagnation avoiding technique~\cite[Section
8]{Hels08a}, and computes $q$ of~(\ref{eq:q}). The user has to specify
the opening angle $\theta$, the parameter $\lambda$, the number
$n_{\rm pan}$ of coarse panels on $\Gamma$, the unit vector $e$ and
the number of subdivisions $n_{\rm sub}$. The opening angle should be
in the interval $\pi/3\le\theta\le 5\pi/3$. We choose $\lambda=0.999$,
$\theta=\pi/2$, $n_{\rm pan}=10$, and $e=(1,0)$. The quantity $q$
converges initially as $n_{\rm sub}$ is increased, but for $n_{\rm
  sub}>44$ the results start to get worse. See Figure~\ref{fig:conv1}.
This is related to the fact, pointed out by Bremer~\cite{Brem12b},
that standard Nyström discretization captures the $L^{\infty}$
behavior of the solution $\rho$, while our $\rho$ is unbounded. See,
further, Appendix E.

\section{Compressed inverse preconditioning}
\label{sec:comp}

Let us split the matrices ${\bf K}_{\rm coa}$ and ${\bf K}_{\rm fin}$
of~(\ref{eq:inteq3a}) and~(\ref{eq:inteq3b}) into two parts each
\begin{align}
{\bf K}_{\rm coa}&={\bf K}_{\rm coa}^\star
                         +{\bf K}_{\rm coa}^\circ\,,
\label{eq:split1}\\
{\bf K}_{\rm fin}&={\bf K}_{\rm fin}^\star
                         +{\bf K}_{\rm fin}^\circ\,.
\label{eq:split2}
\end{align}
Here the superscript $\star$ indicates that only entries of a matrix
$K_{ij}$ whose indices $i$ and $j$ correspond to points $z_i$ and
$z_j$ that both belong to the boundary subset $\Gamma^\star$ are
retained. The remaining entries are zero.

Now we introduce two diagonal matrices ${\bf W}_{\rm coa}$ and ${\bf
  W}_{\rm fin}$ which have the quadrature weights $w_i$ on the
diagonal. Furthermore, we need a prolongation matrix ${\bf P}$ which
interpolates functions known at points on the coarse grid to points on
the fine grid. The construction of ${\bf P}$ relies on panelwise
15-degree polynomial interpolation in parameter using Vandermonde
matrices. We also construct a weighted prolongation matrix ${\bf P}_W$
via
\begin{equation}
{\bf P}_W={\bf W}_{\rm fin}{\bf P}{\bf W}_{\rm coa}^{-1}\,.
\label{eq:PW}
\end{equation}
The matrices ${\bf P}$ and ${\bf P}_W$ share the same sparsity
pattern. They are rectangular matrices, similar to the identity
matrix, but with one full $(4+2n_{\rm sub})16\times 64$ block. Let
superscript {\footnotesize $T$} denote the transpose. Then
\begin{equation}
{\bf P}_W^T{\bf P}={\bf I}_{\rm coa}
\label{eq:PWTP}
\end{equation}
holds exactly. See Appendix A and~\cite[Section 4.3]{Hels11SISC}.

Equipped with ${\bf P}$ and ${\bf P}_W$ we are ready to
compress~(\ref{eq:inteq3b}) on the fine grid to an equation
essentially on the coarse grid. This compression is done without the
loss of accuracy -- the discretization error in the solution is
unaffected and no information is lost. The compression relies on the
variable substitution
\begin{equation}
\left({\bf I}_{\rm fin}+\lambda{\bf K}_{\rm fin}^\star\right)
\boldsymbol{\rho}_{\rm fin}={\bf P}\tilde{\boldsymbol{\rho}}_{\rm coa}\,.
\label{eq:subst}
\end{equation}
Here $\tilde{\boldsymbol{\rho}}_{\rm coa}$ is the discretization of a
piecewise smooth {\it transformed density}. The compression also uses
the low-rank decomposition
\begin{equation}
{\bf K}_{\rm fin}^\circ={\bf P}{\bf K}_{\rm coa}^\circ{\bf P}_W^T\,,
\label{eq:decomp}
\end{equation}
which should hold to about machine precision. 

The compressed version of~(\ref{eq:inteq3b}) reads
\begin{equation}
\left({\bf I}_{\rm coa}+\lambda{\bf K}_{\rm coa}^\circ{\bf R}\right)
\tilde{\boldsymbol{\rho}}_{\rm coa}=\lambda{\bf g}_{\rm coa}\,,
\label{eq:inteq4}
\end{equation}
where the compressed weighted inverse ${\bf R}$ is given by
\begin{equation}
{\bf R}=
{\bf P}_W^T\left({\bf I}_{\rm fin}+\lambda{\bf K}_{\rm fin}^\star\right)^{-1}
{\bf P}\,.
\label{eq:R}
\end{equation}
See Appendix B for details on the derivation. The compressed weighted
inverse ${\bf R}$, for $\Gamma$ of~(\ref{eq:gamma}), is a block
diagonal matrix with one full $64\times 64$ block and the remaining
entries coinciding with those of the identity matrix.

After having solved~(\ref{eq:inteq4}) for
$\tilde{\boldsymbol{\rho}}_{\rm coa}$, the density
$\boldsymbol{\rho}_{\rm fin}$ can easily be reconstructed from
$\tilde{\boldsymbol{\rho}}_{\rm coa}$ in a post-processor, see
Section~\ref{sec:recon}. It is important to observe, however, that
$\boldsymbol{\rho}_{\rm fin}$ is not always needed. For example, the
quantity $q$ of~(\ref{eq:q}) can be computed directly from
$\tilde{\boldsymbol{\rho}}_{\rm coa}$. Let $\boldsymbol{\zeta}_{\rm
  coa}$ be a column vector which contains values of $|\dot{z}_i|$
multiplied with $\Re\left\{\bar{e}z_i\right\}$. Then
\begin{equation}
q=\Re\left\{\boldsymbol{\zeta}_{\rm coa}\right\}^T{\bf W}_{\rm coa}
{\bf R}\tilde{\boldsymbol{\rho}}_{\rm coa}\,.
\end{equation}
See Appendix~C for a proof.

\begin{figure}
\centering 
\centerline{
  \includegraphics[height=36mm, trim=0mm -12mm 0mm 12mm]{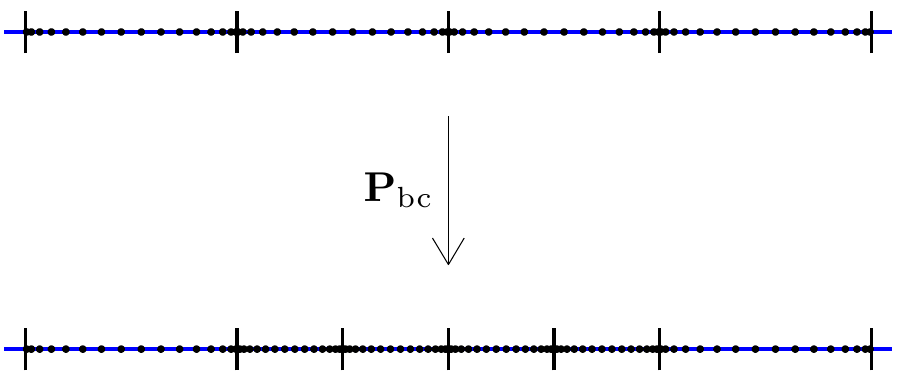}
\hspace{3mm}
  \includegraphics[height=52mm, trim=0mm 10mm 0mm -10mm]{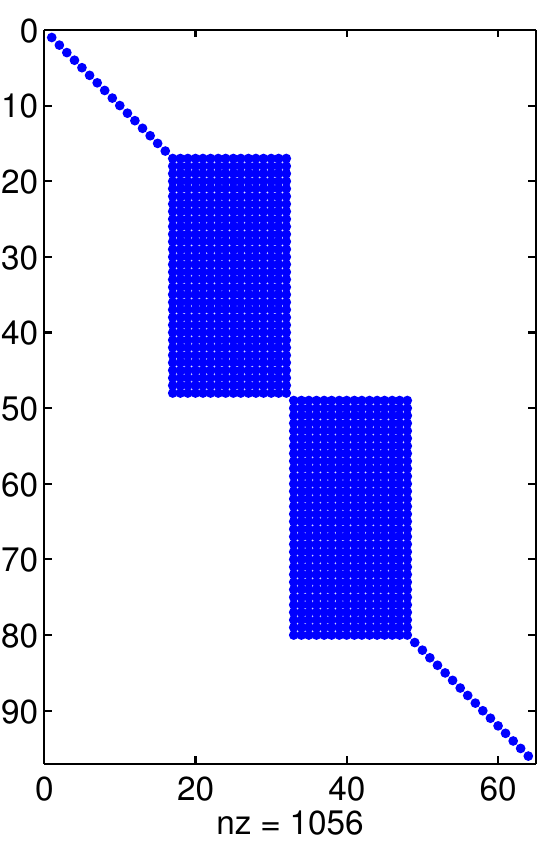}
}
\caption{\sf Left: The prolongation operator ${\bf P}_{\rm bc}$ performs 
  panelwise interpolation from a grid on a four-panel mesh to a grid
  on a six-panel mesh. Right: The sparsity pattern of ${\bf P}_{\rm
    bc}$.}
\label{fig:Pbc}
\end{figure}

\section{The recursion for ${\bf R}$}
\label{sec:rec}

The compressed weighted inverse ${\bf R}$ is costly to compute from
its definition~(\ref{eq:R}). As we saw in Section~\ref{sec:disc}, the
inversion of large matrices $({\bf I}+{\bf K})$ on highly refined
grids could also be unstable. Fortunately, the computation of ${\bf
  R}$ can be greatly sped up and stabilized via a recursion.
In~\cite[Section 7.2]{Hels08b} this recursion is derived in a
roundabout way and uses a refined grid that differs from that of the
present tutorial. A better derivation can be found in~\cite[Section
5]{Hels11SISC}, but there the setting is more general so that text
could be hard to follow. Here we focus on results.

\subsection{Basic prolongation matrices}
\label{sec:basepro}

Let ${\bf P}_{\rm bc}$ be a prolongation matrix, performing panelwise
15-degree polynomial interpolation in parameter from a 64-point grid
on a four-panel mesh to a 96-point grid on a six-panel mesh as shown
in Figure~\ref{fig:Pbc}. Let ${\bf P}_{W{\rm bc}}$ be a weighted
prolongation matrix in the style of~(\ref{eq:PW}). If {\tt T16} and
{\tt W16} are the nodes and weights of 16-point Gauss--Legendre
quadrature on the canonical interval $[-1,1]$, then ${\bf P}_{\rm bc}$
and ${\bf P}_{W{\rm bc}}$ can be constructed as
\begin{verbatim}
  T32=[T16-1;T16+1]/2;
  W32=[W16;W16]/2;
  A=ones(16);
  AA=ones(32,16);
  for k=2:16
    A(:,k)=A(:,k-1).*T16;
    AA(:,k)=AA(:,k-1).*T32;         
  end
  IP=AA/A;
  IPW=IP.*(W32*(1./W16)');
%
  Pbc =blkdiag(eye(16),IP ,IP ,eye(16));
  PWbc=blkdiag(eye(16),IPW,IPW,eye(16));
\end{verbatim}
See~\cite[Appendix A]{Hels08a} for an explanation of why high-degree
polynomial interpolation involving ill-conditioned Vandermonde systems
gives accurate results for smooth functions.

\begin{figure}
\centering 
\includegraphics[height=40mm]{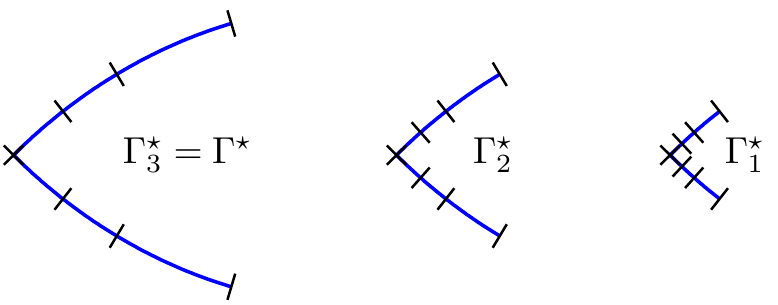}
\caption{\sf The boundary subsets $\Gamma_3^\star$,
  $\Gamma_2^\star$, and $\Gamma_1^\star$ along with their
  corresponding type {\sf b} meshes for $n_{\rm sub}=3$.}
\label{fig:subsets}
\end{figure}

\subsection{Discretization on nested meshes}
\label{sec:nest}

Let $\Gamma_i^\star$, $i=1,2,\ldots,n_{\rm sub}$, be a sequence of
subsets of $\Gamma^\star$ with
$\Gamma_{i-1}^\star\subset\Gamma_i^\star$ and $\Gamma_{n_{\rm
    sub}}^\star=\Gamma^\star$. Let there also be a six-panel mesh and
a corresponding 96-point grid on each $\Gamma_i^\star$. The
construction of the subsets and their meshes should be such that if
$z(s)$, $s\in[-2,2]$, is a local parameterization of $\Gamma_i^\star$,
then the breakpoints (locations of panel endpoints) of its mesh are at
$s\in\{-2,-1,-0.5,0,0.5,1,2\}$ and the breakpoints of the mesh on
$\Gamma_{i-1}^\star$ are at $s=\{-1,-0.5,-0.25,0,-0.25,0.5,1\}$. We
denote this type of nested six-panel meshes {\it type {\sf b}}. The
index $i$ is the {\it level}. An example of a sequence of subsets and
type {\sf b} meshes on $\Gamma^\star$ is shown in
Figure~\ref{fig:subsets} for $n_{\rm sub}=3$. Compare~\cite[Figure
2]{Hels09JCP} and~\cite[Figure 5.1]{Hels11SISC}.

Let ${\bf K}_{i{\rm b}}$ denote the discretization of $K$ on a type
{\sf b} mesh on $\Gamma_i^\star$. In the spirit
of~(\ref{eq:split1},\ref{eq:split2}) we write
\begin{equation}
{\bf K}_{i{\rm b}}={\bf K}_{i{\rm b}}^\star+{\bf K}_{i{\rm b}}^\circ\,,
\end{equation}
where the superscript $\star$ indicates that only entries with both
indices corresponding to points on the four inner panels are retained.

\subsection{The recursion proper}

Now, let ${\bf R}_{n_{\rm sub}}$ denote the full $64\times 64$
diagonal block of ${\bf R}$. The recursion for ${\bf R}_{n_{\rm sub}}$
is derived in Appendix~D and it reads
\begin{align}
&{\bf R}_i={\bf P}^T_{W\rm{bc}}
\left(
\mathbb{F}\{{\bf R}_{i-1}^{-1}\}+{\bf I}_{\rm b}^\circ+\lambda{\bf K}_{i{\rm
    b}}^\circ
\right)^{-1}{\bf P}_{\rm{bc}}\,,\quad i=1,\ldots,n_{\rm sub}\,,
\label{eq:recur}\\
&\mathbb{F}\{{\bf R}_0^{-1}\}=
{\bf I}_{\rm b}^\star+\lambda{\bf K}^\star_{1{\rm b}}\,,
\label{eq:rstart}
\end{align}
where the operator $\mathbb{F}\{\cdot\}$ expands its matrix argument
by zero-padding (adding a frame of zeros of width 16 around it). Note
that the initializer ${\bf R}_0$ of~(\ref{eq:rstart}) makes the
recursion~(\ref{eq:recur}) take the first step
\begin{displaymath}
{\bf R}_1={\bf P}^T_{W\rm{bc}} 
 \left({\bf I}_{\rm b}+\lambda{\bf K}_{1{\rm b}}\right)^{-1}
{\bf P}_{\rm{bc}}\,.
\end{displaymath}

The program {\tt demo2.m} sets up the linear system~(\ref{eq:inteq4}),
runs the recursion~(\ref{eq:recur},\ref{eq:rstart}), and solves the
linear system using the same techniques as {\tt demo1.m}, see
Section~\ref{sec:disc}. In fact, the results produced by the two
programs are very similar, at least up to $n_{\rm sub}=40$. This
supports the claim of Section~\ref{sec:comp} that the discretization
error in the solution is unaffected by compression.

\begin{figure}
\centering 
\noindent\makebox[\textwidth]{
\begin{minipage}{1.1\textwidth}
\includegraphics[height=66mm]{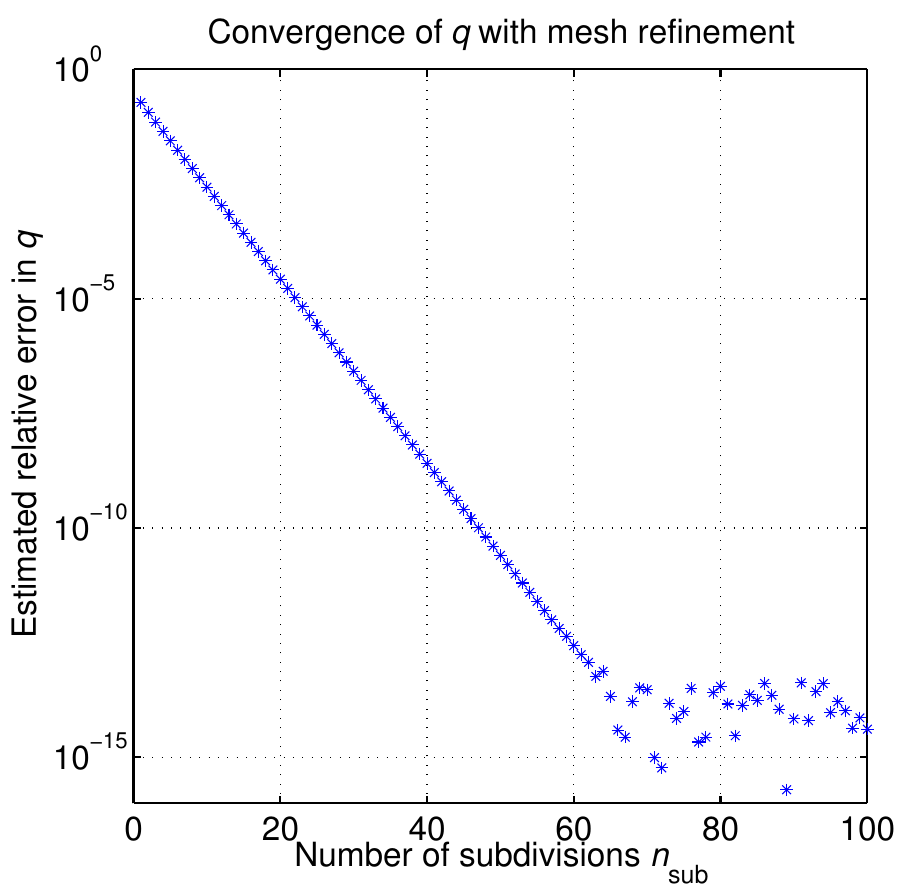}
\includegraphics[height=66mm]{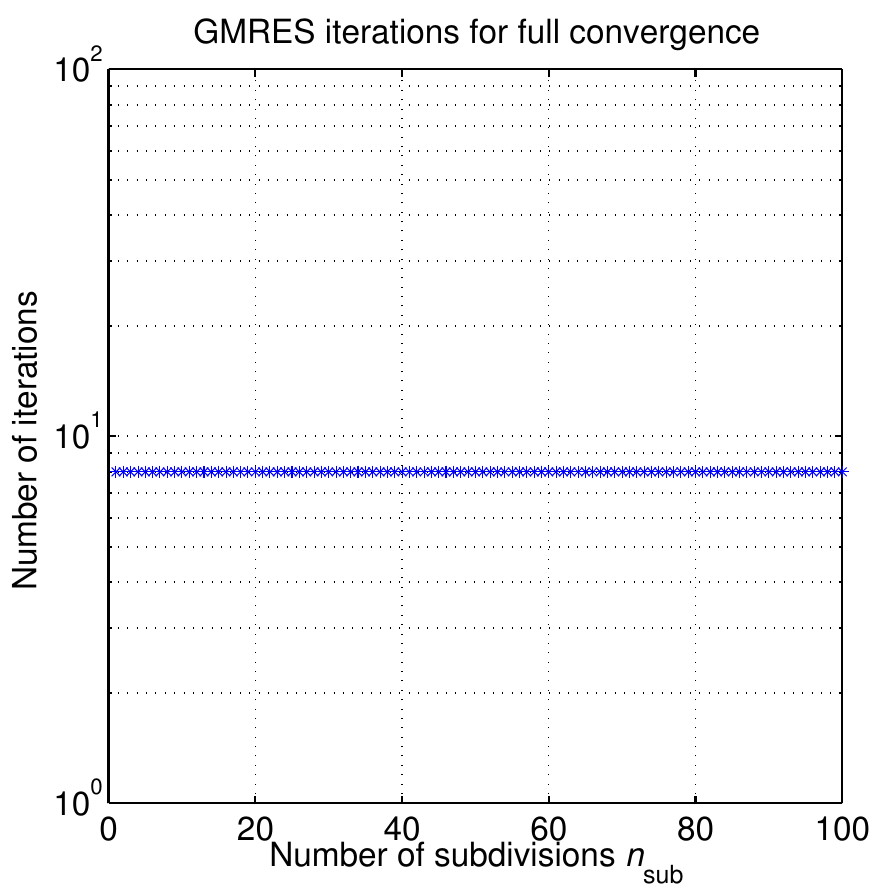}
\end{minipage}}
\caption{\sf Same as Figure~\ref{fig:conv1}, but using~(\ref{eq:inteq4})
  and the program {\tt demo2b.m} (a loop of {\tt demo2.m}). There are
  only $n_{\rm p}=160$ unknowns in the main linear system.}
\label{fig:conv2}
\end{figure}

Figure~\ref{fig:conv2} demonstrates the power of RCIP: fewer unknowns
and faster execution, better conditioning (the number of GMRES
iterations does not grow), and higher achievable accuracy. Compare
Figure~\ref{fig:conv1}. We emphasize that the number $n_{\rm sub}$ of
recursion steps (levels) used in~(\ref{eq:recur}) corresponds to the
number of subdivisions $n_{\rm sub}$ used to construct the fine mesh.

\section{Schur--Banachiewicz speedup of the recursion}
\label{sec:SB}

The recursion~(\ref{eq:recur}) can be sped up using the
Schur--Banachiewicz inverse formula for partitioned
matrices~\cite{Hend81}, which in this context can be
written~\cite[Appendix B]{Hels09IJSS}
\begin{displaymath}
\begin{bmatrix}
     {\bf P}_W^{\star T} & {\bf 0} \\
     {\bf 0}   & {\bf I}
\end{bmatrix} 
\begin{bmatrix}  
     {\bf A}^{-1} & {\bf U} \\
     {\bf V}      & {\bf D}
\end{bmatrix}^{-1}
\begin{bmatrix}
     {\bf P}^\star   & {\bf 0} \\
     {\bf 0}   & {\bf I}
\end{bmatrix} 
=
\end{displaymath}
\begin{equation}
\begin{bmatrix}
{\bf P}_W^{\star T}{\bf A}{\bf P}^\star+
{\bf P}_W^{\star T}{\bf AU}
({\bf D}-{\bf VAU})^{-1}{\bf VA}{\bf P}^\star & 
-{\bf P}_W^{\star T}{\bf AU}({\bf D}-{\bf VAU})^{-1} \\
-({\bf D}-{\bf VAU})^{-1}{\bf VA}{\bf P}^\star & 
({\bf D}-{\bf VAU})^{-1}
\end{bmatrix}\,,
\label{eq:cbana}
\end{equation}
where ${\bf A}$ plays the role of ${\bf R}_{i-1}$, ${\bf P}^\star$ and
${\bf P}_W^\star$ are submatrices of ${\bf P}_{\rm bc}$ and ${\bf
  P}_{W{\rm bc}}$, and ${\bf U}$, ${\bf V}$, and ${\bf D}$ refer to
blocks of ${\bf I}_{\rm b}^\circ+\lambda{\bf K}_{i{\rm b}}^\circ$.

The program {\tt demo3.m} is based on {\tt demo2.m}, but
has~(\ref{eq:cbana}) incorporated. Besides, the integral
equation~(\ref{eq:inteq1a}) is replaced with
\begin{equation}
\rho(r)+2\lambda\int_{\Gamma}\frac{\partial G}{\partial\nu}(r,r')\rho(r')
\,{\rm d}\ell'+\int_{\Gamma}\rho(r')\,{\rm d}\ell'=
2\lambda\left(e\cdot\nu\right)\,,
\quad r\in \Gamma\,,
\label{eq:inteq1b}
\end{equation}
which has the same solution $\rho(r)$ but is more stable for $\lambda$
close to one. For the discretization of~(\ref{eq:inteq1b}) to fit the
form~(\ref{eq:inteq4}), the last term on the left-hand side
of~(\ref{eq:inteq1b}) is added to the matrix $\lambda{\bf K}_{\rm
  coa}^\circ$ of~(\ref{eq:inteq4}).

\begin{figure}
\centering 
\noindent\makebox[\textwidth]{
\begin{minipage}{1.1\textwidth}
\includegraphics[height=66mm]{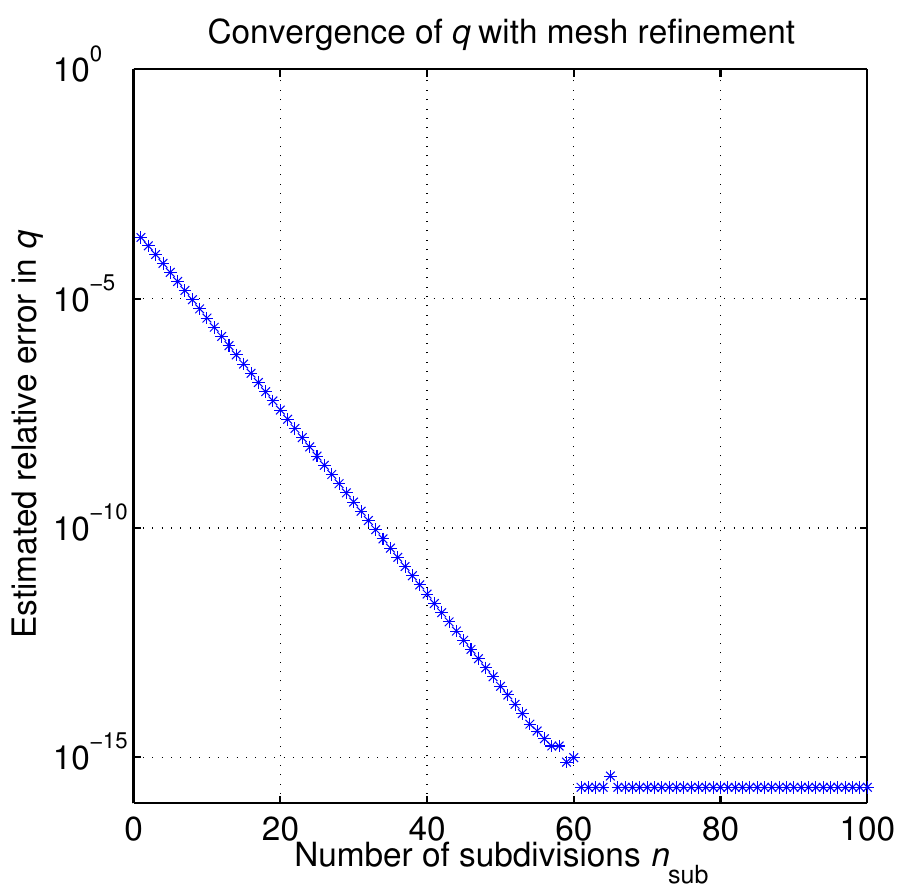}
\includegraphics[height=66mm]{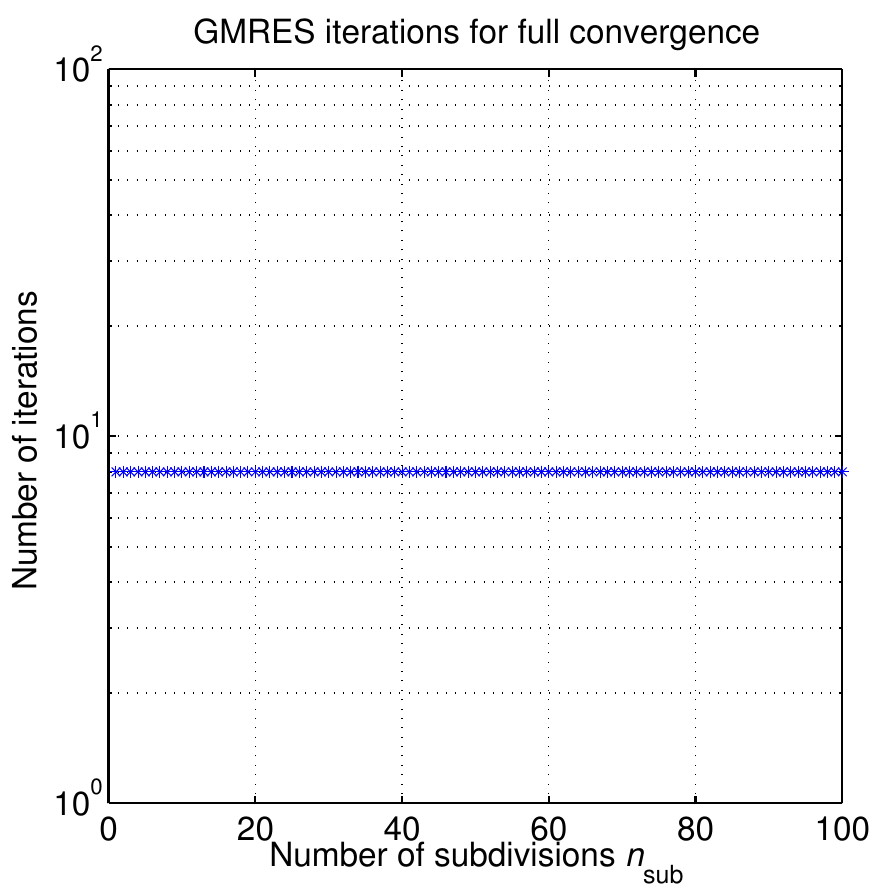}
\end{minipage}}
\caption{\sf Same as Figure~\ref{fig:conv2}, but the program {\tt
    demo3b.m} is used.}
\label{fig:conv3}
\end{figure}

The execution of {\tt demo3.m} is faster than that of {\tt demo2.m}.
Figure~\ref{fig:conv3} shows that a couple of extra digits are gained
by using~(\ref{eq:inteq1b}) rather than~(\ref{eq:inteq1a}) and that
full machine accuracy is achieved for $n_{\rm sub}>60$.

\section{Various useful quantities}
\label{sec:useful}

Let us introduce a new discrete density $\hat{\boldsymbol{\rho}}_{\rm
  coa}$ via
\begin{equation}
\hat{\boldsymbol{\rho}}_{\rm coa}={\bf R}\tilde{\boldsymbol{\rho}}_{\rm coa}.
\label{eq:wcd}
\end{equation}
Rewriting~(\ref{eq:inteq4}) in terms of $\hat{\boldsymbol{\rho}}_{\rm
  coa}$ gives
\begin{equation}
\left({\bf R}^{-1}+\lambda{\bf K}_{\rm coa}^\circ\right)
\hat{\boldsymbol{\rho}}_{\rm coa}=\lambda{\bf g}_{\rm coa}\,,
\label{eq:inteq5}
\end{equation}
which resembles the original equation~(\ref{eq:inteq3a}). We see that
${\bf K}_{\rm coa}^\circ$, which is discretized using Gauss--Legendre
quadrature, acts on $\hat{\boldsymbol{\rho}}_{\rm coa}$. Therefore one
can interpret $\hat{\boldsymbol{\rho}}_{\rm coa}$ as pointwise values
of the original density $\rho(r)$, multiplied with weight corrections
suitable for integration against polynomials. We refer to
$\hat{\boldsymbol{\rho}}_{\rm coa}$ as a {\it weight-corrected
  density}. See, further, Appendix~C.

Assume now that there is a square matrix ${\bf S}$ which maps
$\tilde{\boldsymbol{\rho}}_{\rm coa}$ to discrete values
$\boldsymbol{\rho}_{\rm coa}$ of the original density on the coarse
grid
\begin{equation}
\boldsymbol{\rho}_{\rm coa}={\bf S}\tilde{\boldsymbol{\rho}}_{\rm coa}\,.
\label{eq:Smap}
\end{equation}
The matrix ${\bf S}$ allows us to rewrite~(\ref{eq:inteq4}) as a
system for the original density
\begin{equation}
\left({\bf S}^{-1}+\lambda{\bf K}_{\rm coa}^\circ{\bf R}{\bf S}^{-1}\right)
\boldsymbol{\rho}_{\rm coa}=\lambda{\bf g}_{\rm coa}\,.
\label{eq:inteq6}
\end{equation}
We can interpret the composition ${\bf R}{\bf S}^{-1}$ as a matrix of
multiplicative weight corrections that compensate for the singular
behavior of $\rho(r)$ on $\Gamma^\star$ when Gauss--Legendre
quadrature is used. 

Let ${\bf Y}$ denote the rectangular matrix
\begin{equation}
{\bf Y}=\left({\bf I}_{\rm fin}+\lambda{\bf K}_{\rm fin}^\star\right)^{-1}
{\bf P}\,,
\end{equation}
and let ${\bf Q}$ be a restriction operator which performs panelwise
15-degree polynomial interpolation in parameter from a grid on the
fine mesh to a grid on a the coarse mesh. We see from~(\ref{eq:subst})
that ${\bf Y}$ is the mapping from $\tilde{\boldsymbol{\rho}}_{\rm
  coa}$ to $\boldsymbol{\rho}_{\rm fin}$. Therefore the columns of
${\bf Y}$ can be interpreted as {\it discrete basis functions} for
$\rho(r)$. It holds by definition that
\begin{align}
{\bf Q}{\bf P}&={\bf I}_{\rm coa}\,,
\label{eq:QP}\\
{\bf Q}{\bf Y}&={\bf S}\,.
\label{eq:QY}
\end{align}

The quantities and interpretations of this section come in handy in
various situations, for example in 3D extensions of the RCIP
method~\cite{HelsP12}. An efficient scheme for constructing ${\bf S}$
will be presented in Section~\ref{sec:S}.

\section{Reconstruction of $\boldsymbol{\rho}_{\rm fin}$ from 
  $\tilde{\boldsymbol{\rho}}_{\rm coa}$}
\label{sec:recon}

The action of ${\bf Y}$ on $\tilde{\boldsymbol{\rho}}_{\rm coa}$,
which gives $\boldsymbol{\rho}_{\rm fin}$, can be obtained by, in a
sense, running the recursion~(\ref{eq:recur}) backwards. The process
is described in detail in~\cite[Section 7]{Hels09JCP}. Here we focus
on results.

The backward recursion on $\Gamma^\star$ reads
\begin{equation}
\vec{\boldsymbol{\rho}}_{{\rm coa},i}=
\left[{\bf I}_{\rm b}-\lambda{\bf K}_{i{\rm b}}^\circ
\left(\mathbb{F}\{{\bf R}_{i-1}^{-1}\}+
{\bf I}_{\rm b}^\circ+\lambda{\bf K}_{i{\rm b}}^\circ
\right)^{-1}\right]{\bf P}_{\rm bc}\tilde{\boldsymbol{\rho}}_{{\rm coa},i}\,,
\quad i=n_{\rm sub},\ldots,1\,.
\label{eq:back}
\end{equation}
Here $\tilde{\boldsymbol{\rho}}_{{\rm coa},i}$ is a column vector with
$64$ elements. In particular, $\tilde{\boldsymbol{\rho}}_{{\rm
    coa},n_{\rm sub}}$ is the restriction of
$\tilde{\boldsymbol{\rho}}_{\rm coa}$ to $\Gamma^\star$, while
$\tilde{\boldsymbol{\rho}}_{{\rm coa},i}$ are taken as elements
$\{17:80\}$ of $\vec{\boldsymbol{\rho}}_{{\rm coa},i+1}$ for $i<n_{\rm
  sub}$. The elements $\{1:16\}$ and $\{81:96\}$ of
$\vec{\boldsymbol{\rho}}_{{\rm coa},i}$ are the reconstructed values
of $\boldsymbol{\rho}_{\rm fin}$ on the outermost panels of a type
{\sf b} mesh on $\Gamma_i^\star$. Outside of $\Gamma^\star$,
$\boldsymbol{\rho}_{\rm fin}$ coincides with
$\tilde{\boldsymbol{\rho}}_{\rm coa}$.

When the recursion is completed, the reconstructed values of
$\boldsymbol{\rho}_{\rm fin}$ on the four innermost panels are
obtained from
\begin{equation}
{\bf R}_0\tilde{\boldsymbol{\rho}}_{{\rm coa},0}\,.
\label{eq:recend}
\end{equation}
Should one wish to interrupt the recursion~(\ref{eq:back})
prematurely, at step $i=j$ say, then
\begin{equation}
{\bf R}_{j-1}\tilde{\boldsymbol{\rho}}_{{\rm coa},(j-1)}\,
\label{eq:premat}
\end{equation}
gives values of a weight-corrected density on the four innermost
panels of a type {\sf b} mesh on $\Gamma^\star_j$. That is, we have
a part-way reconstructed weight-corrected density
$\hat{\boldsymbol{\rho}}_{\rm part}$ on a mesh that is $n_{\rm
  sub}-j+1$ times refined. This observation is useful in the context
of evaluating layer potentials close to their sources.

If the memory permits, one can store the matrices ${\bf K}_{i{\rm
    b}}^\circ$ and ${\bf R}_i$ in the forward
recursion~(\ref{eq:recur}) and reuse them in the backward
recursion~(\ref{eq:back}). Otherwise they may be computed afresh.

\begin{figure}
\centering 
\noindent\makebox[\textwidth]{
\begin{minipage}{1.1\textwidth}
\includegraphics[height=66mm]{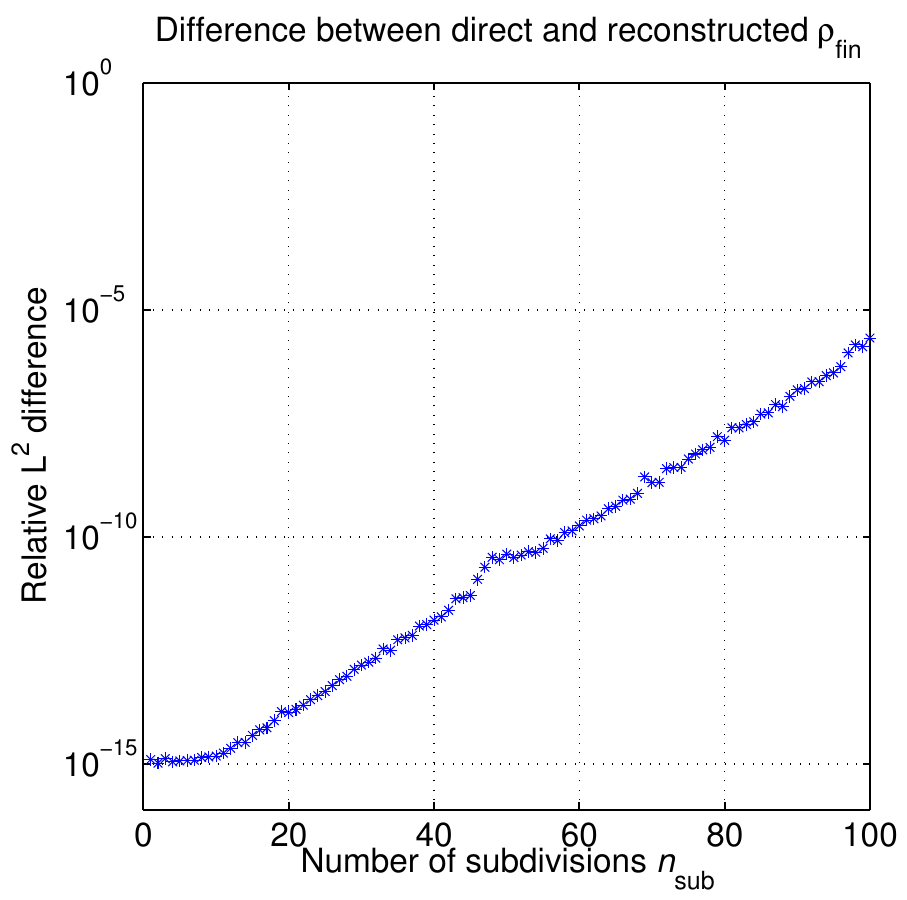}
\includegraphics[height=66mm]{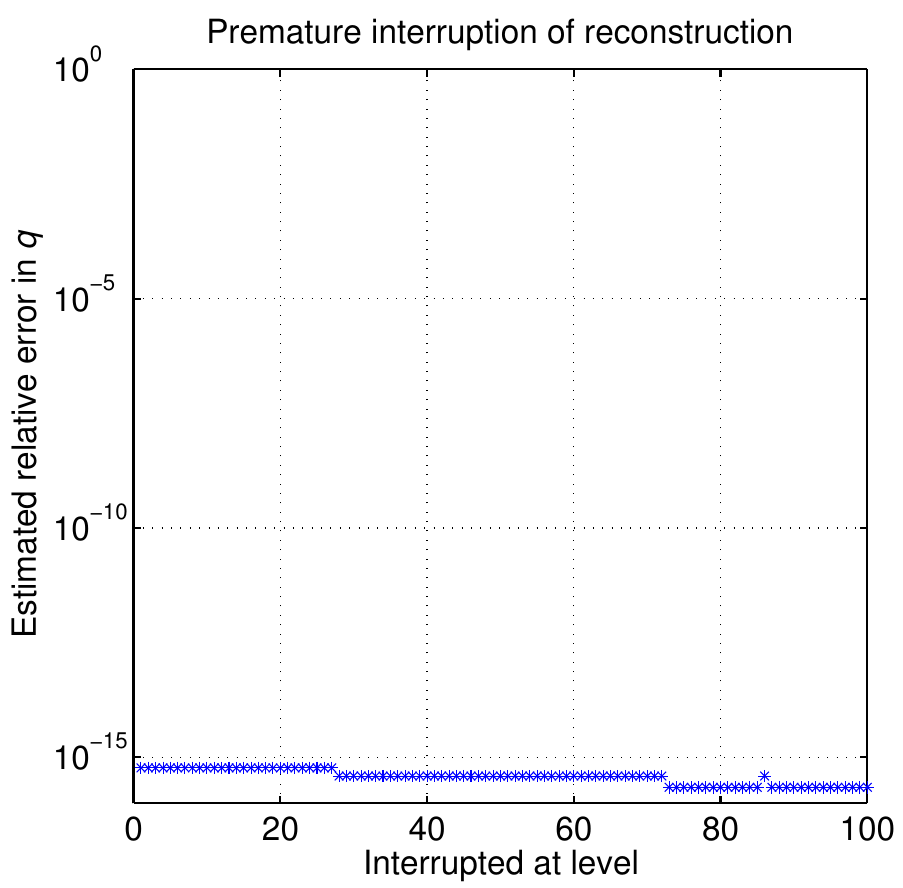}
\end{minipage}}
\caption{\sf Output from {\tt demo4.m} and {\tt demo5.m}. 
  Left: A comparison of $\boldsymbol{\rho}_{\rm fin}$ from the
  unstable equation~(\ref{eq:inteq3b}) and $\boldsymbol{\rho}_{\rm
    fin}$ reconstructed from $\tilde{\boldsymbol{\rho}}_{\rm coa}$
  of~(\ref{eq:inteq4}) via~(\ref{eq:back}). Right: Relative accuracy
  in $q$ of~(\ref{eq:q}) from part-way reconstructed solutions
  $\hat{\boldsymbol{\rho}}_{\rm part}$.}
\label{fig:conv45}
\end{figure}

The program {\tt demo4.m} builds on the program {\tt demo3.m},
using~(\ref{eq:inteq4}) for~(\ref{eq:inteq1b}). After the main linear
system is solved for $\tilde{\boldsymbol{\rho}}_{\rm coa}$, a
postprocessor reconstructs $\boldsymbol{\rho}_{\rm fin}$
via~(\ref{eq:back}). Then a comparison is made with a solution
$\boldsymbol{\rho}_{\rm fin}$ obtained by solving the un-compressed
system~(\ref{eq:inteq3b}). Figure~\ref{fig:conv45} shows that for
$n_{\rm sub}<10$ the results are virtually identical. This verifies
the correctness of~(\ref{eq:back}). For $n_{\rm sub}>10$ the result
start to deviate. That illustrates the instabilities associated with
solving~(\ref{eq:inteq3b}) on a highly refined mesh. Compare
Figure~\ref{fig:conv1}.

The program {\tt demo5.m} investigates the effects of premature
interruption of~(\ref{eq:back}). The number of recursion steps is set
to $n_{\rm sub}=100$ and the recursion is interrupted at different
levels. The density $\boldsymbol{\rho}_{\rm fin}$ is reconstructed on
outer panels down to the level of interruption. Then a
weight-corrected density is produced at the innermost four panels
according to~(\ref{eq:premat}). Finally $q$ of~(\ref{eq:q}) is
computed from this part-way reconstructed solution. The right image of
Figure~\ref{fig:conv45} shows that the quality of $q$ is unaffected by
the level of interruption.

\section{The construction of ${\bf S}$}
\label{sec:S}

This section discusses the construction of ${\bf S}$ and other
auxiliary matrices. Note that in many applications, these matrices are
not needed.

The entries of the matrices ${\bf P}$, ${\bf P}_W$, ${\bf Q}$, ${\bf
  R}$, ${\bf S}$, and ${\bf Y}$ can only differ from those of the
identity matrix when both indices correspond to discretization points
on $\Gamma^\star$. For example, the entries of ${\bf R}$ only differ
from the identity matrix for the $64\times 64$ block denoted ${\bf
  R}_{n_{\rm sub}}$ in~(\ref{eq:recur}). In accordance with this
notation we introduce ${\bf P}_{n_{\rm sub}}$, ${\bf P}_{Wn_{\rm
    sub}}$, ${\bf Q}_{n_{\rm sub}}$, ${\bf S}_{n_{\rm sub}}$ and ${\bf
  Y}_{n_{\rm sub}}$ for the restriction of ${\bf P}$, ${\bf P}_W$,
${\bf Q}$, ${\bf S}$ and ${\bf Y}$ to $\Gamma^\star$. In the codes
of this section we often use this restricted type of matrices, leaving
the identity part out.

We observe that ${\bf S}_{n_{\rm sub}}$ is a square $64\times 64$
matrix; ${\bf P}_{n_{\rm sub}}$, ${\bf P}_{Wn_{\rm sub}}$ and ${\bf
  Y}_{n_{\rm sub}}$ are rectangular $16(4+2n_{\rm sub})\times 64$
matrices; and ${\bf Q}_{n_{\rm sub}}$ is a rectangular $64\times
16(4+2n_{\rm sub})$ matrix. Furthermore, ${\bf Q}_{n_{\rm sub}}$ is
very sparse for large $n_{\rm sub}$. All columns of ${\bf Q}_{n_{\rm
    sub}}$ with column indices corresponding to points on panels that
result from more than eight subdivisions are identically zero.

\begin{figure}
\centering 
\noindent\makebox[\textwidth]{
\begin{minipage}{1.1\textwidth}
\includegraphics[height=66mm]{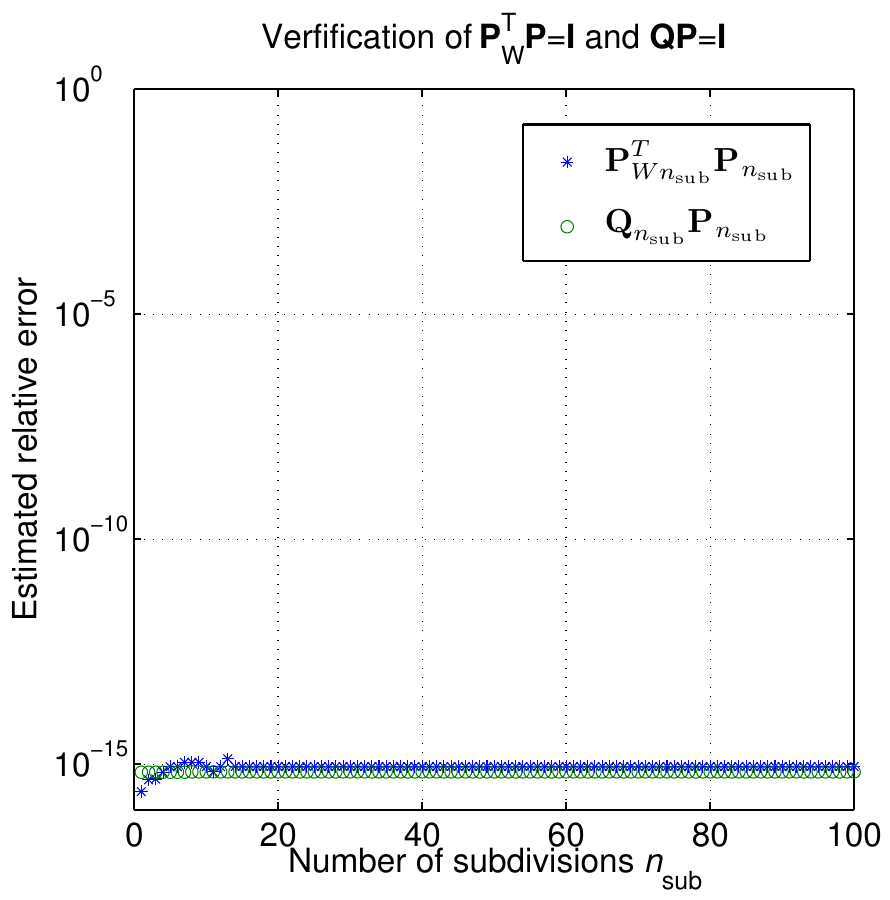}
\includegraphics[height=66mm]{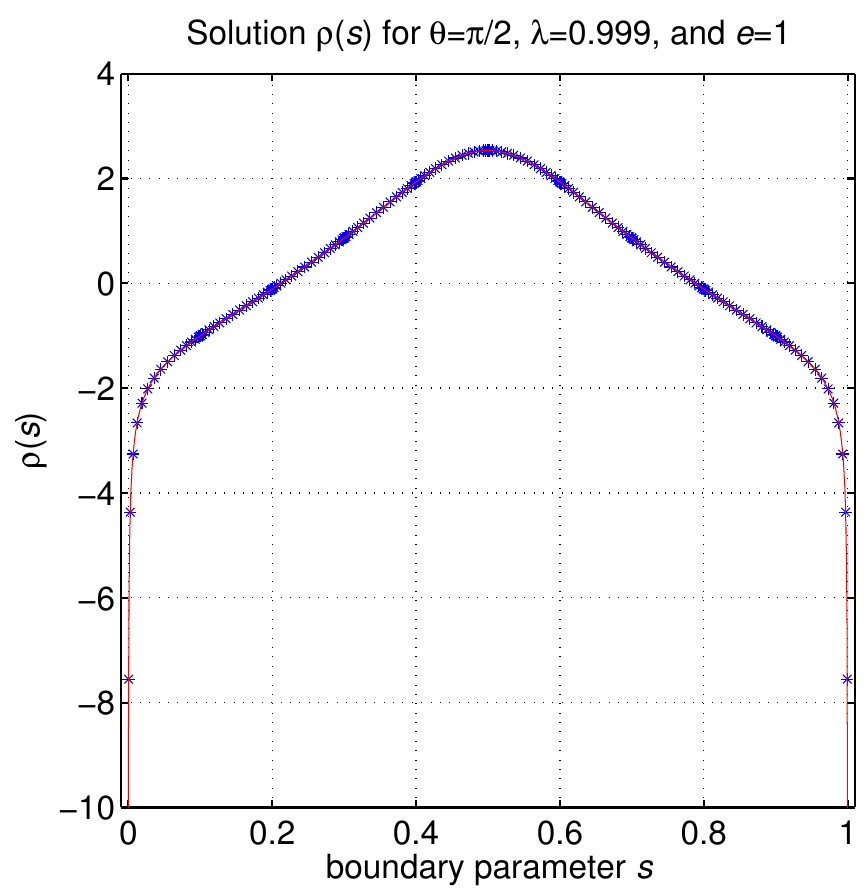}
\end{minipage}}
\caption{\sf Left: The identities~(\ref{eq:PWTP}) and~(\ref{eq:QP})
  hold to high accuracy in our implementation, irrespective of the
  degree of mesh refinement. Right: The solution $\rho$
  to~(\ref{eq:inteq1b}) on~(\ref{eq:gamma}) with parameters as
  specified in Section~\ref{sec:disc}. The solution with
  RCIP~(\ref{eq:inteq4}) and~(\ref{eq:Smap}), shown as blue stars,
  agrees with the solution from~(\ref{eq:inteq3b}), shown as a red
  solid line. The solution diverges in the corner.}
\label{fig:conv67}
\end{figure}

The program {\tt demo6.m} sets up ${\bf P}_{n_{\rm sub}}$, ${\bf
  P}_{Wn_{\rm sub}}$ and ${\bf Q}_{n_{\rm sub}}$, shows their sparsity
patterns, and verifies the identities~(\ref{eq:PWTP})
and~(\ref{eq:QP}). The implementations for ${\bf P}_{n_{\rm sub}}$ and
${\bf P}_{Wn_{\rm sub}}$ rely on repeated interpolation from coarser
to finer intermediate grids. The implementation of ${\bf Q}_{n_{\rm
    sub}}$ relies on keeping track of the relation between points on
the original coarse and fine grids. Output from {\tt demo6.m} is
depicted in the left image of Figure~\ref{fig:conv67}. Note that the
matrices ${\bf P}_{n_{\rm sub}}$ and ${\bf P}_{Wn_{\rm sub}}$ are
never needed in applications.

We are now ready to construct ${\bf S}$. Section~\ref{sec:recon}
presented a scheme for evaluating the action of ${\bf Y}_{n_{\rm
    sub}}$ on discrete functions on the coarse grid on $\Gamma^\star$.
The matrix ${\bf Y}_{n_{\rm sub}}$, itself, can be constructed by
applying this scheme to a $64\times 64$ identity matrix. The matrix
${\bf Q}_{n_{\rm sub}}$ was set up in {\tt demo6.m}. Composing these
two matrices gives ${\bf S}_{n_{\rm sub}}$, see~(\ref{eq:QY}). This is
done in the program {\tt demo7.m}, where the identity part is added as
to get the entire matrix ${\bf S}$. In previous work on RCIP we have
found use for ${\bf S}$ in complex situations where~(\ref{eq:inteq6})
is preferable over~(\ref{eq:inteq4}), see~\cite[Section 9]{HelsP12}.
If one merely needs $\boldsymbol{\rho}_{\rm coa}$ from
$\tilde{\boldsymbol{\rho}}_{\rm coa}$ in a post-processor, setting up
${\bf S}$ and using~(\ref{eq:Smap}) is not worthwhile. It is cheaper
to let ${\bf Y}$ act on $\tilde{\boldsymbol{\rho}}_{\rm coa}$ and then
let ${\bf Q}$ act on the resulting vector. Anyhow, {\tt demo7.m}
builds on {\tt demo4.m} and gives as output $\boldsymbol{\rho}_{\rm
  coa}$ computed via~(\ref{eq:Smap}), see the right image of
Figure~\ref{fig:conv67}. For comparison, $\boldsymbol{\rho}_{\rm
  fin}$, computed from~(\ref{eq:inteq3b}), is also shown.

\section{Initiating ${\bf R}$ using fixed-point iteration}
\label{sec:init}

It often happens that $\Gamma^\star_i$ is wedge-like. A corner of a
polygon, for example, has wedge-like $\Gamma^\star_i$ at all levels.
If $\Gamma^\star$ is merely piecewise smooth, then the
$\Gamma^\star_i$ are wedge-like to double precision accuracy for
$n_{\rm sub}-i\gg 1$.

Wedge-like sequences of $\Gamma^\star_i$ open up for simplifications
and speedup in the recursion~(\ref{eq:recur},\ref{eq:rstart}).
Particularly so if the kernel of the integral operator $K$
of~(\ref{eq:inteq3}) is scale invariant on wedges. Then the matrix
${\bf K}^\circ_{i{\rm b}}$ becomes independent of $i$. It can be
denoted by ${\bf K}^\circ_{\rm b}$ and needs only to be constructed
once. Furthermore, the recursion~(\ref{eq:recur},\ref{eq:rstart})
assumes the form of a fixed-point iteration
\begin{align}
&{\bf R}_i={\bf P}^T_{W\rm{bc}}
\left(
\mathbb{F}\{{\bf R}_{i-1}^{-1}\}+{\bf I}_{\rm b}^\circ+\lambda{\bf K}_{\rm
    b}^\circ
\right)^{-1}{\bf P}_{\rm{bc}}\,, \quad i=1,\ldots
\label{eq:fixed}\\
&\mathbb{F}\{{\bf R}_0^{-1}\}=
{\bf I}_{\rm b}^\star+\lambda{\bf K}^\star_{\rm b}\,.
\label{eq:fstart}
\end{align}
The iteration~(\ref{eq:fixed}) can be run until ${\bf R}_i$ reaches
its converged value ${\bf R}_*$. One need not know in advance how many
iterations this takes. Choosing the number $n_{\rm sub}$ of levels
needed, in order to meet a beforehand given tolerance in ${\bf
  R}_{n_{\rm sub}}$, is otherwise a problem in connection
with~(\ref{eq:recur},\ref{eq:rstart}) and non-wedge-like
$\Gamma^\star$. This number has no general upper bound.

\begin{figure}
\centering 
\noindent\makebox[\textwidth]{
\begin{minipage}{1.1\textwidth}
\includegraphics[height=66mm]{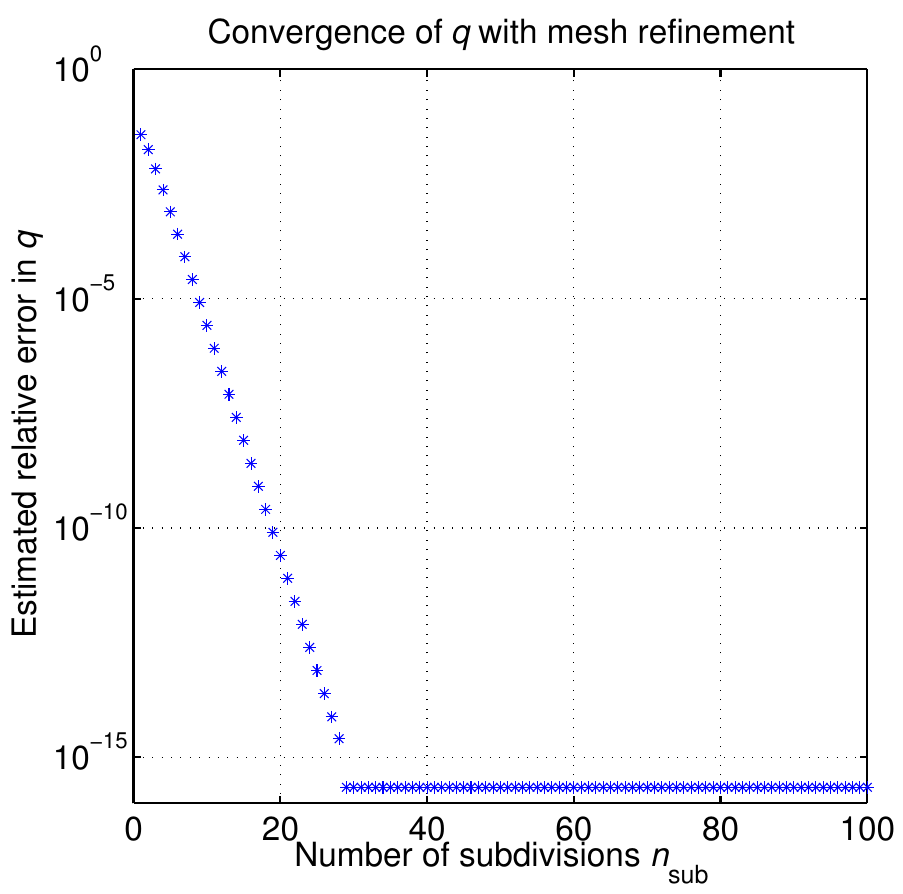}
\includegraphics[height=66mm]{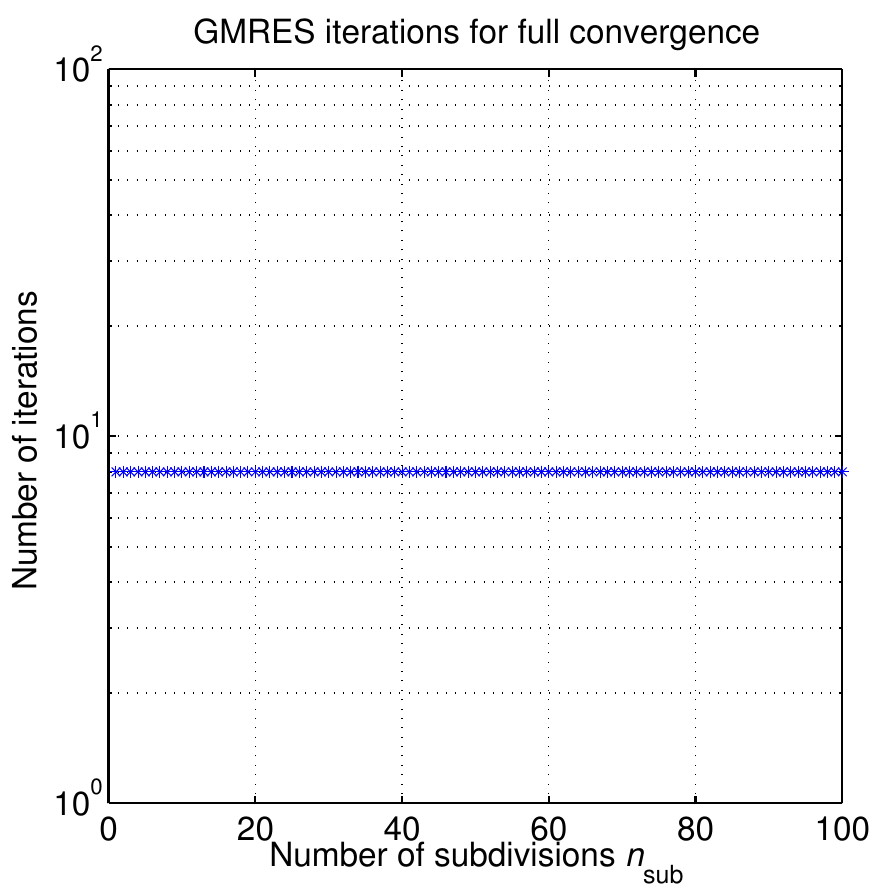}
\end{minipage}}
\caption{\sf Same as Figure~\ref{fig:conv3}, but the program {\tt
    demo8b.m} is used.}
\label{fig:conv8}
\end{figure}

Assume now that the kernel of $K$ is scale invariant on wedges. If all
$\Gamma_i^\star$ are wedge-like, then~(\ref{eq:fixed},\ref{eq:fstart})
replaces~(\ref{eq:recur},\ref{eq:rstart}) entirely. If $\Gamma^\star$
is merely piecewise smooth, then~(\ref{eq:fixed},\ref{eq:fstart}) can
be run on a wedge with the same opening angle as $\Gamma^\star$, to
produce an initializer to~(\ref{eq:recur}). That initializer could
often be more appropriate than ${\bf R}_0$ of~(\ref{eq:rstart}), which
is plagued with a very large discretization error
whenever~(\ref{eq:inteq2c}) is used. The fixed-point initializer ${\bf
  R}_*$ is implemented in the program {\tt demo8b.m}, which is an
upgrading of {\tt demo3b.m}, and produces Figure~\ref{fig:conv8}. A
comparison of Figure~\ref{fig:conv8} with Figure~\ref{fig:conv3} shows
that the number $n_{\rm sub}$ of levels needed for full convergence
with the initializer ${\bf R}_*$ is halved compared to when using the
initializer ${\bf R}_0$ of~(\ref{eq:rstart}).

There are, generally speaking, several advantages with using the
initializer ${\bf R}_*$, rather than ${\bf R}_0$ of~(\ref{eq:rstart}),
in~(\ref{eq:recur}) on a non-wedge-like $\Gamma^\star$: First, the
number of different matrices ${\bf R}_i$ and ${\bf K}^\circ_{i{\rm
    b}}$ needed in~(\ref{eq:recur}) and in~(\ref{eq:back}) is reduced
as the recursions are shortened. This means savings in storage.
Second, the number $n_{\rm sub}$ of levels needed for full convergence
in~(\ref{eq:recur}) seems to always be bounded. The hard work is done
in~(\ref{eq:fixed}). Third, Newton's method can be used to
accelerate~(\ref{eq:fixed}). That is the topic of
Section~\ref{sec:Newton}.

\section{Newton acceleration}
\label{sec:Newton}

When solving integral equations stemming from particularly challenging
elliptic boundary value problems with solutions $\rho(r)$ that are
barely absolutely integrable, the fixed-point
iteration~(\ref{eq:fixed},\ref{eq:fstart}) on wedge-like
$\Gamma^\star$ may need a very large number of steps to reach full
convergence. See~\cite[Section 6.3]{Hels11JCPb} for an example where
$2\cdot 10^5$ steps are needed.

Fortunately,~(\ref{eq:fixed}) can be cast as a non-linear matrix
equation
\begin{equation}
{\bf G}({\bf R}_*)\equiv
{\bf P}^T_{W\rm{bc}}{\bf A}({\bf R}_*){\bf P}_{\rm{bc}}-{\bf R}_*=0\,,
\label{eq:Newton}
\end{equation}
where ${\bf R}_*$, as in Section~\ref{sec:init}, is the fixed-point
solution and
\begin{equation}
{\bf A}({\bf R}_*)=\left(\mathbb{F}\{{\bf R}_*^{-1}\}
+{\bf I}_{\rm b}^\circ+\lambda{\bf K}_{\rm b}^\circ\right)^{-1}\,.
\end{equation}
The non-linear equation~(\ref{eq:Newton}), in turn, can be solved for
${\bf R}_*$ with a variant of Newton's method. Let ${\bf X}$ be a
matrix-valued perturbation of ${\bf R}_*$ and expand ${\bf G}({\bf
  R}_*+{\bf X})=0$ to first order in ${\bf X}$. This gives a
Sylvester-type matrix equation
\begin{equation}
{\bf X}-{\bf P}^T_{W\rm{bc}}{\bf A}({\bf R}_*)\mathbb{F}\{{\bf R}_*^{-1}
{\bf X}{\bf R}_*^{-1}\}{\bf A}({\bf R}_*){\bf P}_{\rm{bc}}={\bf G}({\bf R}_*)
\label{eq:Sylvester}
\end{equation}
for the Newton update ${\bf X}$. One can use the {\sc Matlab} built-in
function {\tt dlyap} for~(\ref{eq:Sylvester}), but GMRES seems to be
more efficient and we use that method. Compare~\cite[Section
6.2]{Hels11JCPb}.

\begin{figure}
\centering 
\includegraphics[height=74mm]{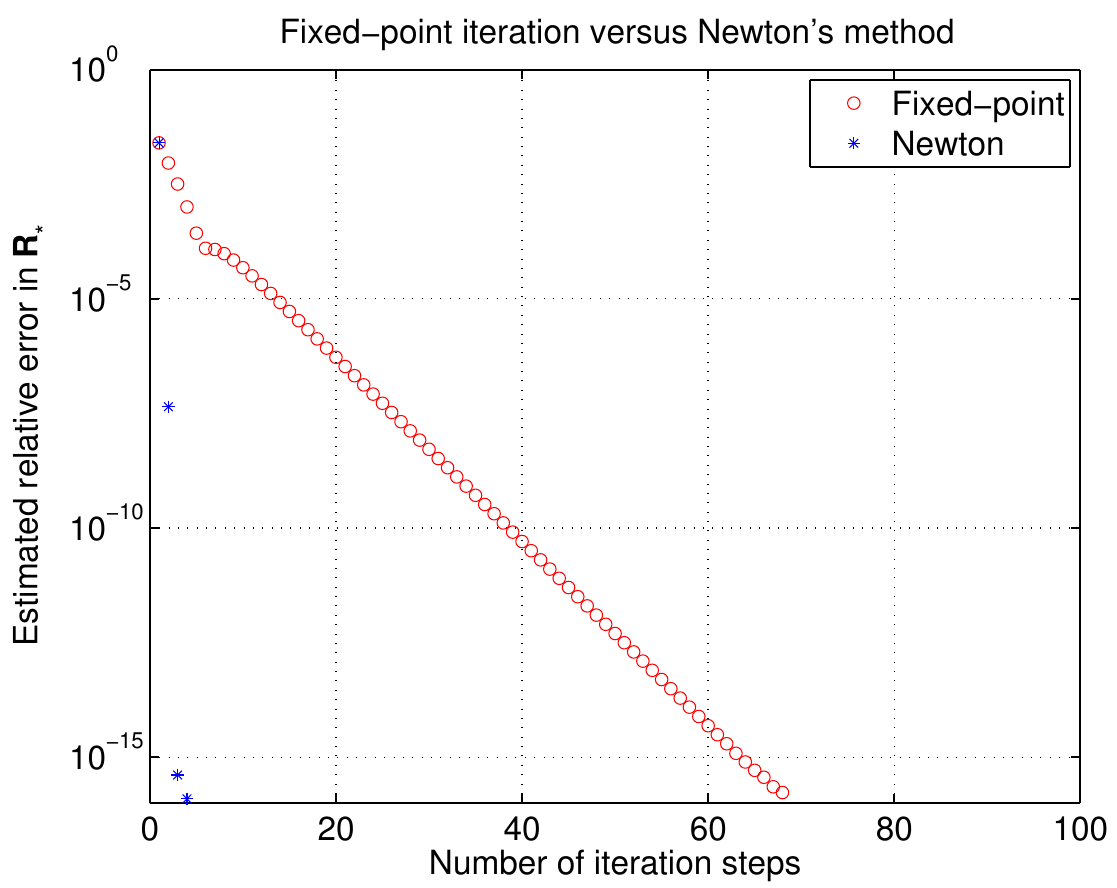}
\caption{\sf Output from the program {\tt demo9.m}. The fixed-point 
  iteration (\ref{eq:fixed},\ref{eq:fstart}) is compared to Newton's
  method for~(\ref{eq:Newton}).}
\label{fig:conv9}
\end{figure}

Figure~\ref{fig:conv9} shows a comparison between the fixed-point
iteration~(\ref{eq:fixed},\ref{eq:fstart}) and Newton's method for
computing the fixed-point solution ${\bf R}_*$ to~(\ref{eq:Newton}) on
a wedge-like $\Gamma^\star$. The program {\tt demo9.m} is used and it
incorporates Schur--Banachiewicz speedup in the style of
Section~\ref{sec:SB}. The wedge opening angle is $\theta=\pi/2$, The
integral operator $K$ is the same as in~(\ref{eq:inteq3}), and
$\lambda=0.999$. The relative difference between the two converged
solutions ${\bf R}_*$ is $5\cdot 10^{-16}$. Figure~\ref{fig:conv9}
clearly shows that~(\ref{eq:fixed},\ref{eq:fstart}) converges linearly
(in 68 iterations), while Newton's method has quadratic convergence.
Only four iterations are needed. The computational cost per iteration
is, of course, higher for Newton's method than for the fixed-point
iteration, but it is the same at each step. Recall that the size of
the underlying matrix ${\bf I}_{\rm fin}+\lambda{\bf K}_{\rm
  fin}^\star$, that is inverted according to~(\ref{eq:R}), grows
linearly with the number of steps needed in the fixed-point iteration.
This example therefore demonstrates that one can invert and compress a
linear system of the type~(\ref{eq:R}) in sub-linear time.

\section{The asymptotics of $\rho(r)$ in the corner}
\label{sec:asymp}

The recursion~(\ref{eq:back}) provides a powerful tool for computing
the asymptotics of $\rho(r)$ close to the corner vertex: Deep in the
corner, for large $n_{\rm sub}$ and small $i$, the matrices ${\bf
  R}_{i-1}$ and ${\bf K}^\circ_{i{\rm b}}$ can be replaced with their
asymptotic counterparts ${\bf R}_*$ and ${\bf K}^\circ_{\rm b}$, see
Section~\ref{sec:init}, so that~(\ref{eq:back}) reads
\begin{equation}
\vec{\boldsymbol{\rho}}_{{\rm coa},i}=
{\bf C}\tilde{\boldsymbol{\rho}}_{{\rm coa},i}\,,
\quad i=n_{\rm asm},\ldots,1\,, \quad n_{\rm sub}-n_{\rm asm}\gg 1\,,
\label{eq:steady1}
\end{equation}
where ${\bf C}$ is the constant $96\times 64$ matrix
\begin{equation}
{\bf C}=\left[{\bf I}_{\rm b}-\lambda{\bf K}_{\rm b}^\circ
\left(\mathbb{F}\{{\bf R}_*^{-1}\}+
{\bf I}_{\rm b}^\circ+\lambda{\bf K}_{\rm b}^\circ
\right)^{-1}\right]{\bf P}_{\rm bc}\,.
\label{eq:steady2}
\end{equation}
Each step in~(\ref{eq:steady1}) reconstructs $\rho(r)$ on the
outermost panels of a mesh on a subset $\Gamma_i^\star$ that is half
the size of the subset $\Gamma_{i+1}^\star$ in the previous step.
Furthermore, the evolution of $\tilde{\boldsymbol{\rho}}_{{\rm
    coa},i}$ is determined by power iteration applied to a submatrix
of ${\bf C}$ given by row indices $\{17:80\}$ and all columns. We
denoted this submatrix by ${\bf C}^\star$. The eigenvalues of ${\bf
  C}^\star$ control the behavior of $\rho(r)$ as the distance to the
corner vertex is halved. In particular, if $\gamma$ is the arc length
distance to the vertex then the leading asymptotic behavior is
$\rho(\gamma)\propto\gamma^\beta$ with
\begin{equation}
\beta=-\log_2{\left(d_{\rm max}\right)}\,,
\label{eq:beta1}
\end{equation}
where $d_{\rm max}$ is the largest eigenvalue in modulus of ${\bf
  C}^\star$.

\begin{figure}
\centering 
\noindent\makebox[\textwidth]{
\begin{minipage}{1.1\textwidth}
\hspace{2mm}
\includegraphics[height=66mm]{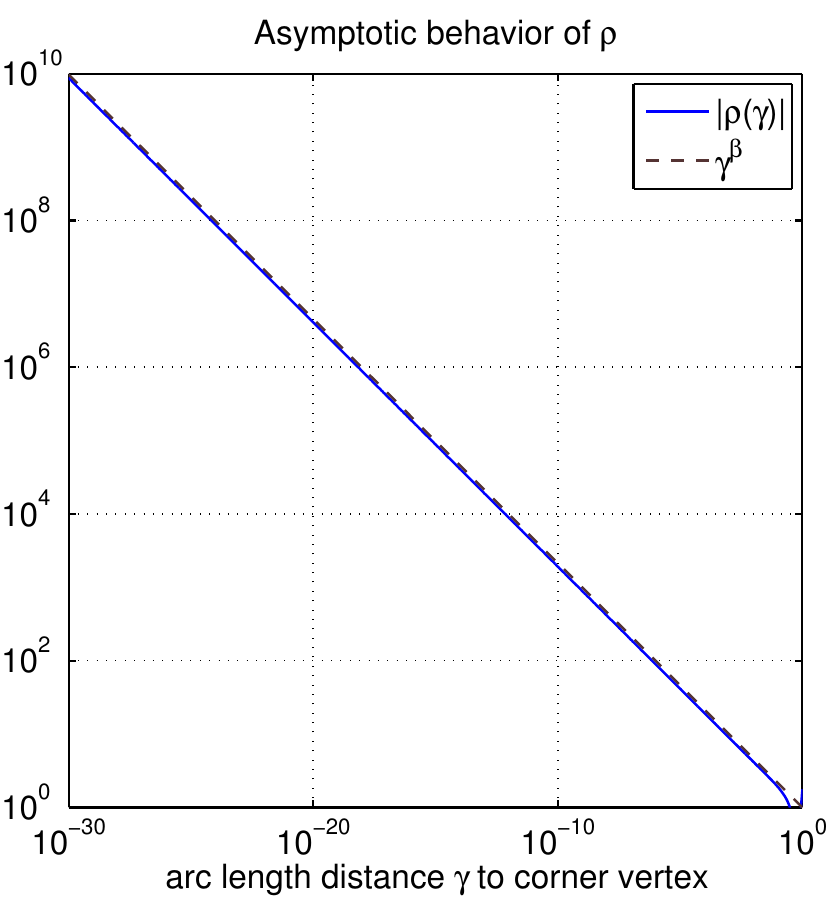}
\includegraphics[height=66mm]{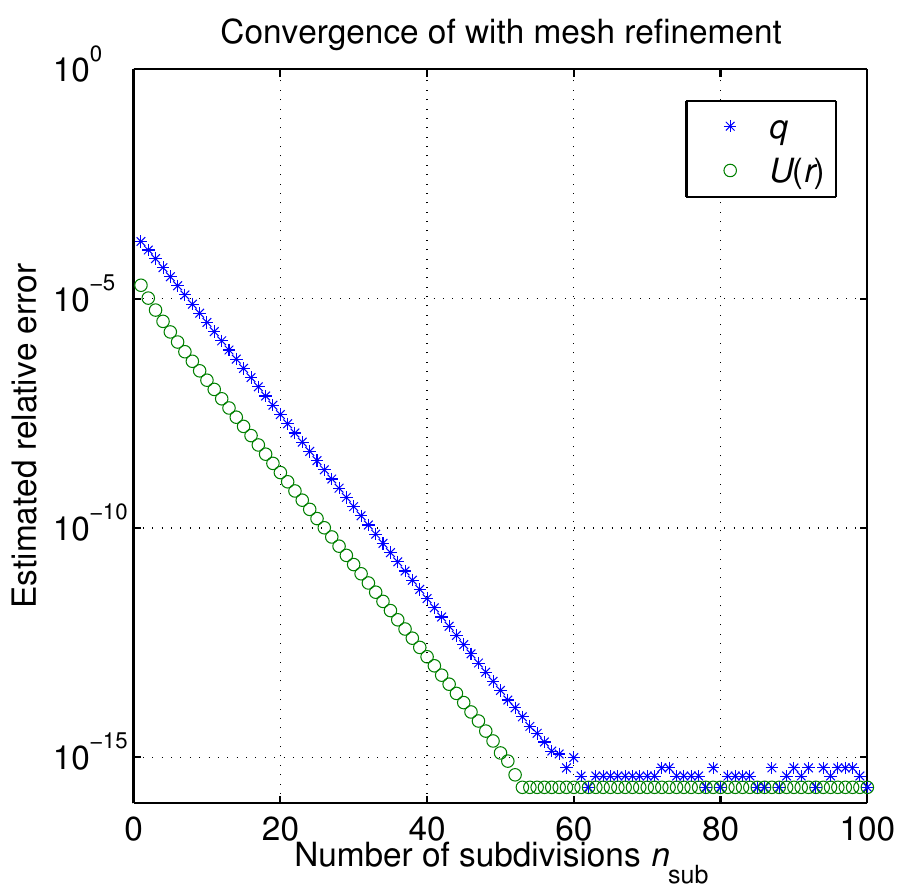}
\end{minipage}}
\caption{\sf Left: $\rho(\gamma)$ from {\tt demo8c.m} compared to
  the asymptotic behavior $\gamma^\beta$ with $\beta$
  from~(\ref{eq:beta2}). Right: similar as in Figure~\ref{fig:conv3},
  but from {\tt demo3c.m} and with {\tt npan=14}. The potential $U(r)$
  at $r=(0.4,0.1)$ is evaluated via~(\ref{eq:UofR}).}
\label{fig:conv3c}
\end{figure}

For the opening angle $\theta=\pi/2$ in the model problem of
Section~\ref{sec:model} it is possible to derive the closed-form
expression~\cite[Eq.~(13)]{Hels11NJP}
\begin{equation}
\beta=\frac{2}{\pi}\arccos{\left(\frac{\lambda}{2}\right)}-1\,.
\label{eq:beta2}
\end{equation}
The program {\tt demo8c.m} compares $\beta$ computed
from~(\ref{eq:beta1}) to $\beta$ from~(\ref{eq:beta2}) with
$\lambda=0.999$. The relative difference is a mere $2\cdot 10^{-15}$,
which means that RCIP provides a competitive alternative to
traditional techniques, such as separation of variables~\cite[Section
2]{Hels00}, also for asymptotic studies. The left image of
Figure~\ref{fig:conv3c} shows the asymptotic behavior of
$\rho(\gamma)$ in the corner.

\section{On the accuracy of ``the solution''}
\label{sec:acc}

The integral equation~(\ref{eq:inteq1a}) comes from a boundary value
problem for Laplace's equation where the potential field $U(r)$ at a
point $r$ in the plane is related to $\rho(r)$ via
\begin{equation}
U(r)=(e\cdot r)-\int_{\Gamma}G(r,r')\rho(r')\,{\rm d}\ell'\,,
\label{eq:UofR}
\end{equation}
see~\cite[Section 2.1]{Hels08b}. The right image of
Figure~\ref{fig:conv3c} shows how $U(r)$ converges with mesh
refinement at a point $r=(0.4,0.1)$ inside the contour $\Gamma$. We
see that the accuracy in $U(r)$ is slightly better than the accuracy
of the dipole moment $q$ of~(\ref{eq:q}). One can say that measuring
the field error at a point some distance away from the corner is more
forgiving than measuring the dipole moment error. It is possible to
construct examples where the difference in accuracy between field
solutions and moments of layer densities are more pronounced and this
raises the question of how the accuracy of integral equation solvers
best should be measured.

\section{Composed integral operators}
\label{sec:compose}

Assume that we have a modification of~(\ref{eq:inteq3}) which reads
\begin{equation}
\left(I+MK\right)\rho_1(r)=g(r)\,,\quad z\in \Gamma\,.
\label{eq:inteq7}
\end{equation}
Here $K$ and $g$ are as in~(\ref{eq:inteq3}), $M$ is a new, bounded,
integral operator, and $\rho_1$ is an unknown layer density to be
solved for. This section shows how to apply RCIP to~(\ref{eq:inteq7})
using a simplified version of the scheme in~\cite{Hels11SISC}.

Let us, temporarily, expand~(\ref{eq:inteq7}) into a system of
equations by introducing a new layer density $\rho_2(r)=K\rho_1(r)$.
Then
\begin{align}
  \rho_1(r)+       M&\rho_2(r)=g(r)\,,\\
-K\rho_1(r)+\quad\; &\rho_2(r)=0\,,
\end{align}
and after discretization on the fine mesh
\begin{equation}
\left(
\begin{bmatrix}
{\bf I}_{\rm fin} & {\bf 0}_{\rm fin} \\
{\bf 0}_{\rm fin} & {\bf I}_{\rm fin}
\end{bmatrix} 
+
\begin{bmatrix}
 {\bf 0}_{\rm fin} & {\bf M}_{\rm fin} \\
-{\bf K}_{\rm fin} & {\bf 0}_{\rm fin}
\end{bmatrix} 
\right)
\begin{bmatrix}
 \boldsymbol{\rho}_{1{\rm fin}} \\ 
 \boldsymbol{\rho}_{2{\rm fin}}
\end{bmatrix}
=
\begin{bmatrix}
   {\bf g}_{\rm fin} \\ 0 
\end{bmatrix}\,.
\label{eq:disc1}
\end{equation}
Standard RCIP gives
\begin{equation}
\left(
\begin{bmatrix}
{\bf I}_{\rm coa} & {\bf 0}_{\rm coa} \\
{\bf 0}_{\rm coa} & {\bf I}_{\rm coa}
\end{bmatrix} 
+
\begin{bmatrix}
 {\bf 0}_{\rm coa}         & {\bf M}_{\rm coa}^\circ \\
-{\bf K}_{\rm coa}^\circ & {\bf 0}_{\rm coa}
\end{bmatrix} 
\begin{bmatrix}
 {\bf R}_1  & {\bf R}_3 \\
 {\bf R}_2  & {\bf R}_4
\end{bmatrix} 
\right)
\begin{bmatrix}
 \tilde{\boldsymbol{\rho}}_{1{\rm coa}} \\ 
 \tilde{\boldsymbol{\rho}}_{2{\rm coa}}
\end{bmatrix}
=
\begin{bmatrix}
   {\bf g}_{\rm coa} \\ 
    0 
\end{bmatrix}\,,
\label{eq:disc2}
\end{equation}
where the compressed inverse ${\bf R}$ is partitioned into four
equi-sized blocks.

\begin{figure}[t]
\centering 
\noindent\makebox[\textwidth]{
\begin{minipage}{1.1\textwidth}
\includegraphics[height=66mm]{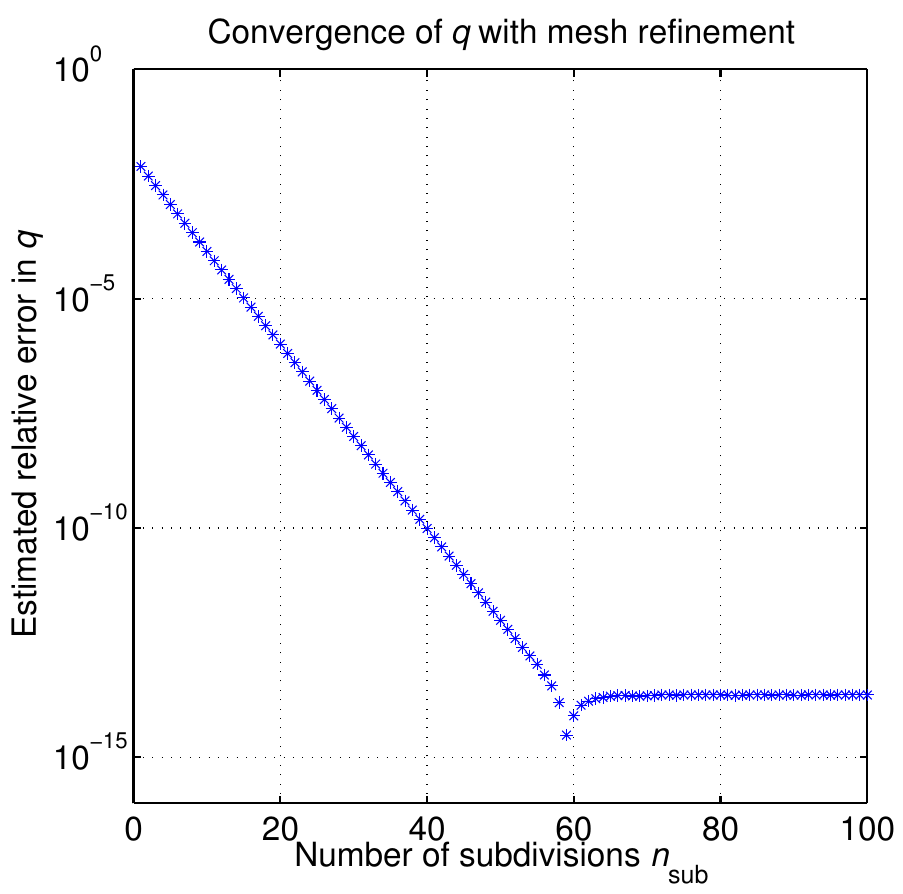}
\includegraphics[height=66mm]{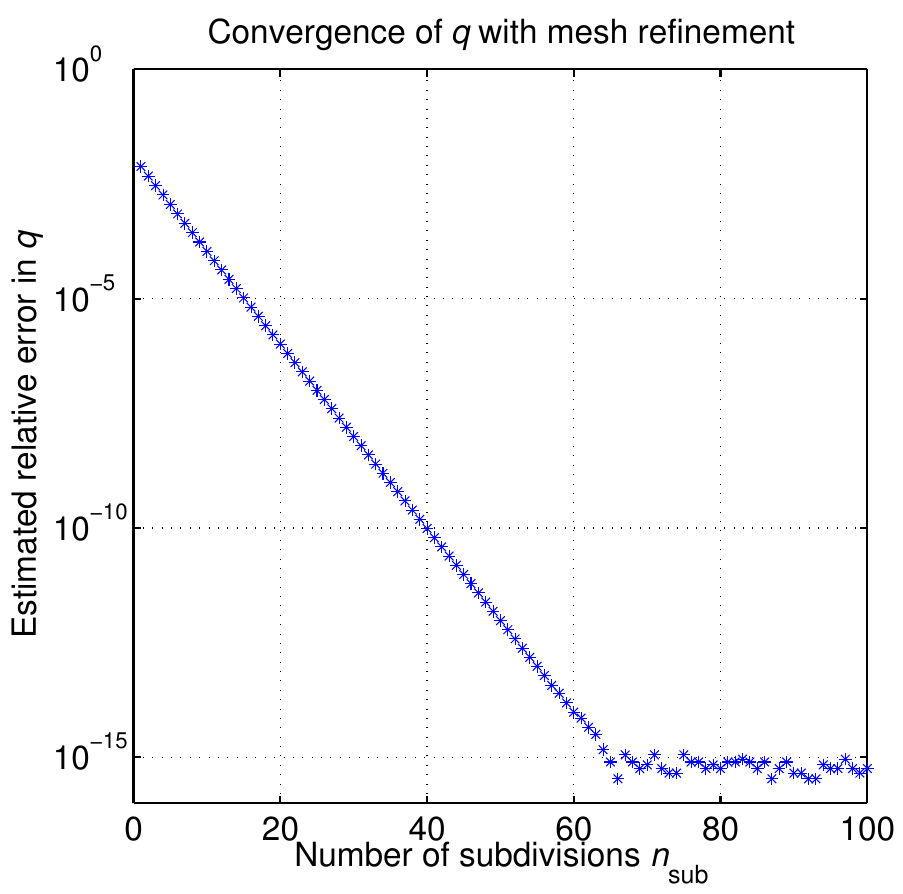}
\end{minipage}}
\caption{\sf Convergence for $q$ of~(\ref{eq:q})
  with $\rho=\rho_1$ from~(\ref{eq:inteq7}). The curve $\Gamma$ is as
  in~(\ref{eq:gamma}) and {\tt theta=pi/2}, {\tt npan=11}, and {\tt
    evec=1}. The reference values is taken as $q=1.95243329423584$.
  Left: Results from the inner product preserving scheme of Appendix~E
  produced with {\tt demo10.m}. Right: Results with RCIP according
  to~(\ref{eq:disc3},\ref{eq:hat1}) produced with {\tt demo10b.m}}.
\label{fig:conv10}
\end{figure}

Now we replace $\tilde{\boldsymbol{\rho}}_{1{\rm coa}}$ and
$\tilde{\boldsymbol{\rho}}_{2{\rm coa}}$ with a single unknown
$\tilde{\boldsymbol{\rho}}_{\rm coa}$ via
\begin{align}
  \tilde{\boldsymbol{\rho}}_{1{\rm coa}}&=
  \tilde{\boldsymbol{\rho}}_{\rm coa}-
  {\bf R}_1^{-1}{\bf R}_3{\bf K}_{\rm coa}^\circ{\bf R}_1
  \tilde{\boldsymbol{\rho}}_{\rm coa}\,,
\label{eq:single1}\\
  \tilde{\boldsymbol{\rho}}_{2{\rm coa}}&=
  {\bf K}_{\rm coa}^\circ{\bf R}_1
  \tilde{\boldsymbol{\rho}}_{\rm coa}\,.
\label{eq:single2}
\end{align}
The change of variables~(\ref{eq:single1},\ref{eq:single2}) is chosen
so that the second block-row of~(\ref{eq:disc2}) is automatically
satisfied. The first block-row of~(\ref{eq:disc2}) becomes
\begin{multline}
\left[{\bf I}_{\rm coa}+{\bf M}_{\rm coa}^\circ
\left({\bf R}_4-{\bf R}_2{\bf R}_1^{-1}{\bf R}_3\right)
{\bf K}_{\rm coa}^\circ{\bf R}_1+{\bf M}_{\rm coa}^\circ{\bf R}_2
\right.\\
\left.-{\bf R}_1^{-1}{\bf R}_3{\bf K}_{\rm coa}^\circ{\bf R}_1
\right]\tilde{\boldsymbol{\rho}}_{\rm coa}={\bf g}_{\rm coa}\,.
\label{eq:disc3}
\end{multline}

When~(\ref{eq:disc3}) has been solved for
$\tilde{\boldsymbol{\rho}}_{\rm coa}$, the weight-corrected version of
the original density $\rho_1$ can be recovered as
\begin{equation}
\hat{\boldsymbol{\rho}}_{1{\rm coa}}=
{\bf R}_1\tilde{\boldsymbol{\rho}}_{\rm coa}\,.
\label{eq:hat1}
\end{equation}

Figure~\ref{fig:conv10} shows results for~(\ref{eq:inteq7}) with $M$
being the double layer potential
\begin{equation}
M\rho(r)\equiv
-2\int_{\Gamma}\frac{\partial G}{\partial\nu'}(r,r')\rho(r')
\,{\rm d}\ell'=\frac{1}{\pi}\int_{\Gamma}\rho(\tau)
\Im\left\{\frac{{\rm d}\tau}{\tau-z}\right\}\,.
\end{equation}
The left image shows the convergence of $q$ of~(\ref{eq:q}) with
$n_{\rm sub}$ using the inner product preserving discretization scheme
of Appendix~E for~(\ref{eq:inteq7}) as implemented in {\tt demo10.m}.
The right image shows $q$ produced with RCIP according
to~(\ref{eq:disc3},\ref{eq:hat1}) as implemented in {\tt demo10b.m}.
The reference value for $q$ is computed with the program {\tt
  demo10c.m}, which uses inner product preserving discretization
together with compensated summation~\cite{High96,Kaha65} in order to
enhance the achievable accuracy. One can see that, in addition to
being faster, RCIP gives and extra digit of accuracy. Actually, it
seems as if the scheme in {\tt demo10.m} converges to a $q$ that is
slightly wrong.

In conclusion, in this example and in terms of stability, the RCIP
method is better than standard inner product preserving discretization
and on par with inner product preserving discretization enhanced with
compensated summation. In terms of computational economy and speed,
RCIP greatly outperforms the two other schemes.

\section{Nyström discretization of singular kernels}
\label{sec:sing}
  
The Nyström scheme of Section~\ref{sec:disc}
discretizes~(\ref{eq:inteq3}) using composite 16-point Gauss--Legendre
quadrature. This works well when the kernel of the integral operator
$K$ is smooth on smooth $\Gamma$. When the kernel is not smooth on
smooth $\Gamma$, then the quadrature fails and something better is
needed. See~\cite{Hao11} for a comparison of the performance of
various modified high-order accurate Nyström discretizations for
weakly singular kernels and~\cite{Kloc12} for a high-order general
approach to the evaluation of layer potentials.

We are not sure what modified discretization is optimal in every
situation. When nearly singular, weakly singular, and singular
operators need to be discretized in the following, we chiefly use a
modification to composite Gauss--Legendre quadrature called {\it local
  panelwise evaluation}. See~\cite[Section 2]{Hels09JCP}
and~\cite{HelsHols15,HelsKarl18} for a description of this technique.

\section{The exterior Dirichlet Helmholtz problem}
\label{sec:HelmDiri}

Let $D$ be the domain enclosed by the curve $\Gamma$ and let $E$ be
the exterior to the closure of $D$. The exterior Dirichlet problem for
the Helmholtz equation
\begin{align}
  \Delta U(r)+\omega^2 U(r)&=0\,,
  \quad r\in E\,,\\
\lim_{E\ni r\to r^\circ}
  U(r)&=g(r^\circ)\,,\quad r^\circ\in\Gamma\,,\label{eq:HelmBV1}\\
  \lim_{|r|\to\infty}
  \sqrt{|r|}\left(\frac{\partial}{\partial|r|}-{\rm
        i}\omega\right)U(r)&=0\,,
\label{eq:goout}
\end{align}
has a unique solution $U(r)$ under mild assumptions on $\Gamma$ and
$g(r)$~\cite{Mitr96} and can be modeled using a combined integral
representation~\cite[Chapter 3]{Colt98}
\begin{equation}
U(r)=\int_\Gamma\frac{\partial\Phi_\omega}{\partial\nu'}(r,r')\rho(r')
\,{\rm d}\ell'-
\frac{\rm i\omega}{2}\int_\Gamma\Phi_\omega(r,r')\rho(r')
\,{\rm d}\ell'\,, \quad r\in\mathbb{R}^2\setminus \Gamma\,,
\label{eq:Helmrep1}
\end{equation}
where $\Phi_{\omega}(r,r')$ is the fundamental solution to the
Helmholtz equation in two dimensions
\begin{equation}
\Phi_\omega(r,r')=\frac{\rm i}{4}H_0^{(1)}(\omega|r-r'|)\,.
\end{equation}
Here $H_0^{(1)}(\cdot)$ is the zeroth order Hankel function of the
first kind. Insertion of~(\ref{eq:Helmrep1}) into~(\ref{eq:HelmBV1})
gives the combined field integral equation
\begin{equation}
\left(I+K_\omega-\frac{{\rm i}\omega}{2}S_\omega\right)\rho(r)=2g(r)\,, 
\quad r\in\Gamma\,,
\label{eq:DiriHelm}
\end{equation}
where
\begin{align}
  K_\omega\rho(r)&=
  2\int_\Gamma\frac{\partial\Phi_\omega}{\partial\nu'}(r,r')\rho(r')
  \,{\rm d}\ell'\,,
  \label{eq:Koper}\\
  S_\omega\rho(r)&=2\int_\Gamma\Phi_\omega(r,r')\rho(r')\,
  {\rm d}\ell'\,.
  \label{eq:Soper}
\end{align}

\begin{figure}[t]
\centering 
\noindent\makebox[\textwidth]{
\begin{minipage}{1.1\textwidth}
\includegraphics[height=66mm]{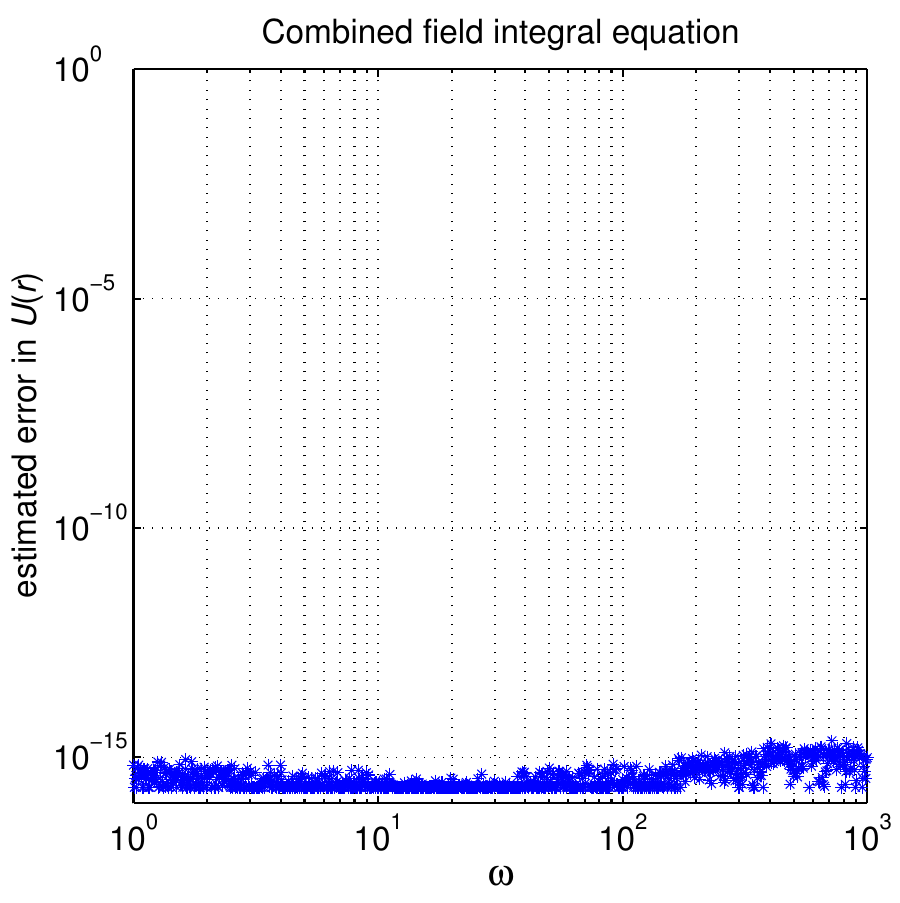}
\includegraphics[height=66mm]{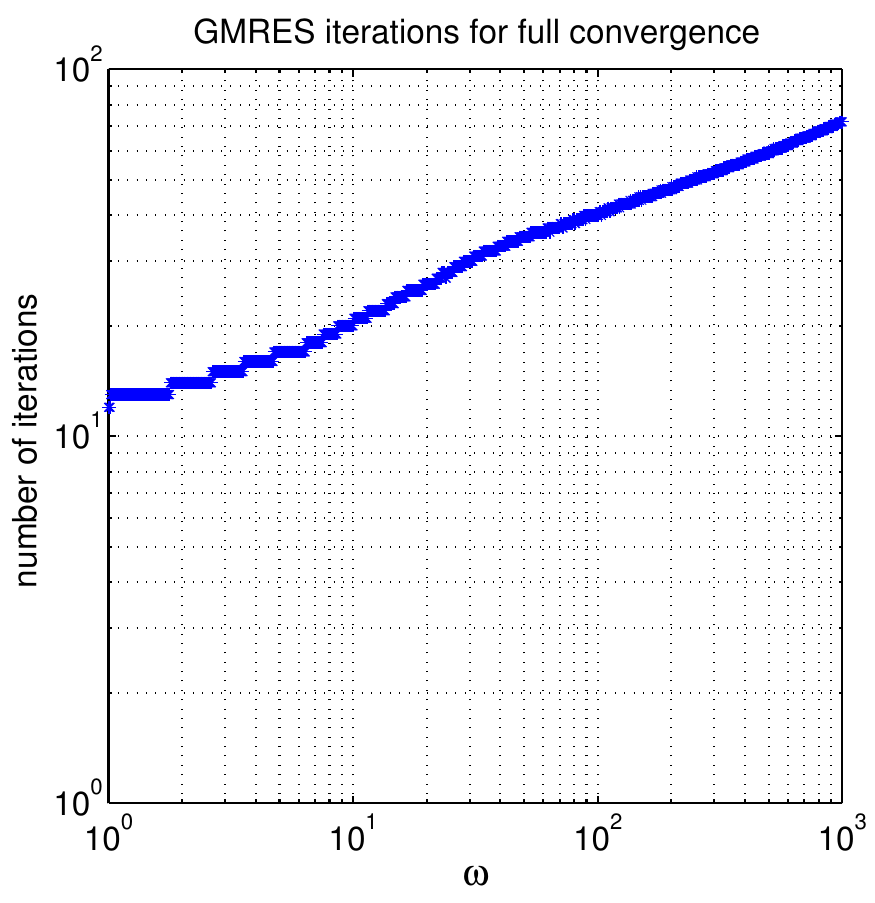}
\end{minipage}}
\caption{\sf The exterior Dirichlet problem for Helmholtz equation
  with RCIP applied to~(\ref{eq:DiriHelm}). The program {\tt demo11.m}
  is used with $\Gamma$ as in~(\ref{eq:gamma}) and $\theta=\pi/2$. The
  boundary condition $g(r)$ of~(\ref{eq:HelmBV1}) is generated by a
  point source at $(0.3,0.1)$. Left: the absolute error in $U(r)$ at
  $r=(-0.1,0.2)$. Right: the number of GMRES iterations needed to meet
  an estimated relative residual of $\epsilon_{\rm mach}$.}
\label{fig:conv11}
\end{figure}

Figure~\ref{fig:conv11} shows the performance of RCIP applied
to~(\ref{eq:DiriHelm}) for 1000 different values of
$\omega\in[1,10^3]$. The program {\tt demo11.m} is used. This program
has a fixed-point initializer ${\bf R}_*$, see Section~\ref{sec:init},
whose construction only takes the leading asymptotic behavior of
$I+K_\omega-{\rm i}\omega S_\omega/2$ at the corner vertex into
account. The boundary $\Gamma$ is as in~(\ref{eq:gamma}) with
$\theta=\pi/2$ and the boundary conditions are chosen as
$g(r)=H_0^{(1)}(r-r')$ with $r'=(0.3,0.1)$ inside $\Gamma$. The error
in $U(r)$ of~(\ref{eq:Helmrep1}) is evaluated at $r=(-0.1,0.2)$
outside $\Gamma$. Since the magnitude of $U(r)$ varies with $\omega$,
peaking at about unity, the absolute error is shown rather than the
relative error. The number of panels on the coarse mesh is chosen as
{\tt npan=0.6*omega+18} rounded to the nearest integer.

\section{The exterior Neumann Helmholtz problem}
\label{sec:HelmNeu}

The exterior Neumann problem for the Helmholtz equation
\begin{align}
  \Delta U(r)+\omega^2 U(r)&=0\,, \quad r\in E\,,\\
  \lim_{E\ni r\to r^\circ}\nu^\circ\cdot\nabla U(r)&=g(r^\circ)\,,
  \quad r^\circ\in\Gamma\,,\label{eq:HelmBV2}\\
  \lim_{|r|\to\infty}
  \sqrt{|r|}\left(\frac{\partial}{\partial|r|}-{\rm
      i}\omega\right)U(r)&=0\,,
\end{align}
has a unique solution $U(r)$ under mild assumptions on $\Gamma$ and
$g(r)$~\cite{Mitr96} and can be modeled as an integral equation in
several ways. We shall consider two options: an ``analogy with the
standard approach for Laplace's equation'', which is not necessarily
uniquely solvable for all $\omega$, and a ``regularized combined field
integral equation'' which is always uniquely solvable. See,
further,~\cite{Brem12a,Brun12}.

\subsection{An analogy with the standard Laplace approach}
\label{sec:analogy}

\begin{figure}[!t]
\centering 
\noindent\makebox[\textwidth]{
\begin{minipage}{1.1\textwidth}
\includegraphics[height=66mm]{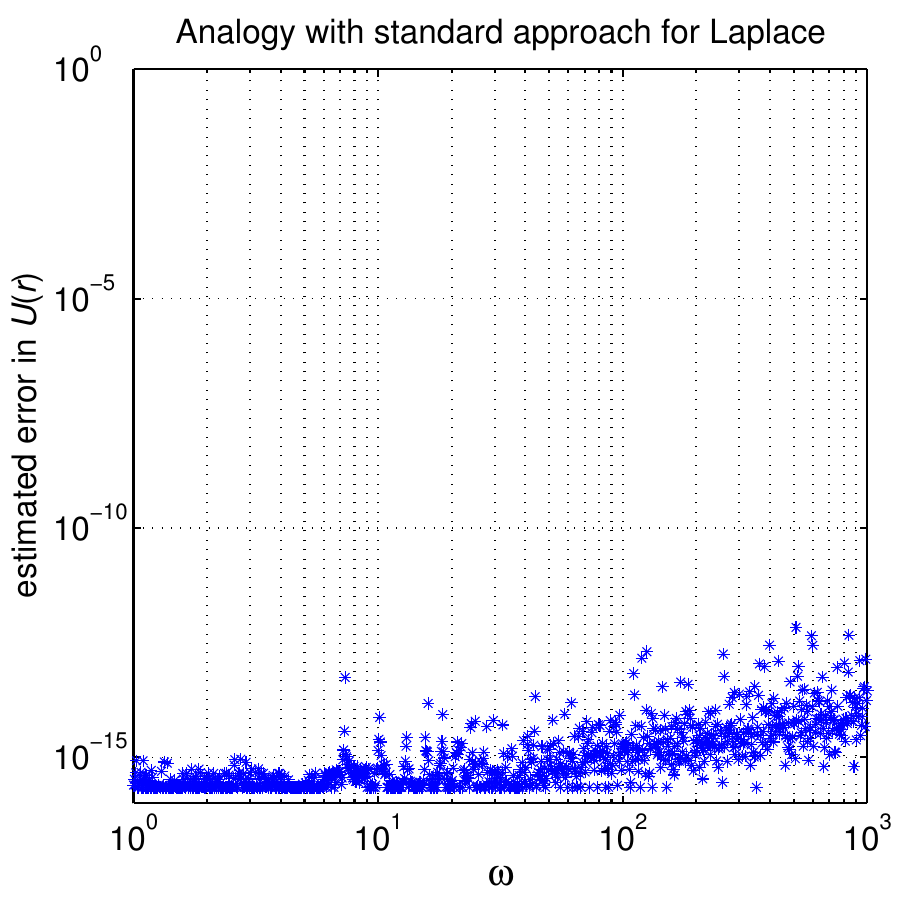}
\includegraphics[height=66mm]{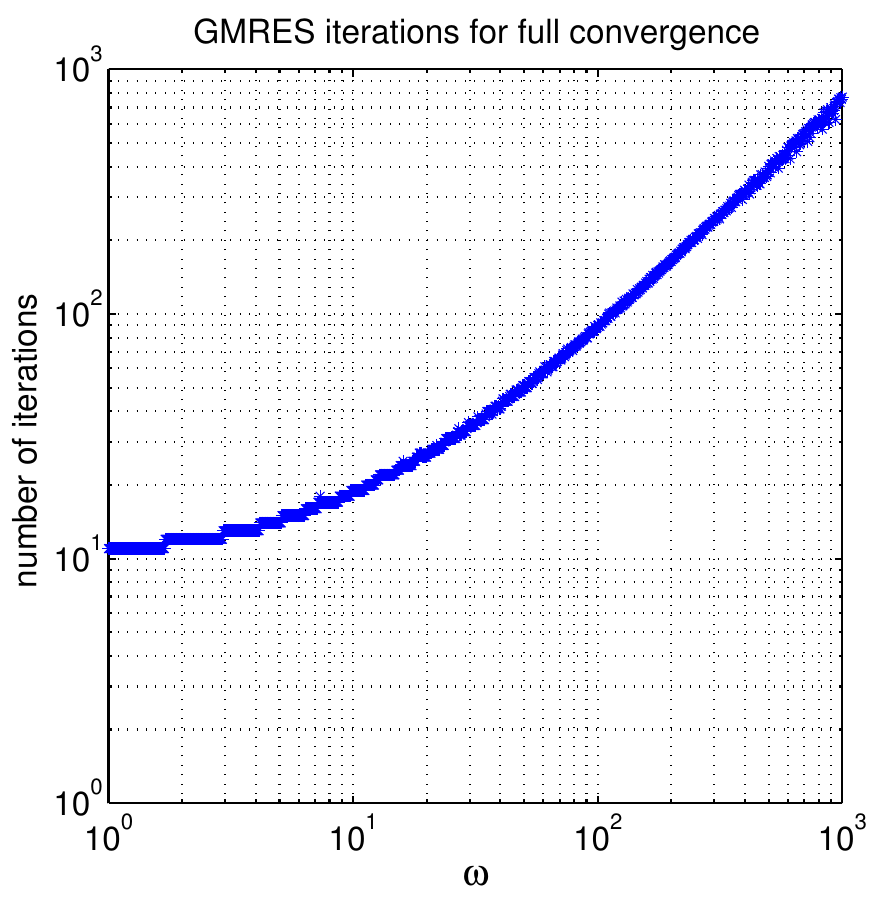}
\end{minipage}}
\caption{\sf The exterior Neumann problem for Helmholtz equation
  with RCIP applied to~(\ref{eq:NeuHelm1}). The program {\tt demo12.m}
  is used with $\Gamma$ as in~(\ref{eq:gamma}) and $\theta=\pi/2$. The
  boundary condition $g(r)$ of~(\ref{eq:HelmBV2}) is generated by a
  point source at $(0.3,0.1)$. Left: the absolute error in $U(r)$ at
  $r=(-0.1,0.2)$. Right: the number of GMRES iterations needed to meet
  an estimated relative residual of $\epsilon_{\rm mach}$.}
\label{fig:conv12}
\end{figure}

Let $K^{\rm A}_{\omega}$ be the adjoint to the double-layer integral
operator $K_{\omega}$ of~(\ref{eq:Koper})
\begin{equation}
  K^{\rm A}_{\omega}\rho(r)=
  2\int_{\Gamma}\frac{\partial\Phi_{\omega}}{\partial\nu}(r,r')\rho(r')
  \,{\rm d}\ell'\,.
\end{equation}
Insertion of the integral representation
\begin{equation}
U(r)=\int_{\Gamma}\Phi_{\omega}(r,r')\rho(r')
\,{\rm d}\ell'\,, \quad r\in\mathbb{R}^2\setminus \Gamma
\label{eq:Helmrep2}
\end{equation}
into~(\ref{eq:HelmBV2}) gives the integral equation
\begin{equation}
\left(I-K^{\rm A}_{\omega}\right)\rho(r)=-2g(r)\,, \quad r\in\Gamma\,.
\label{eq:NeuHelm1}
\end{equation}

Figure~\ref{fig:conv12} shows the performance of RCIP applied
to~(\ref{eq:NeuHelm1}). The program {\tt demo12.m} is used and the
setup is the same as that for the Dirichlet problem in
Section~\ref{sec:HelmDiri}. A comparison between
Figure~\ref{fig:conv12} and Figure~\ref{fig:conv11} shows that the
number of GMRES iterations needed for full convergence now grows much
faster with $\omega$. Furthermore, the relative error in the solution
to the Neumann problem is larger, particularly so when $\omega$
happens to be close to values for which the operator $I-K^{\rm
  A}_{\omega}$ in~(\ref{eq:NeuHelm1}) has a nontrivial null space.
Recall that~(\ref{eq:DiriHelm}) is always uniquely solvable
while~(\ref{eq:NeuHelm1}) is not.

\subsection{A regularized combined field integral equation}
\label{sec:regul}

The literature on regularized combined field integral equations for
the exterior Neumann problem is rich and several formulations have
been suggested. We shall use the representation~\cite{Brun12}
\begin{equation}
U(r)=\int_{\Gamma}\Phi_{\omega}(r,r')\rho(r')
\,{\rm d}\ell'+
{\rm i}\int_{\Gamma}\frac{\partial\Phi_{\omega}}{\partial\nu'}(r,r')
\left(S_{{\rm i}\omega}\rho\right)(r')\,
{\rm d}\ell'\,, \quad r\in\mathbb{R}^2\setminus\Gamma\,,
\label{eq:Helmrep3}
\end{equation}
which after insertion into~(\ref{eq:HelmBV2}) gives the integral
equation
\begin{equation}
\left(I-K^{\rm A}_{\omega}
-{\rm i}T_{\omega}S_{{\rm i}\omega}\right)\rho(r)=-2g(r)\,, 
\quad r\in\Gamma\,,
\label{eq:NeuHelm2}
\end{equation}
where
\begin{equation}
T_{\omega}\rho(r)=
2\int_{\Gamma}\frac{\partial^2\Phi_{\omega}}{\partial\nu\partial\nu'}(r,r')
   \rho(r')\,{\rm d}\ell'\,.
\label{eq:T}
\end{equation}

\begin{figure}[t]
\centering 
\noindent\makebox[\textwidth]{
\begin{minipage}{1.1\textwidth}
\includegraphics[height=66mm]{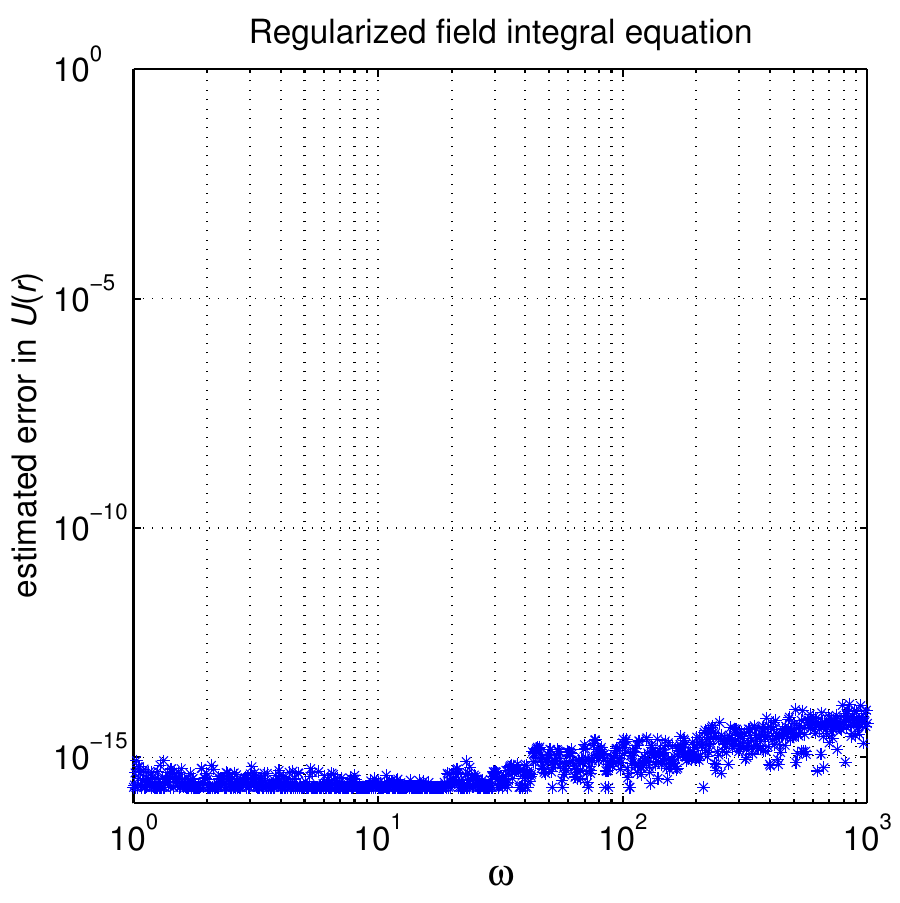}
\includegraphics[height=66mm]{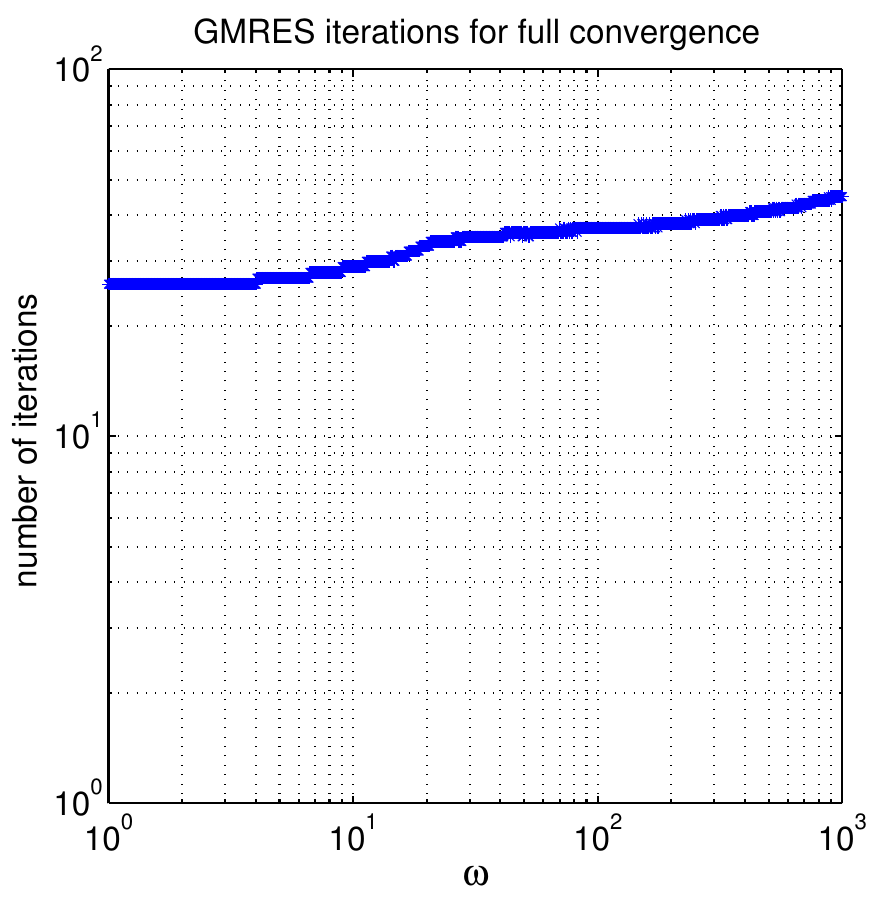}
\end{minipage}}
\caption{\sf The same exterior Neumann problem for Helmholtz equation
  as in Figure~\ref{fig:conv12}, but RCIP is now applied
  to~(\ref{eq:NeuHelm3}). The program {\tt demo13b.m} is used.}
\label{fig:conv13}
\end{figure}

The hypersingular operator $T_{\omega}$ of~(\ref{eq:T}) can be
expressed as a sum of a simple operator and an operator that requires
differentiation with respect to arc length only~\cite{Kress95}
\begin{equation}
T_{\omega}\rho(r)=
2\omega^2\int_{\Gamma}\Phi_{\omega}(r,r')(\nu\cdot\nu')\rho(r')\,
{\rm d}\ell'+
2\int_{\Gamma}\frac{{\rm d}\Phi_{\omega}}{{\rm d}\ell}
(r,r')\frac{{\rm d}\rho}{{\rm d}\ell'}(r')\,
{\rm d}\ell'\,.
\end{equation}
This makes it possible to write~(\ref{eq:NeuHelm2}) in the form
\begin{equation}
\left(I+A+B_1B_2+C_1C_2\right)\rho(r)=-2g(r)\,, 
\quad r\in\Gamma\,,
\label{eq:NeuHelm3}
\end{equation}
where $A=-K^{\rm A}_\omega$, $B_2=S_{{\rm i}\omega}$, and the action
of the operators $B_1$, $C_1$, and $C_2$ is given by
\begin{align}
  B_1\rho(r)&=-2{\rm i}\omega^2\int_{\Gamma}\Phi_{\omega}(r,r')
  (\nu\cdot\nu')\rho(r')\,{\rm d}\ell'\,,\\
  C_1\rho(r)&=-2{\rm i}\int_{\Gamma}\frac{{\rm d}\Phi_{\omega}}{{\rm d}\ell}
  (r,r')\rho(r')\,{\rm d}\ell'\,,
  \label{eq:C1}\\
  C_2\rho(r)&= 2\int_{\Gamma}\frac{{\rm d}\Phi_{{\rm i}\omega}}{{\rm d}\ell}
  (r,r')\rho(r')\,{\rm d}\ell'\,.
  \label{eq:C2}
\end{align}
All integral operators in~(\ref{eq:NeuHelm3}) are such that their
discretizations admit the low-rank decomposition~(\ref{eq:decomp}).
We use the temporary expansion technique of Section~\ref{sec:compose}
for~(\ref{eq:NeuHelm3}), with two new layer densities that are later
eliminated, to arrive at a single compressed equation analogous
to~(\ref{eq:disc3}). That equation involves nine equi-sized blocks of
the compressed inverse ${\bf R}$.

Solving the problem in the example of Section~\ref{sec:analogy} again,
we now take the number of panels on the coarse mesh as {\tt
  npan=0.6*omega+48} rounded to the nearest integer.
Figure~\ref{fig:conv13} shows results from the program {\tt
  demo13b.m}. The resonances, visible in Figure~\ref{fig:conv12}, are
now gone. It is interesting to observe in Figure~\ref{fig:conv13}
that, despite the presence of several singular operators and
compositions in~(\ref{eq:NeuHelm3}), the results produced with RCIP
are essentially fully accurate and the number of GMRES iterations
needed for convergence grows very slowly with $\omega$.

The program {\tt demo13c.m} differs from {\tt demo13b.m} in that it
uses {\it local regularization}~\cite[Section 2.1]{Hels09JCP} for the
Cauchy-singular operators of~(\ref{eq:C1}) and~(\ref{eq:C2}) rather
than local panelwise evaluation. The results produced by the two
programs are virtually identical. We do not show yet another figure.

\begin{figure}[!t]
  \centering \includegraphics[height=80mm]{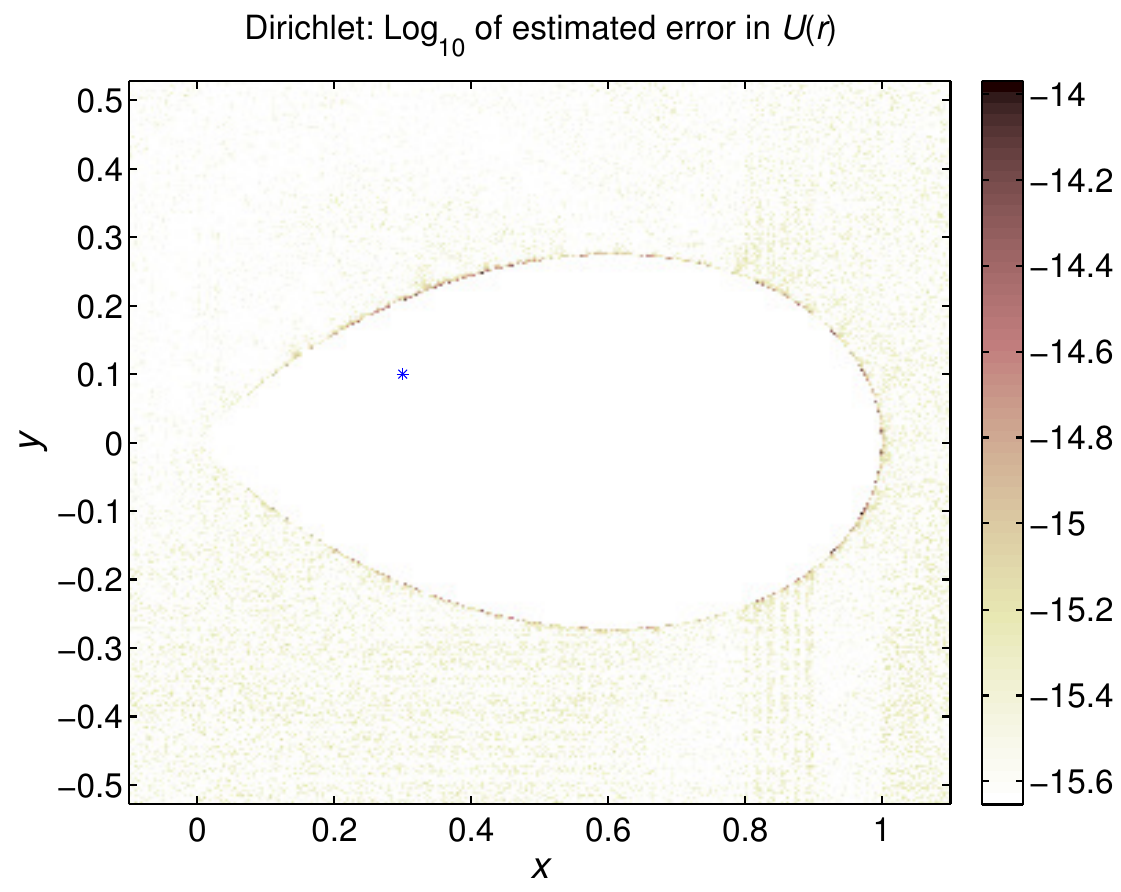}
\caption{\sf The error in the solution $U(r)$ to the exterior Dirichlet
  Helmholtz problem. The coarse grid on $\Gamma$ has 896
  discretization points and $U(r)$ is evaluated at 62392 points on a
  Cartesian grid in $E$ using {\tt demo11b.m} with $\omega=10$. The
  source at $r'=(0.3,0.1)$ is shown as a blue star.}
\label{fig:Dirifield}
\end{figure}

\begin{figure}[!ht]
  \centering \includegraphics[height=80mm]{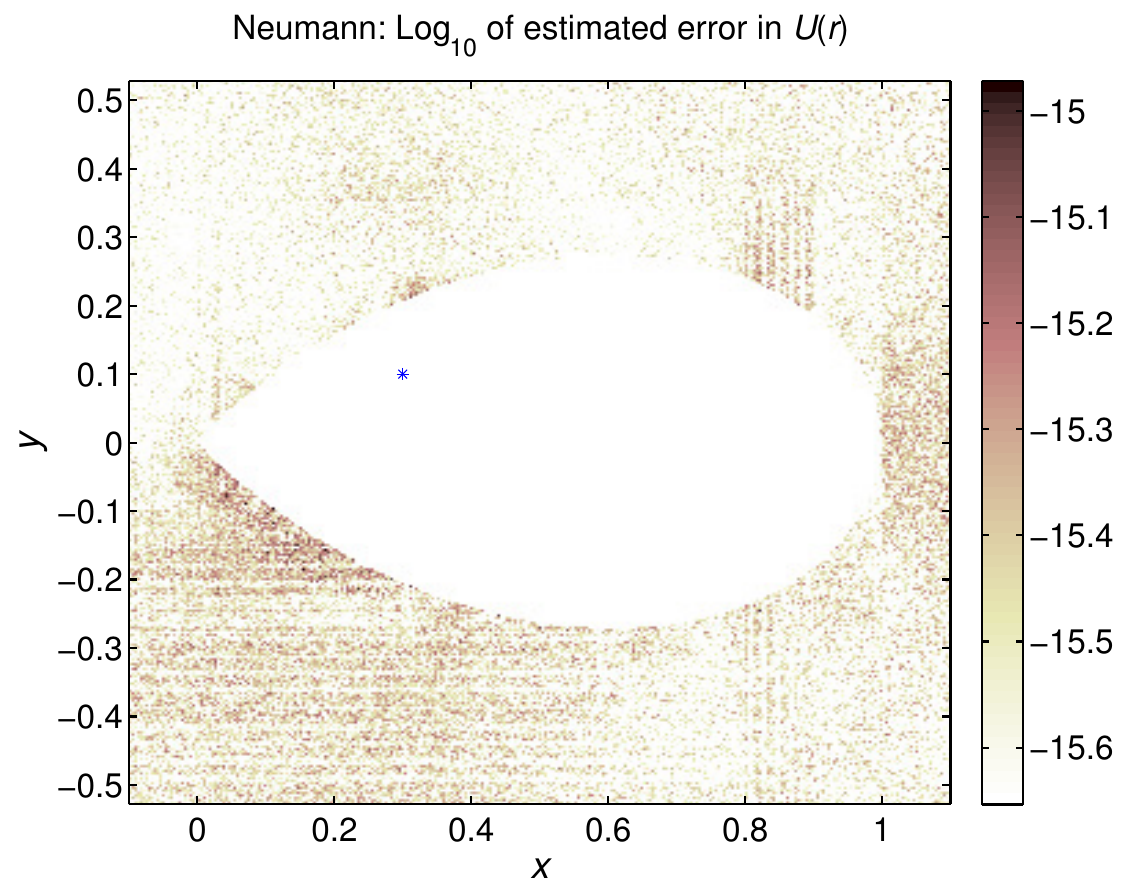}
\caption{\sf Same as Figure~\ref{fig:Dirifield}, but the exterior Neumann
  Helmholtz problem is solved using {\tt demo13d.m}. The accuracy is
  even higher than in Figure~\ref{fig:Dirifield}.}
\label{fig:Neufield}
\end{figure}

\begin{figure}[!t]
  \centering \includegraphics[height=80mm]{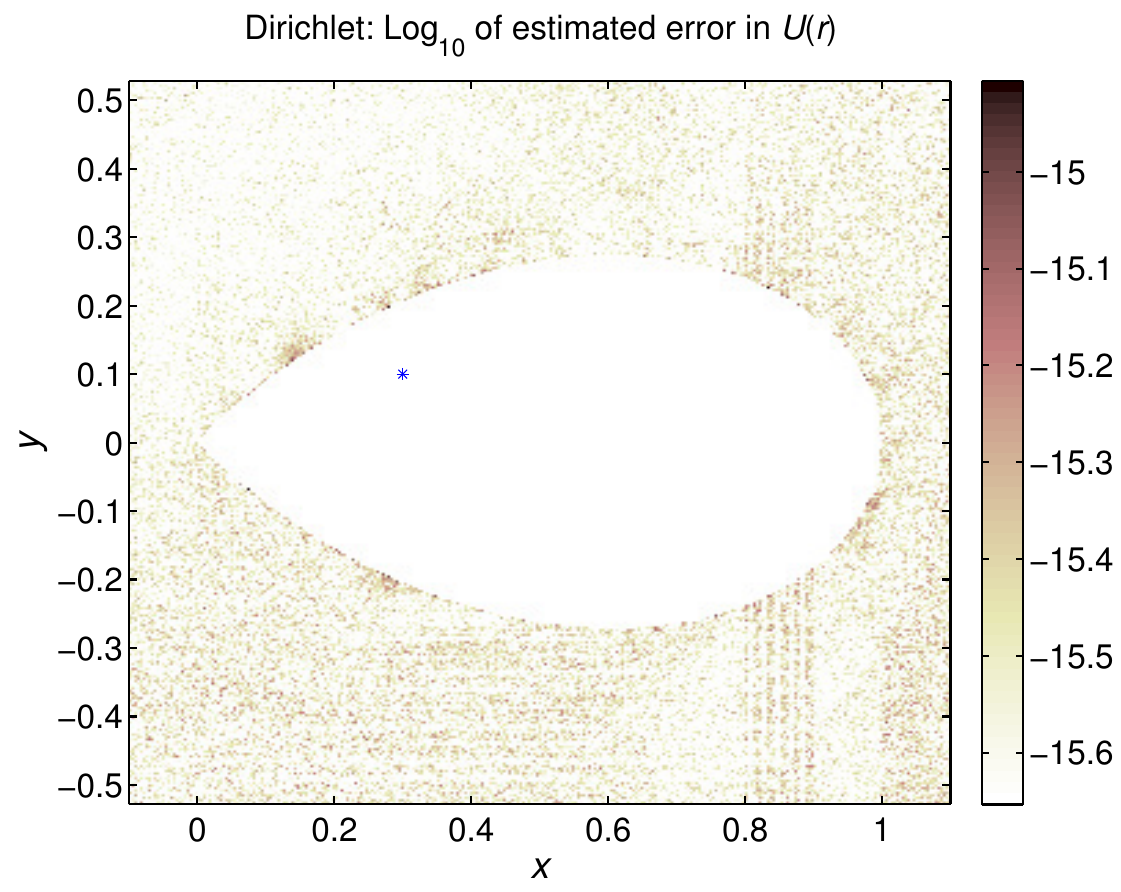}
\caption{\sf Same as Figure~\ref{fig:Dirifield}, but with {\tt
    demo11e.m}. Local coordinates are used on $\Gamma$ and the
  accuracy in $U(r)$ is improved with up to one digit.}
\label{fig:DirifieldPL}
\end{figure}

\section{Field evaluations}
\label{sec:HelmField}

Strictly speaking, a boundary value problem is not properly solved
until its solution can be accurately evaluated in the {\it entire}
computational domain. The program {\tt demo11b.m} is a continuation of
{\tt demo11.m} which, after solving~(\ref{eq:DiriHelm}) for
$\tilde{\boldsymbol{\rho}}_{\rm coa}$ with RCIP and forming
$\hat{\boldsymbol{\rho}}_{\rm coa}$ via~(\ref{eq:wcd}), computes the
solution $U(r)$ via~(\ref{eq:Helmrep1}) using three slightly different
discretizations:
\begin{itemize}
\item[(i)] When $r$ is away from $\Gamma$, 16-point Gauss--Legendre
  quadrature is used in~(\ref{eq:Helmrep1}) on all quadrature panels.
\item[(ii)] When $r$ is close to $\Gamma$, but not close to a panel
  neighboring a corner, 16-point Gauss--Legendre quadrature is used
  in~(\ref{eq:Helmrep1}) on panels away from $r$ and local panelwise
  evaluation is used for panels close to $r$.
\item[(iii)] When $r$ is close to a panel neighboring a corner, the
  density $\tilde{\boldsymbol{\rho}}_{\rm coa}$ is first used to
  reconstruct $\hat{\boldsymbol{\rho}}_{\rm part}$ according to
  Section~\ref{sec:recon}. Then 16-point Gauss--Legendre quadrature is
  used in~(\ref{eq:Helmrep1}) on panels away from $r$ and local
  panelwise evaluation is used for panels close to $r$.
\end{itemize}
The first two discretizations only use the coarse grid on $\Gamma$.
The third discretization needs a grid on a partially refined mesh on
$\Gamma$.

The program {\tt demo13d.m} is a continuation of {\tt demo13b.m}
which, after solving~(\ref{eq:NeuHelm2}) with RCIP as described in
Section~\ref{sec:regul}, computes the solution $U(r)$
via~(\ref{eq:Helmrep3}) using the three discretizations of the
previous paragraph.

Figure~\ref{fig:Dirifield} and~\ref{fig:Neufield} show that RCIP in
conjunction with the quadrature of~\cite[Section 2]{Hels09JCP} can
produce very accurate solutions to exterior Helm\-holtz problems in,
essentially, the entire computational domain.

The main source of error in the computed field $U(r)$ of
Figure~\ref{fig:Dirifield} is cancellation in the evaluation of the
difference $r-r'$ for $r\in\Gamma$. The program {\tt demo11b.m} needs
such differences in the discretized kernels of~(\ref{eq:Koper})
and~(\ref{eq:Soper}) and the vectors $r$ and $r'$ are individually
evaluated in global coordinates via~(\ref{eq:gamma}). The program {\tt
  demo11e.m} is the same as {\tt demo11b.m}, but with $r-r'$ computed
in local coordinates whenever $r'$ is close to $r\in\Gamma$. A
comparison of Figure~\ref{fig:DirifieldPL} with
Figure~\ref{fig:Dirifield} shows that the use of local coordinates on
$\Gamma$ lead to an improved quality in $\rho(r)$ which, in turn,
affects $U(r)$. The improvement is most pronounced for $U(r)$ close to
$\Gamma$.

Further examples of Helmholtz problems in non-smooth exterior domains
and more details on the discretization of Hankel kernels are found
in~\cite{HelsHols15,HelsKarl13}.

\section{A Helmholtz transmission problem}
\label{sec:HelmTrans}

This section reviews some results from~\cite{HelsKarl18}. A
transmission problem for the Helmholtz equation is formulated as
\begin{align}
  \Delta U(r)+\omega_1^2 U(r)&=0\,, \quad r\in E\,,\\
  \Delta U(r)+\omega_2^2 U(r)&=0\,, \quad r\in D\,,\\
  \lim_{E\ni r\to r^\circ} U(r)&=
  \lim_{D\ni r\to r^\circ} U(r)\,,
  \quad r^\circ\in\Gamma\,,\label{eq:HelmTr1}\\
  \lim_{E\ni r\to r^\circ}\varepsilon\nu^\circ\cdot\nabla U(r)&=
  \lim_{D\ni r\to r^\circ}\nu^\circ\cdot\nabla U(r)\,,
  \quad r^\circ\in\Gamma\,,\label{eq:HelmTr2}
\end{align}
where $\varepsilon$ is a material parameter and
$\omega_2=\sqrt{\varepsilon}\omega_1$. We separate $U(r)$ into an
incident field $U^{\rm in}(r)$ and a scattered field, represented by
two layer densities $\mu$ and $\rho$ and a uniqueness parameter $c$,
so that for $r\in E$
\begin{equation}
U(r)=U^{\rm in}(r)
 +\int_{\Gamma}\frac{\partial\Phi_{\omega_1}}{\partial\nu'}
                       (r,r')\mu(r')\,{\rm d}\ell'
 +\int_{\Gamma}\Phi_{\omega_1}(r,r')\rho(r')\,{\rm d}\ell'
\label{eq:Hrep1}
\end{equation}
and for $r\in D$
\begin{equation}
U(r)=\varepsilon\int_{\Gamma}\frac{\partial\Phi_{\omega_2}}{\partial\nu'}
                        (r,r')\mu(r')\,{\rm d}\ell'
              +c\int_{\Gamma}\Phi_{\omega_2}(r,r')\rho(r')\,{\rm d}\ell'\,.
\label{eq:Hrep2}
\end{equation}
By this, the scattered field satisfies the outgoing radiation
condition~(\ref{eq:goout}).

Insertion of~(\ref{eq:Hrep1}) and~(\ref{eq:Hrep2})
into~(\ref{eq:HelmTr1}) and~(\ref{eq:HelmTr2}) gives the system of
integral equations~\cite[Eq.~(4.2)]{KleiMart88} with compact
(differences of) operators
\begin{equation}
\begin{bmatrix}
 I-\alpha_2K_{\omega_2}+\alpha_1K_{\omega_1}
& -\alpha_1(cS_{\omega_2}-S_{\omega_1}) \\
   \alpha_4(T_{\omega_2}-T_{\omega_1})
&I+c\alpha_3K^{\rm A}_{\omega_2}-\alpha_4K^{\rm A}_{\omega_1} 
\end{bmatrix}
\begin{bmatrix}
\mu(r)\\
\rho(r)
\end{bmatrix}
=
\begin{bmatrix}
 g_1(r)\\
 g_2(r)
\end{bmatrix}\,,
\label{eq:Mullsys}
\end{equation}
and with
\begin{gather}
g_1(r)=-2\alpha_1U^{\rm in}(r)\,, \qquad
g_2(r)= 2\alpha_4\frac{\partial U^{\rm in}}{\partial\nu}(r)\,,\\
\alpha_1=\frac{1}{1+\varepsilon}\,,\quad
\alpha_2=\frac{\varepsilon}{1+\varepsilon}\,,\quad
\alpha_3=\frac{1}{c+\varepsilon}\,,\quad
\alpha_4=\frac{\varepsilon}{c+\varepsilon}\,.
\end{gather}

\begin{figure}[t]
\centering 
\noindent\makebox[\textwidth]{
\begin{minipage}{1.1\textwidth}
  \includegraphics[height=52mm]{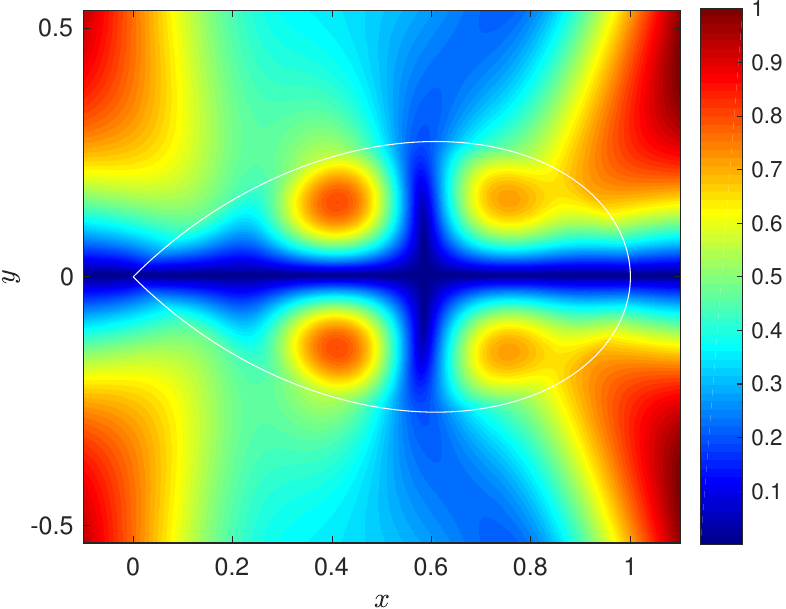}
  \includegraphics[height=52mm]{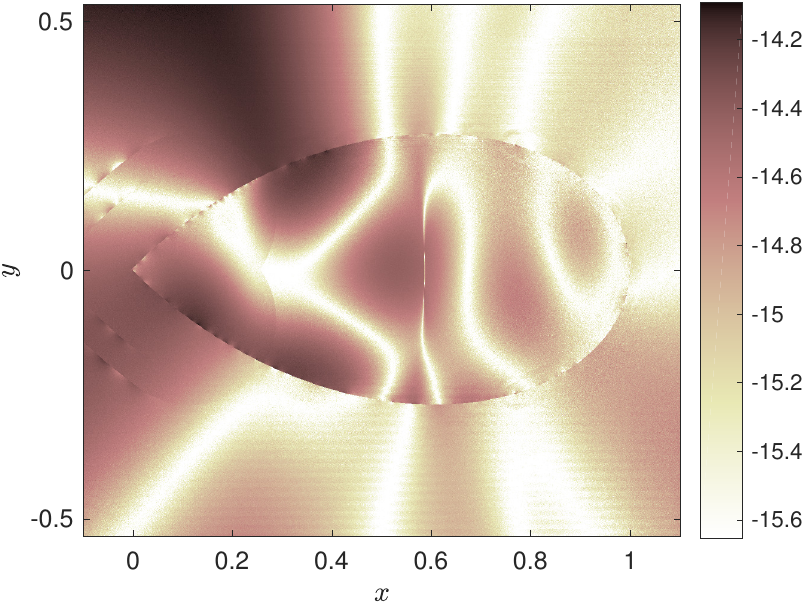}
\end{minipage}}
\caption{\sf A normalized eigenfield $|U(r)|$. The
  eigenwavenumber is $\omega_1=9.701129417644246-2.000374579086419{\rm
    i}$ and $\varepsilon=2.25$. The coarse grid on $\Gamma$ has 320
  discretization points. $U(r)$ is evaluated at $10^6$ field points on
  a (rectangular) Cartesian grid. Right: $\log_{10}$ of estimated
  absolute error in $|U(r)|$. The program {\tt demo19.m} is used.}
\label{fig:Helmtrans1}
\end{figure}

\begin{figure}[t!]
\centering 
\noindent\makebox[\textwidth]{
\begin{minipage}{1.1\textwidth}
  \includegraphics[height=52mm]{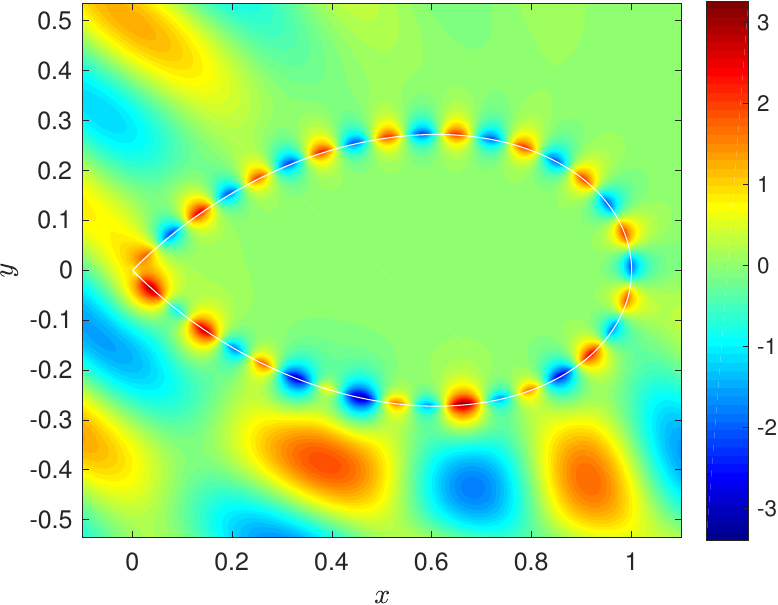}
  \includegraphics[height=52mm]{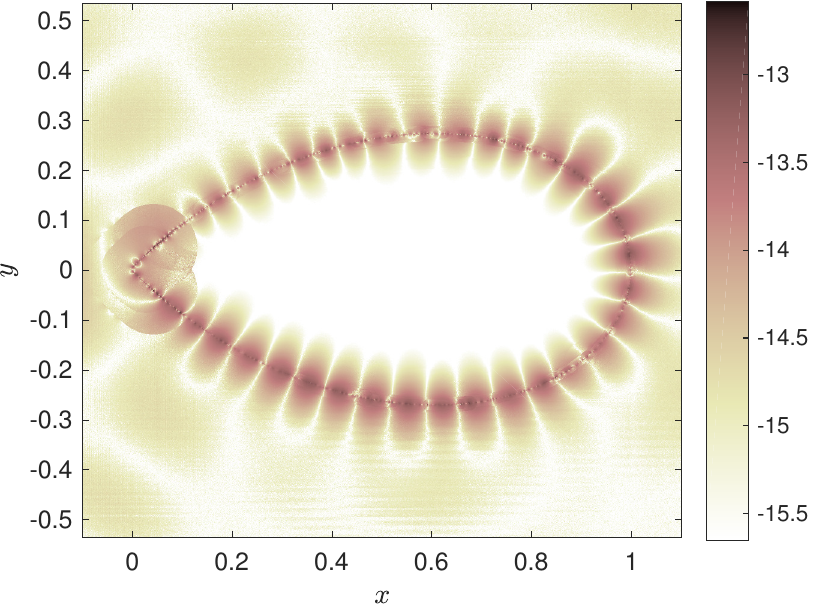}
\end{minipage}}
\caption{\sf The field $\Re\{U^+(r)\}$ with $\omega_1=18$ and 
  $\varepsilon=-1.1838$. The coarse grid on $\Gamma$ has 800
  discretization points. Right: $\log_{10}$ of estimated absolute
  error in $U^+(r)$. The program {\tt demo19b.m} is used.}
\label{fig:Helmtrans2}
\end{figure}

We first apply RCIP to~(\ref{eq:Mullsys}) for the purpose of computing
eigenfields. The boundary $\Gamma$ is as in~(\ref{eq:gamma}) with
$\theta=\pi/2$. We set $\varepsilon=2.25$, $U^{\rm in}(r)=0$, and
$c=1$ (which is a common choice in the literature) and look for
$\omega_1$, $\mu$, $\rho$ that are non-trivial solutions to the
homogeneous system~(\ref{eq:Mullsys}). Unfortunately, the
system~(\ref{eq:Mullsys}) admits false eigenwavenumbers, that is
non-trivial solutions with $\Im\{\omega_1\}<0$ whose corresponding
$\mu$ and $\rho$ generate fields $U(r)$ that vanish when inserted
in~(\ref{eq:Hrep1}) and~(\ref{eq:Hrep2}). Nevertheless, a true
eigenfield is found at
$\omega_1=9.701129417644246-2.000374579086419{\rm i}$ and shown in
Figure~\ref{fig:Helmtrans1} along with estimated field error. The
coarse grid on $\Gamma$ has 320 discretization points. The reference
solution is computed with 50 per cent more points. The program {\tt
  demo19.m} is used.

We then compute the field $\Re\{U(r)\}$ in the limit of $\varepsilon$
approaching the point $\varepsilon=-1.1838$ from above in the complex
$\varepsilon$-plane. We set $\omega_1=18$, $c=-{\rm i}$, $U^{\rm
  in}(r)=e^{{\rm i}\omega_1(r\cdot d)}$ with
$d=\left(\cos(\pi/4),\sin(\pi/4)\right)$, and use 800 discretization
points on the coarse grid on $\Gamma$. The program, {\tt demo19b.m},
is an extension of {\tt demo19.m}: the construction of the initializer
${\bf R}_*$ is accelerated using Newton's method, as described in
Section~\ref{sec:Newton}, and a homotopy method is used for the limit
$\Im\{\varepsilon\}\to 0^+$, see~\cite[Section 6.3]{Hels11JCPb}.
Figure~\ref{fig:Helmtrans2} shows results.

\begin{figure}[!b]
\centering 
\includegraphics[height=74mm]{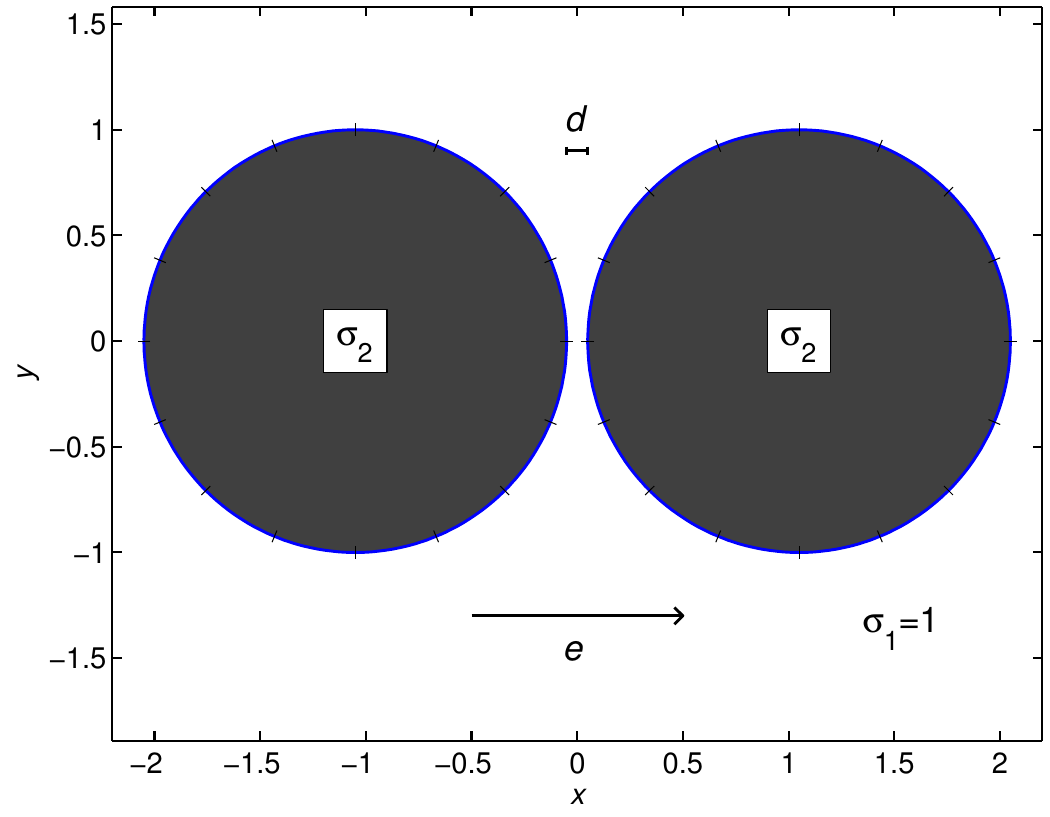}
\caption{\sf Two unit disks with conductivity $\sigma_2>1$ are separated
  by a distance $d$ and embedded in an infinite plane with
  conductivity $\sigma_1=1$. A unit field $e$ is applied in the
  $x$-direction. Equi-sized quadrature panels are placed on the
  circles in such a way that there are breakpoints at $r=(-d/2,0)$ and
  $r=(d/2,0)$.}
\label{fig:twocirc}
\end{figure}

\section{Close-to-touching objects}
\label{sec:ctt}

The usefulness of RCIP is not restricted to corner problems. RCIP
works well also in more general contexts where solutions to integral
equations exhibit some sort of (near) singularities. This section is
about two such problems. First we compute the polarizability $q$ of a
pair of close-to-touching and highly conducting unit disks embedded in
a background unit medium. Then we proceed to doubly periodic boundary
conditions and compute the effective conductivity $\sigma_{\rm eff}$
of a square array of conducting disks.

\subsection{The two-disk problem}

The setup is shown in Figure~\ref{fig:twocirc}. This problem can be
modeled with~(\ref{eq:inteq1a}) and~(\ref{eq:q}) and
\begin{equation}
\lambda=\frac{\sigma_2-\sigma_1}{\sigma_2+\sigma_1}\,.
\end{equation}
To avoid stability problems for $\lambda$ close to one, we instead use
an alternative formulation which in complex notation
reads~\cite[Eqs.~(9,10)]{Hels08b}
\begin{gather}
\mu(z)+\frac{\lambda}{\pi}\int_{\Gamma}\mu(\tau)
\Im\left\{\frac{{\rm d}\tau}{\tau-z}\right\}=
2\lambda\Im\left\{\bar{e}z\right\}\,, \quad z\in\Gamma\,,
\label{eq:inteq1c}\\
q=-\sigma_1\int_{\Gamma}\mu(z)\Re\left\{\bar{e}\,{\rm d}z\right\}\,.
\label{eq:q3}
\end{gather}

\begin{figure}[t]
\centering \includegraphics[height=33mm]{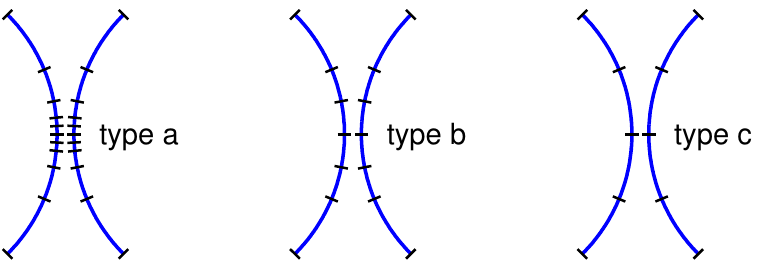}
\caption{\sf Meshes of type {\sf a}, type {\sf b}, and type {\sf c} on
  the boundary subset $\Gamma_i^\star$ for the two-disk problem and
  for $i=n_{\rm sub}=3$. The type {\sf a} mesh has $8+4i$ panels. The
  type {\sf b} mesh has twelve panels. The type {\sf c} mesh has eight
  panels. The type {\sf a} mesh is the restriction of the fine mesh to
  $\Gamma_i^\star$. For $i=n_{\rm sub}$, the type {\sf c} mesh is the
  restriction of the coarse mesh to $\Gamma^\star$. The type {\sf a}
  mesh and the type {\sf b} mesh coincide for $i=1$.}
\label{fig:typesc}
\end{figure}

RCIP can now be applied by considering $r=(-d/2,0)$ and $r=(d/2,0)$ to
be singular boundary points treated in tandem. The subset
$\Gamma^\star$ then covers the eight panels (four on each disk
boundary) that are closest to the origin. Families of twelve-panel
type {\sf b} meshes are constructed in analogy with the procedure in
Section~\ref{sec:nest}. The superscript $\star$ in ${\bf K}_{i{\rm
    b}}^\star$ indicates that only entries with both indices
corresponding to points on the eight innermost panels of a type {\sf
  b} mesh are retained. The derivation of the recursion in Appendix~D
uses meshes of type {\sf a} and type {\sf c} with twice the number of
panels compared to the single corner case. See Figure~\ref{fig:typesc}
and compare Figure~\ref{fig:types}.

The two-disk problem of Figure~\ref{fig:twocirc} is, in a sense,
harder to solve than the one-corner model problem of
Section~\ref{sec:model}. The reason being that the fine mesh on
$\Gamma$ for the two-disk problem has many panels that lie close to
each other and where special quadratures techniques, see
Section~\ref{sec:sing}, need to be activated in the discretization
of~(\ref{eq:inteq1c}). This, in turn, slows down convergence and may
even endanger the validity of the basic assumptions~(\ref{eq:decomp})
and~(\ref{eq:decomp2}) upon which the entire RCIP scheme rests. The
prolongation in~(\ref{eq:decomp}) and~(\ref{eq:decomp2}) only holds on
panels where standard quadrature is sufficient. In the one-corner
model problem, on the other hand, special quadrature is barely needed.
The basic assumptions~(\ref{eq:decomp}) and~(\ref{eq:decomp2}) hold
with, say, standard 16-point Gauss--Legendre quadrature provided that
the opening angle $\theta$ is not too small.

The families of meshes introduced in Figure~\ref{fig:typesc} are
constructed with the validity of~(\ref{eq:decomp})
and~(\ref{eq:decomp2}) in mind. When constructing ${\bf K}_{i{\rm
    b}}^\circ$ in the recursion~(\ref{eq:recur}) and
in~(\ref{eq:decomp2}), special quadrature may be activated in the
discretization on meshes of type {\sf b} if needed. Note, however,
that the need for special quadrature will only arise for source points
on the eight panels farthest away from the origin. For source points
on the innermost four panels of type {\sf b} meshes, standard
quadrature is enough. The same is true for ${\bf K}_{i{\rm a}}^\circ$
in~(\ref{eq:decomp2}): special quadrature on meshes of type {\sf a}
needs only to be activated on the eight panels farthest away from the
origin which are common to type {\sf a} and type {\sf b} meshes and
where no prolongation takes place. Therefore~(\ref{eq:decomp2}) holds
provided the order of the Gauss--Legendre quadrature is sufficiently
high. Numerical experiments indicate that 22-point quadrature is
sufficient.

\begin{figure}[t]
\centering 
\noindent\makebox[\textwidth]{
\begin{minipage}{1.1\textwidth}
\includegraphics[height=66mm]{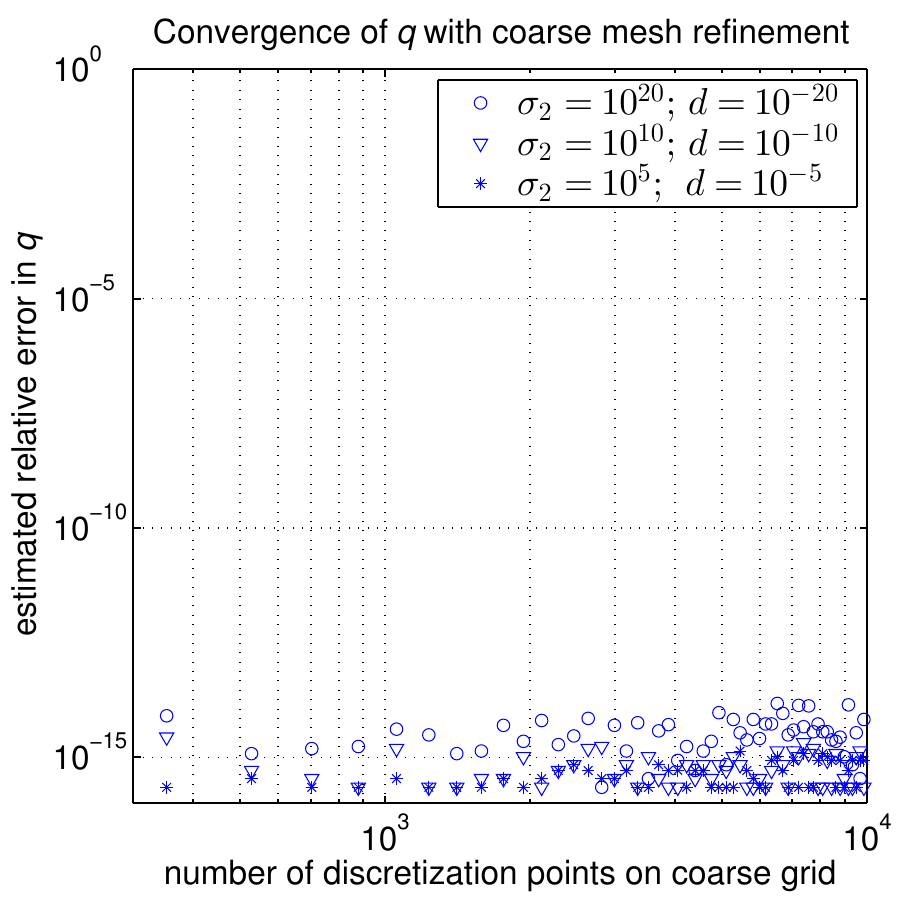}
\includegraphics[height=66mm]{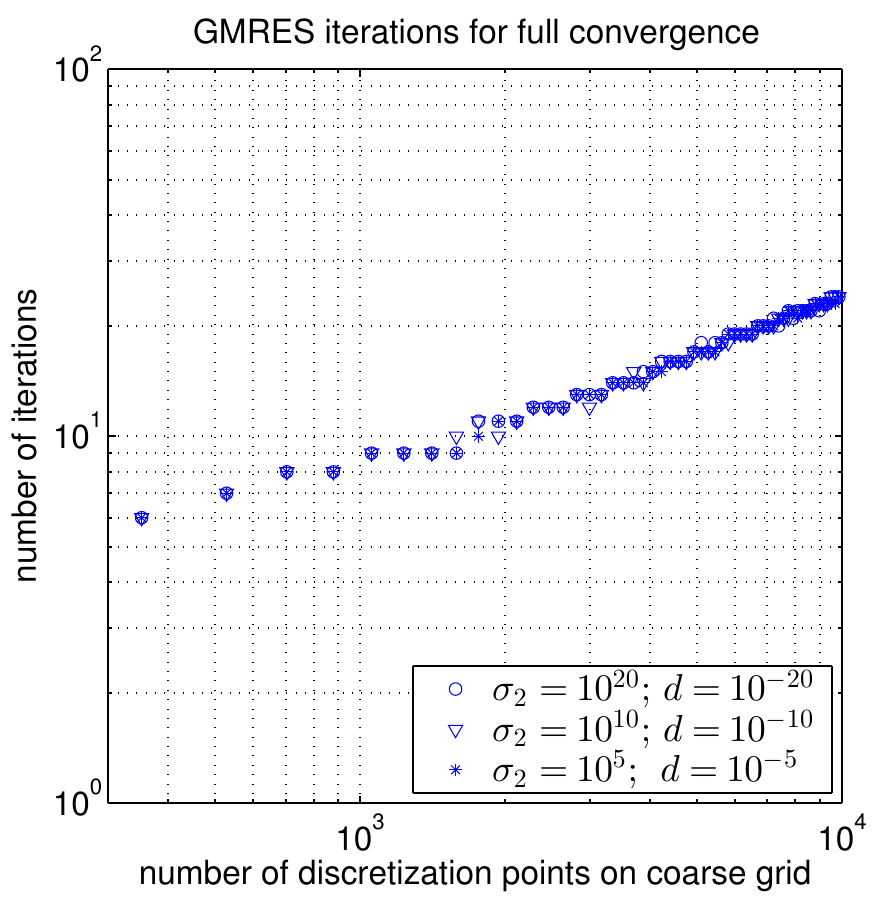}
\end{minipage}}
\caption{\sf The two-disk problem of Figure~\ref{fig:twocirc} for the 
  electrostatic equation and with RCIP applied to~(\ref{eq:inteq1c})
  and~(\ref{eq:q3}). The program {\tt demo14.m} is used. Left: the
  estimated relative error in the polarizability $q$ for various disks
  separation distances $d$ and disk conductivities $\sigma_2$. Right:
  the number of GMRES iterations needed to meet an estimated relative
  residual of $\epsilon_{\rm mach}$.}
\label{fig:conv14}
\end{figure}

Figure~\ref{fig:conv14} illustrates the performance of RCIP applied
to~(\ref{eq:inteq1c}) and~(\ref{eq:q3}) for the two-disk problem. The
program {\tt demo14.m} is used. Convergence is immediate and it
appears as if rather extreme cases can be treated accurately.
Reference values for $q$ can be found in {\tt demo14.m}.

\medskip\noindent 
{\bf Remark:} The two-disk problem was addressed
in~\cite[Section~10.3]{Hels08b}, but not solved with RCIP in its
entirety due to the too simplistic mesh construction technique used
in~\cite{Hels08b}.

\begin{figure}[t]
\centering 
\noindent\makebox[\textwidth]{
\begin{minipage}{1.1\textwidth}
\includegraphics[height=66mm]{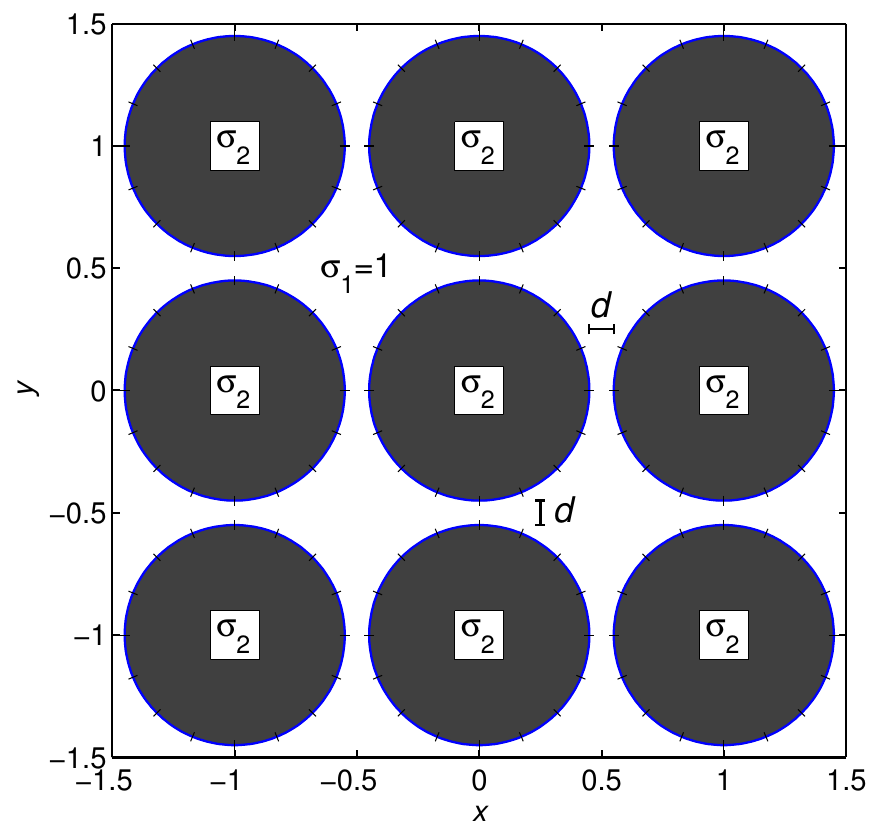}
\includegraphics[height=66mm]{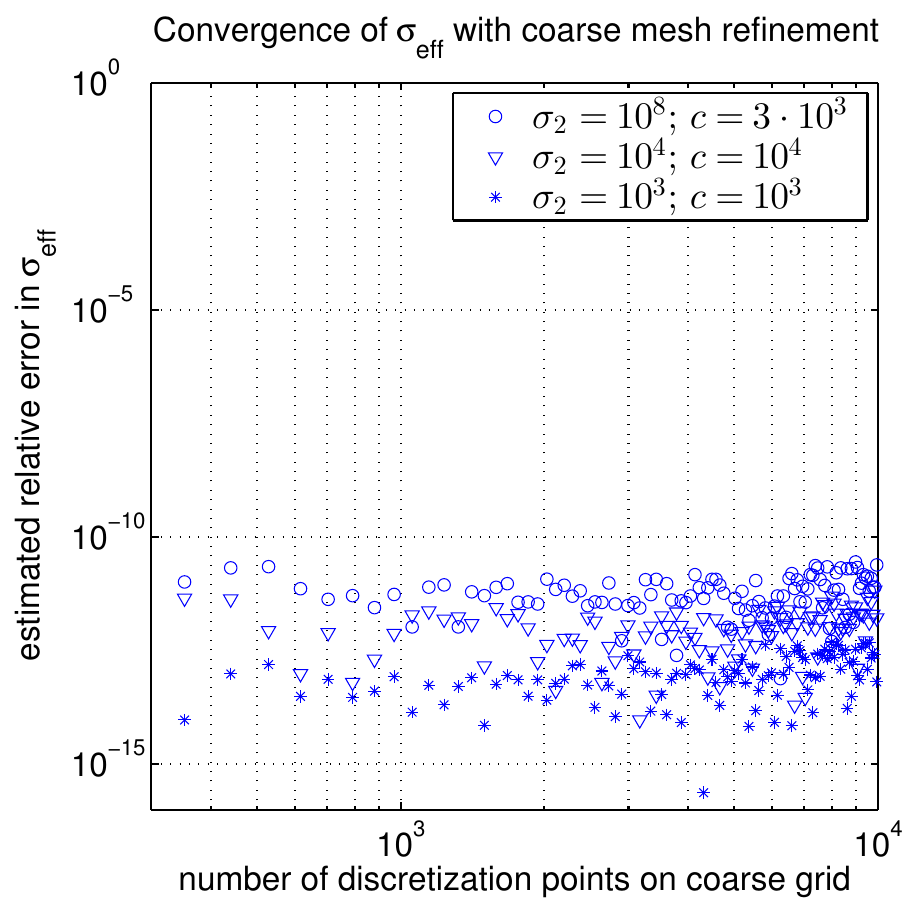}
\end{minipage}}
\caption{\sf Left: a nine-unit-cell cutout from a square array of
  disks with conductivities $\sigma_2$. The background medium has
  $\sigma_1=1$. Right: the estimated relative error in $\sigma_{\rm
    eff}$ of~(\ref{eq:q4}) for three setups with different $\sigma_2$
  and separation distances $d=1/(c^2+c\sqrt{c^2-1})$. The program {\tt
    demo14b.m} is used.}
\label{fig:ninedisk}
\end{figure}

\subsection{The square array of disks}

The left image of Figure~\ref{fig:ninedisk} shows the geometry of this
classic problem. The following modification of~(\ref{eq:inteq1c})
and~(\ref{eq:q3}) is used for modeling
\begin{gather}
\mu(z)+\frac{\lambda}{\pi}\int_{\Gamma_{\rm per}}\mu(\tau)
\Im\left\{\frac{{\rm d}\tau}{\tau-z}\right\}=
2\lambda\Im\left\{\bar{e}z\right\}\,, \quad z\in\Gamma_{\rm unit}\,,
\label{eq:inteq1d}\\
\sigma_{\rm eff}=\sigma_1-\sigma_1\int_{\Gamma_{\rm unit}}
\mu(z)\Re\left\{\bar{e}\,{\rm d}z\right\}\,,
\label{eq:q4}
\end{gather}
where $\Gamma_{\rm per}$ refers to all disk interfaces in the plane,
$\Gamma_{\rm unit}$ refers to the disk interface in the unit cell,
see~\cite[Eqs.~(13) and (14)]{Hels98}, and $\sigma_{\rm eff}$ is the
effective conductivity. The applied electric field is chosen as
$e=(1,0)$.

The disk separation distance $d$ of Figure~\ref{fig:ninedisk} may be
expressed in terms of a parameter $c$ as
\begin{equation}
d=\frac{1}{c^2+c\sqrt{c^2-1}}\,.
\end{equation}
The higher the number $\min\{c,\sigma_2/\sigma_1\}$, the more
difficult it is to compute $\sigma_{\rm eff}$ via traditional
numerical methods~\cite{McPh88}.

We solve~(\ref{eq:inteq1d},\ref{eq:q4}) for three setups:
$\sigma_2=10^8$ and $c=3\cdot 10^3$, which corresponds to the most
extreme parameter choices in~\cite{Chen98}; $\sigma_2=10^4$ and
$c=10^4$, which is the hardest test case of~\cite[Table~2]{Hels94};
and $\sigma_2=10^3$ and $c=10^3$, which is used both
in~\cite[Table~2]{Hels94} and~\cite[Table~1]{Hels96}. The right image
of Figure~\ref{fig:ninedisk} shows that the RCIP-accelerated Nyström
solver {\tt demo14b.m} resolves $\sigma_{\rm eff}$ to full achievable
accuracy already at 352 discretization points on the coarse grid on
$\Gamma_{\rm unit}$ and that the results are stable under mesh
refinement. The number of converged digits compares favorably to what
is reported in~\cite{Chen98,Hels94,Hels96}. Reference values and a
uniformly valid asymptotic expression~\cite{McPh88} for $\sigma_{\rm
  eff}$ are contained in {\tt demo14b.m}.

\section{Mixed boundary conditions}
\label{sec:zaremba}

Elliptic PDEs with mixed boundary conditions, that is, Dirichlet
conditions on parts of the boundary and Neumann conditions on the
remaining contiguous parts (also known as Zaremba boundary conditions)
can often be modeled using Fredholm second kind integral equations
with operators that are smooth away from the points where the boundary
conditions change type. In this context, too, RCIP improves the
stability and greatly reduces the computational cost of Nyström
discretization schemes.

The paper~\cite{Hels09JCP} shows how to apply RCIP to mixed planar
harmonic- and biharmonic problems. In this section we simply repeat
two of the experiments in~\cite{Hels09JCP} for the purpose of
disseminating the underlying {\sc Matlab} programs ({\tt demo15.m} and
{\tt demo15b.m}).

\begin{figure}[t]
  \centering 
\includegraphics[height=74mm]{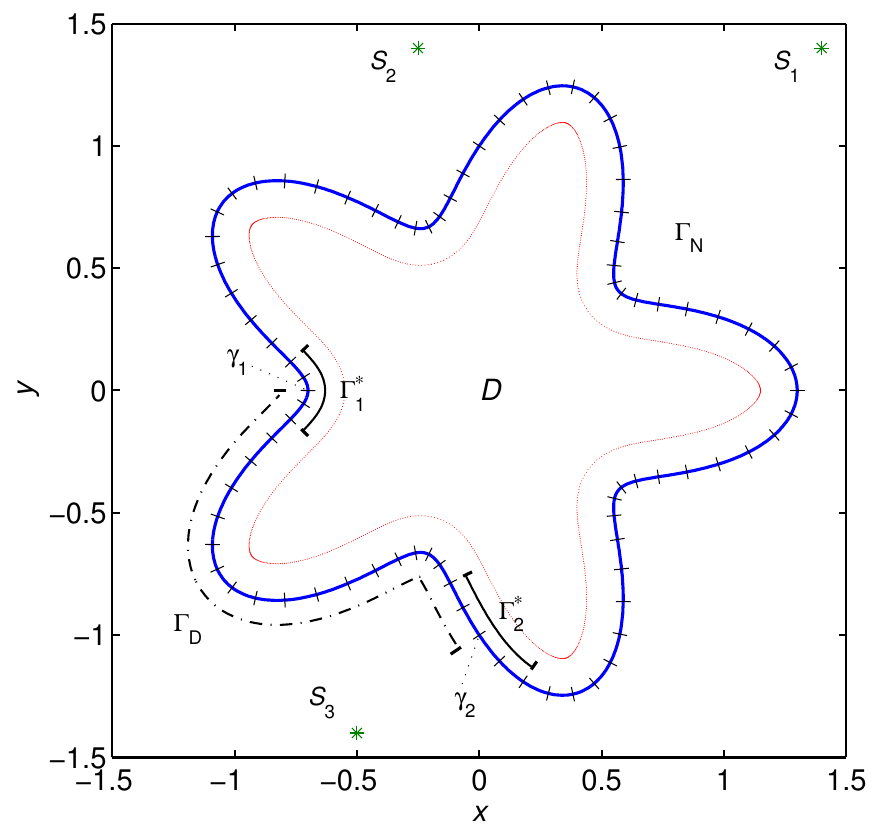}
\caption{\sf Interior domain $D$ with boundary 
  $\Gamma=\Gamma_{\rm D}\cup\Gamma_{\rm N}$ given
  by~(\ref{eq:starfish}) and~(\ref{eq:GDN}). A coarse mesh is
  constructed on $\Gamma$. Two parts of the boundary,
  $\Gamma_1^{\star}$ and $\Gamma_2^{\star}$, cover the four coarse
  panels closest to the singular boundary points $\gamma_1$ and
  $\gamma_2$ where the boundary conditions change type. The three
  sources $S_k$ of~(\ref{eq:bndrylap}), for the generation of boundary
  conditions, are marked by green stars. Thousand target points in $D$
  are marked by tiny red dots.}
\label{fig:amfig}
\end{figure}

The interior mixed problem for Laplace's equation is solved on the
domain $D$ bounded by the contour $\Gamma$ with the parameterization
\begin{equation}
r(s) = (1+0.3\cos(5s))(\cos(s),\sin(s))\,,\quad -\pi\le s\le\pi\,.
\label{eq:starfish}
\end{equation}

We seek a function $U(r)$, harmonic in $D$, such that
\begin{align}
\lim_{D\ni r\to r^\circ}U(r)&=g_{\rm D}(r^\circ)\,,
\quad r^\circ\in\Gamma_{\rm D}\,,
\label{eq:pot}\\
\lim_{D\ni r\to r^\circ}\nu^\circ\cdot\nabla U(r)&=g_{\rm N}(r^\circ)\,, 
\quad r^\circ\in\Gamma_{\rm N}\,,
\label{eq:flow}
\end{align}
where $g_{\rm D}(r)$ is Dirichlet data on the boundary part
$\Gamma_{\rm D}$, $g_{\rm N}(r)$ is Neumann data on the boundary part
$\Gamma_{\rm N}$, and $\Gamma_{\rm D}\cup\Gamma_{\rm N}=\Gamma$. See
Figure~\ref{fig:amfig}.

The solution $U(r)$, $r\in D\cup\Gamma_{\rm N}$, is represented by a
density $\rho(r)$, $r\in\Gamma$,
\begin{equation}
U(z)=
\frac{1}{\pi}\int_{\Gamma_{\rm D}}
\rho(\tau)\Im\left\{\frac{{\rm d}\tau}{\tau-z}\right\}-
\frac{1}{\pi}\int_{\Gamma_{\rm N}}
\rho(\tau)\log|\tau-z|\,{\rm d}|\tau|\,, \quad z\in
D\cup\Gamma_{\rm N}\,.
\label{eq:Urep}
\end{equation}

Insertion of~(\ref{eq:Urep}) into~(\ref{eq:pot}) and~(\ref{eq:flow})
gives the system
\begin{gather}
\rho(z)+
\frac{1}{\pi}\int_{\Gamma_{\rm D}}
\rho(\tau)\Im\left\{\frac{{\rm d}\tau}{\tau-z}\right\}-
\frac{1}{\pi}\int_{\Gamma_{\rm N}}
\rho(\tau)\log|\tau-z|\,{\rm d}|\tau|=g_{\rm D}(z)\,, 
\quad z\in \Gamma_{\rm D}\,,
\label{eq:lapsysa}\\
\rho(z)+
\frac{1}{\pi}\int_{\Gamma_{\rm D}}
\rho(\tau)\Im\left\{\frac{n_z\,{\rm d}\tau}{(\tau-z)^2}\right\}+
\frac{1}{\pi}\int_{\Gamma_{\rm N}}
\rho(\tau)\Re\left\{\frac{n_z\,{\rm d}|\tau|}{\tau-z}\right\}=g_{\rm N}(z)\,,
\quad z\in \Gamma_{\rm N}\,.
\label{eq:lapsysb}
\end{gather}

The boundary parts $\Gamma_{\rm D}$ and $\Gamma_{\rm N}$ are taken as
\begin{equation}
r(s)\in\Gamma_{\rm D}\,, \quad -\pi<s<-\frac{\pi}{2}\,,
\quad {\rm and} \quad
r(s)\in\Gamma_{\rm N}\,, \quad -\frac{\pi}{2}<s<\pi\,,
\label{eq:GDN}
\end{equation}
and the boundary conditions $g_{\rm D}(r)$ and $g_{\rm N}(r)$ are
constructed from a closed form reference solution
\begin{equation}
U_{\rm ref}(z)=\Re\left\{\sum_{k=1}^3 \frac{1}{z-S_k}\right\}\,,
\label{eq:bndrylap}
\end{equation}
where $S_1=1.4+1.4{\rm i}$, $S_2=-0.25+1.4{\rm i}$, and
$S_3=-0.5-1.4{\rm i}$ are sources outside of $D$, see
Figure~\ref{fig:amfig}.

\begin{figure}[!t]
\centering 
\noindent\makebox[\textwidth]{
\begin{minipage}{1.1\textwidth}
\includegraphics[height=66mm]{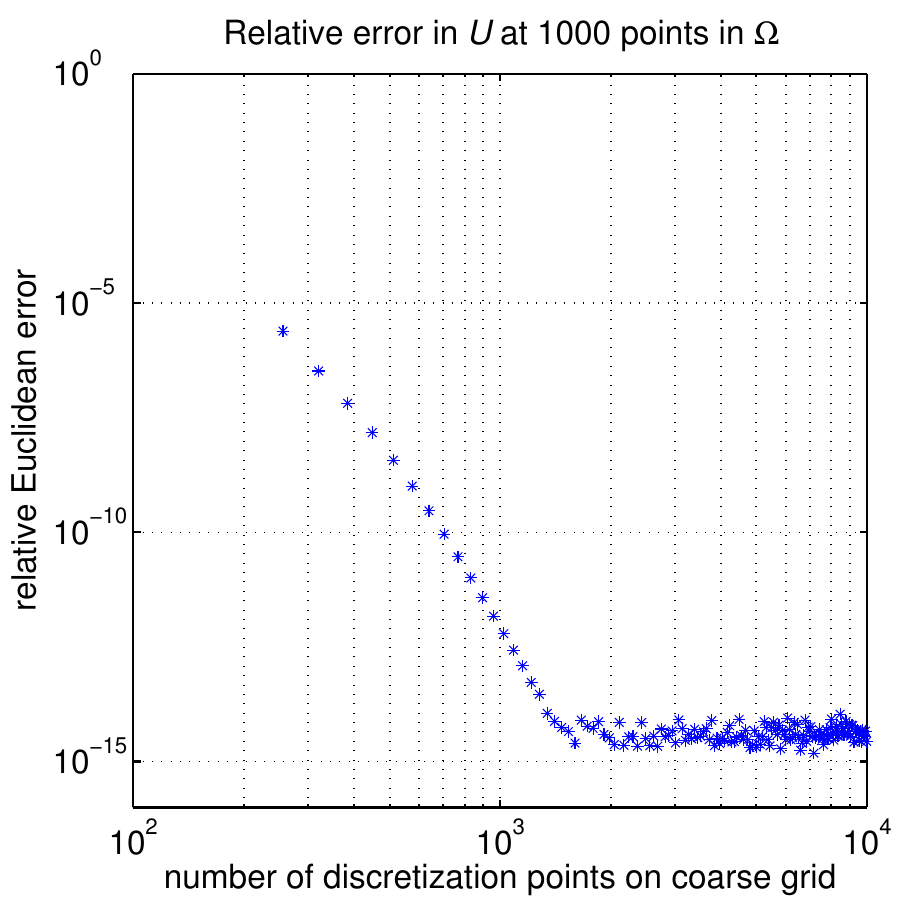}
\includegraphics[height=66mm]{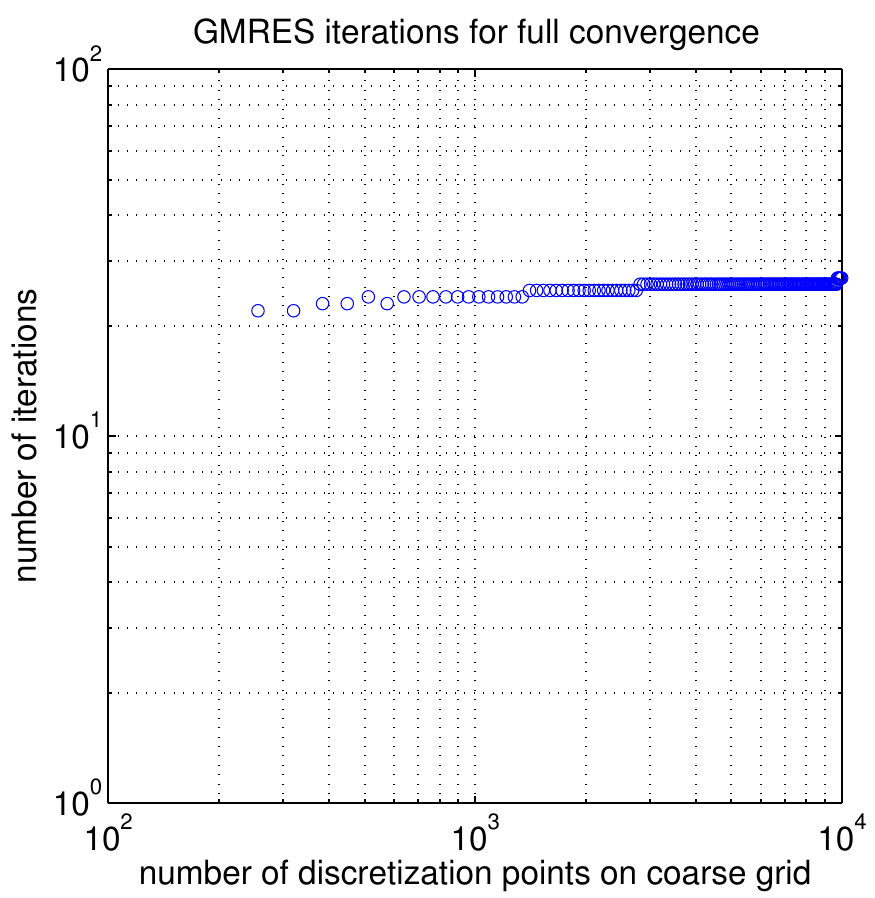}
\end{minipage}}
\caption{\sf The mixed problem for Laplace's equation and with RCIP 
  applied to~(\ref{eq:lapsysa}) and~(\ref{eq:lapsysb}). The program
  {\tt demo15.m} is used. Left: convergence of $U(r)$ at 1000 points
  $r\in D$ with coarse mesh refinement. Right: the number of GMRES
  iterations needed to meet an estimated relative residual of
  $\epsilon_{\rm mach}$.}
\label{fig:conv15}
\end{figure}

\begin{figure}[!ht]
\centering 
\noindent\makebox[\textwidth]{
\begin{minipage}{1.1\textwidth}
\includegraphics[height=66mm]{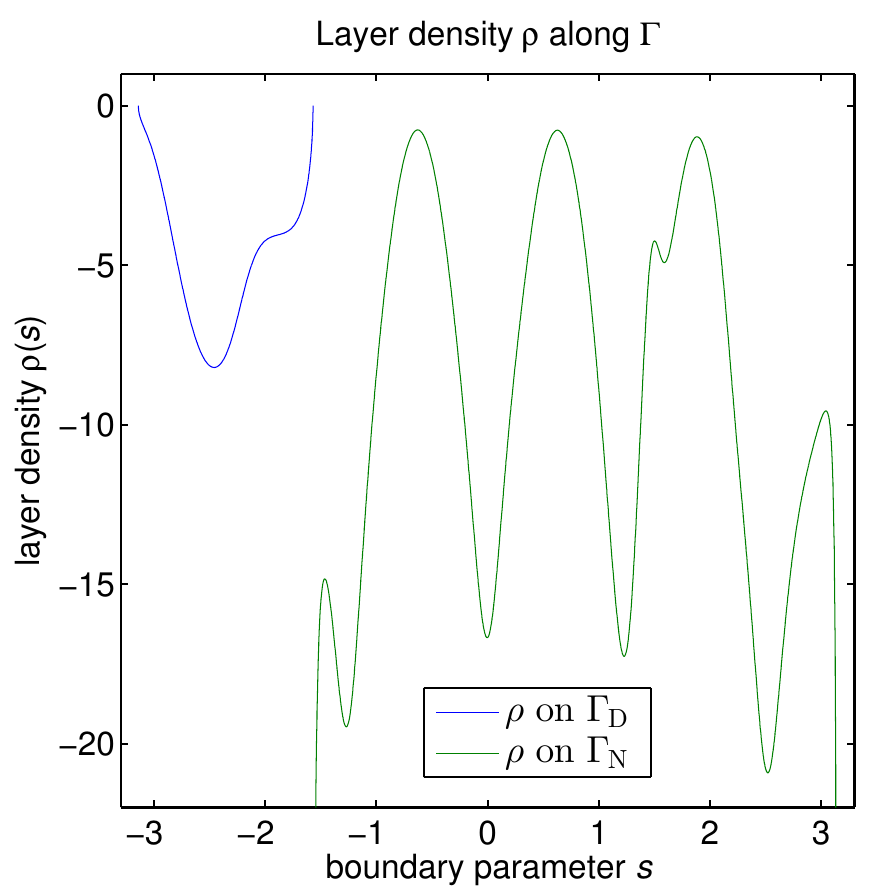}
\includegraphics[height=66mm]{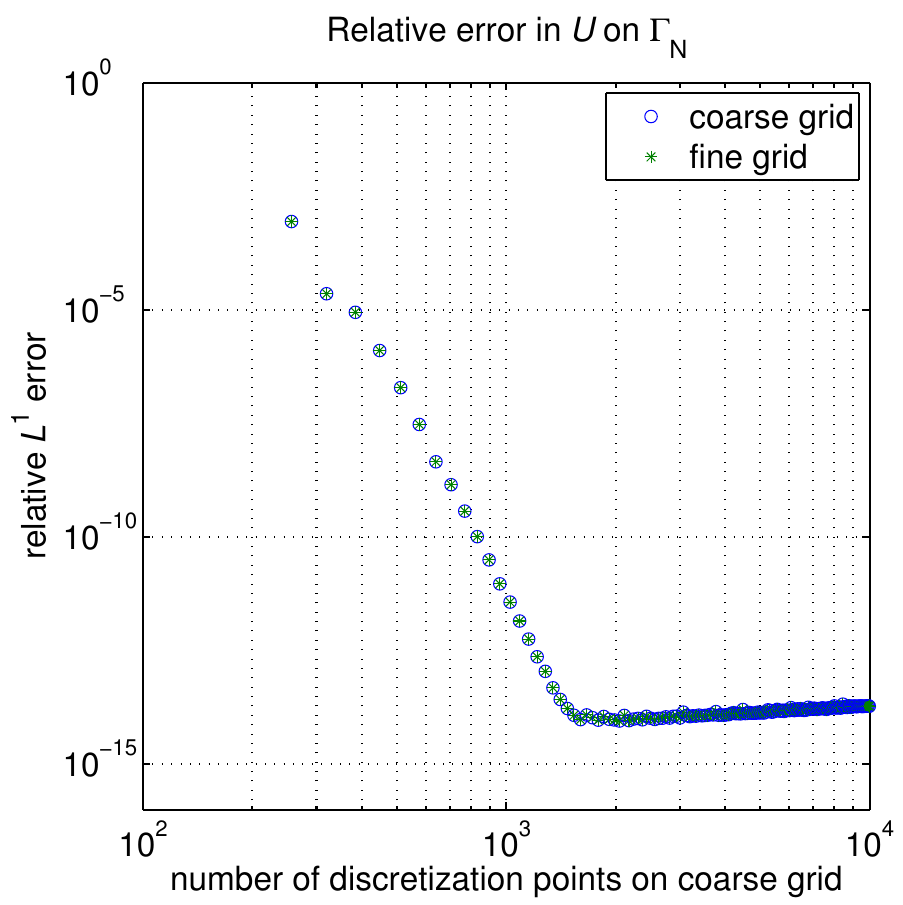}
\end{minipage}}
\caption{\sf Same as in Figure~\ref{fig:conv15} but {\tt demo15b.m} 
  is used. Left: reconstruction of the density $\rho$ on the fine grid
  on $\Gamma$. Right: error in the reconstruction of the solution
  $U(r)$, $r\in\Gamma_{\rm N}$, on the coarse grid and on the fine
  grid.}
\label{fig:conv15b}
\end{figure}

Figure~\ref{fig:conv15} illustrates the performance of RCIP applied
to~(\ref{eq:lapsysa}) and~(\ref{eq:lapsysb}). The program {\tt
  demo15.m} is used. The solution $U(r)$ is evaluated
via~(\ref{eq:Urep}) at the 1000 target points in $D$ indicated by red
dots in Figure~\ref{fig:amfig}. The rapid convergence and high
achievable accuracy seen in Figure~\ref{fig:conv15} means that RCIP
resolves the mixed problem very well.

The program {\tt demo15b.m} is about reconstruction. It is a
simplified version of a program used in~\cite{Hels09JCP}. Once the
solution $\tilde{\boldsymbol{\rho}}_{\rm coa}$ is obtained, the
discrete density $\boldsymbol{\rho}_{\rm fin}$ is reconstructed on the
fine grid on $\Gamma$ using~(\ref{eq:back}) and~(\ref{eq:recend}). The
program {\tt demo15b.m} also constructs $U(r)$ on the fine grid on
$\Gamma_{\rm N}$, using~\cite[Eqs.~(39),~(40), and~(49)]{Hels09JCP},
and then restricts $U(r)$ to the coarse grid. Figure~\ref{fig:conv15b}
shows results. The convergence and the achievable accuracy for $U(r)$,
$r\in\Gamma_{\rm N}$, is similar to that of $U(r)$ with $r$ some
distance away from $\Gamma$. Compare the right image of
Figure~\ref{fig:conv15b} with the left image of
Figure~\ref{fig:conv15}.

\section{Steklov eigenvalue problems}
\label{sec:steklov}

An interesting problem arises if the boundary
conditions~(\ref{eq:pot}) and~(\ref{eq:flow}) of the interior mixed
problem for Laplace's equation are changed into
\begin{align}
 \lim_{D\ni r\to r^\circ}U(r)&=0\,,\quad r^\circ\in \Gamma_{\rm D}\,,
\label{eq:potst}\\
 \lim_{D\ni r\to r^\circ}\nu^\circ\cdot\nabla U(r)&=\varsigma U(r^\circ)\,,
  \quad r^\circ\in\Gamma_{\rm S}\,.
\label{eq:flowst}
\end{align}
Here the condition~(\ref{eq:flowst}) on the boundary part $\Gamma_{\rm
  S}$ is called a Steklov boundary condition and $\Gamma=\Gamma_{\rm
  D}\cup\Gamma_{\rm S}$. Finding nontrivial harmonic solutions $U(r)$
in $D$ satisfying~(\ref{eq:potst}) and~(\ref{eq:flowst}), along with
associated values $\varsigma$, is called a Steklov eigenvalue problem.

\begin{figure}[t]
\centering 
\noindent\makebox[\textwidth]{
\begin{minipage}{1.1\textwidth}
\hspace{-2mm}
\includegraphics[height=66mm]{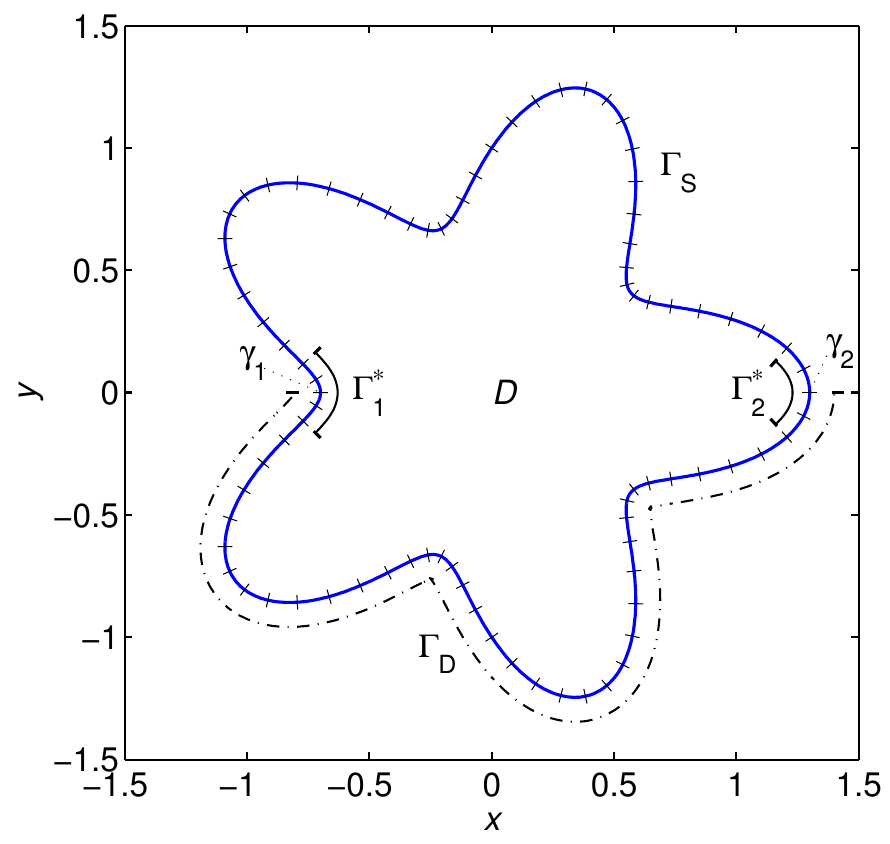}
\includegraphics[height=66mm]{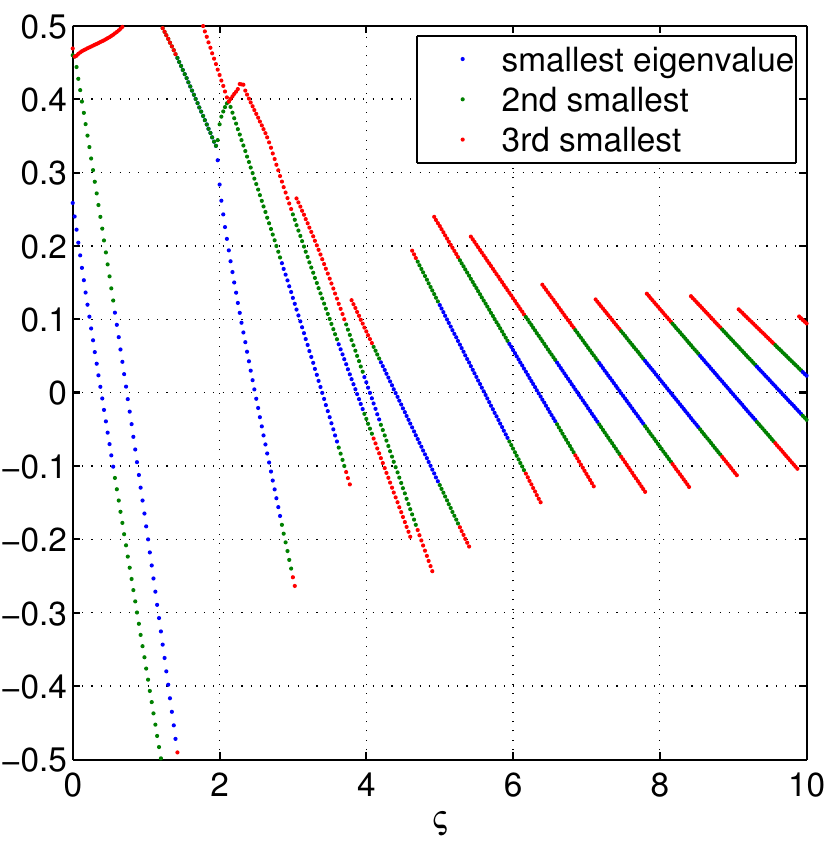}
\end{minipage}}
\caption{\sf Left: similar to Figure~\ref{fig:amfig}, but with
  $\Gamma=\Gamma_{\rm D}\cup\Gamma_{\rm S}$ according
  to~(\ref{eq:GDS}). Right: Flow of the three system matrix
  eigenvalues of~(\ref{eq:inteqst}) with the smallest magnitude as a
  function of the Steklov parameter $\varsigma$. The program {\tt
    demo16c.m} is used. The Steklov eigenvalues are those values
  $\varsigma$ for which the smallest system matrix eigenvalue is
  zero.}
\label{fig:steklov}
\end{figure}

We solve the mixed Steklov eigenvalue problem on the smooth domain $D$
given by~(\ref{eq:starfish}) using the same representation for $U(r)$
as in~(\ref{eq:Urep})
\begin{equation}
U(z)=
\frac{1}{\pi}\int_{\Gamma_{\rm D}}
\rho(\tau)\Im\left\{\frac{{\rm d}\tau}{\tau-z}\right\}-
\frac{1}{\pi}\int_{\Gamma_{\rm S}}
\rho(\tau)\log|\tau-z|\,{\rm d}|\tau|\,, \quad z\in
D\cup\Gamma_{\rm S}\,.
\label{eq:Urepst}
\end{equation}
Insertion of~(\ref{eq:Urepst}) into~(\ref{eq:potst})
and~(\ref{eq:flowst}) gives the homogeneous system
\begin{align}
\rho(z)&+
\frac{1}{\pi}\int_{\Gamma_{\rm D}}
\rho(\tau)\Im\left\{\frac{{\rm d}\tau}{\tau-z}\right\}-
\frac{1}{\pi}\int_{\Gamma_{\rm S}}
\rho(\tau)\log|\tau-z|\,{\rm d}|\tau|=0\,, 
\quad z\in \Gamma_{\rm D}\,,
\label{eq:lapsyssta}\\
\rho(z)&+
\frac{1}{\pi}\int_{\Gamma_{\rm D}}
\rho(\tau)\Im\left\{\frac{n_z\,{\rm d}\tau}{(\tau-z)^2}\right\}+
\frac{1}{\pi}\int_{\Gamma_{\rm S}}
\rho(\tau)\Re\left\{\frac{n_z\,{\rm d}|\tau|}{\tau-z}\right\}
\nonumber\\
&-\frac{\varsigma}{\pi}\int_{\Gamma_{\rm D}}
\rho(\tau)\Im\left\{\frac{{\rm d}\tau}{\tau-z}\right\}+
\frac{\varsigma}{\pi}\int_{\Gamma_{\rm S}}
\rho(\tau)\log|\tau-z|\,{\rm d}|\tau|=0\,,
\quad z\in \Gamma_{\rm S}\,.
\label{eq:lapsysstb}
\end{align}
The boundary parts $\Gamma_{\rm D}$ and $\Gamma_{\rm S}$ are taken
from~(\ref{eq:starfish}) as
\begin{equation}
r(s)\in\Gamma_{\rm D}\,, \quad -\pi<s<0\,,
\quad {\rm and} \quad
r(s)\in\Gamma_{\rm S}\,, \quad 0<s<\pi \,,
\label{eq:GDS}
\end{equation}
see the left image of Figure~\ref{fig:steklov}.

Discretization of~(\ref{eq:lapsyssta}) and~(\ref{eq:lapsysstb})
together with RCIP leads to a linear system
\begin{equation}
\left({\bf R}^{-1}+{\bf K}_{{\rm coa}1}^\circ-
\varsigma{\bf K}_{{\rm coa}2}^\circ\right)
\hat{\boldsymbol{\rho}}_{\rm coa}={\bf 0}\,,
\label{eq:inteqst}
\end{equation}
where ${\bf R}$ depends on $\varsigma$, the matrix ${\bf K}_{{\rm
    coa}1}^\circ$ contains entries coming from the discretization of
the integral operators in~(\ref{eq:lapsyssta})
and~(\ref{eq:lapsysstb}) that are not multiplied with $\varsigma$, and
the entries of ${\bf K}_{{\rm coa}2}^\circ$ come from the
discretization of the remaining operators. Values of $\varsigma$ that
correspond to a zero eigenvalue of the system matrix
in~(\ref{eq:inteqst}) are solutions to the Steklov eigenvalue problem.

The right image of Figure~\ref{fig:steklov}, produced by the program
{\tt demo16c.m}, shows the three smallest system matrix eigenvalues
of~(\ref{eq:inteqst}) as a function of $\varsigma$. The program {\tt
  demo16d.m} uses an eigenvalue search
algorithm~\cite[Section~9.1]{HelsKarl16} to generate a table of the
first 50 Steklov eigenvalues. The estimated relative accuracy is about
$10^{-14}$.

\subsection{Pure Steklov eigenvalue problem on a square}

Setups where $\Gamma=\Gamma_{\rm S}$, that is $\Gamma_{\rm
  D}=\emptyset$, and where $\Gamma$ is only piecewise smooth are of
particular interest in spectral theory. Recently some fascinating open
problems have emerged~\cite{Girou17}. The program {\tt demo16b.m}
computes the 20 first pure Steklov eigenvalues on the square
$D=(-1,1)\times(-1,1)$ using a RCIP-accelerated solver very similar to
that of {\tt demo16d}. The numerical results are compared with results
from the semi-analytic expressions of~\cite[Section 3.1]{Girou17}. The
estimated relative accuracy is on the order of $\epsilon_{\rm mach}$.

\section{Limit polarizability}
\label{sec:polarize}

Let us return to~(\ref{eq:inteq3}) and write it in the form
\begin{equation}
\left(K-w\right)\rho(r)=g(r)\,,
\quad r\in \Gamma\,,
\label{eq:inteq8}
\end{equation}
where
\begin{equation}
w\equiv u+{\rm i}v=-1/\lambda
\end{equation}
is a new complex variable. Values of $w$ for which (\ref{eq:inteq8})
has no solution are points in the spectrum of $K$. The precise nature
of this spectrum depends both on $\Gamma$ and on the function space
considered~\cite{HelsP12,HelsP17}. On the ``energy space''
$H^{-1/2}(\Gamma)$, the spectrum of $K$ is real and may have both
discrete and continuous parts.

\begin{figure}[t]
\centering 
\includegraphics[height=67mm]{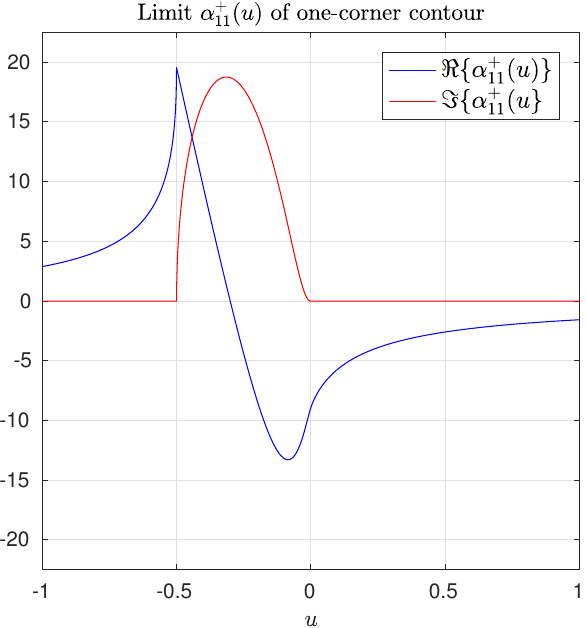}
\includegraphics[height=67mm]{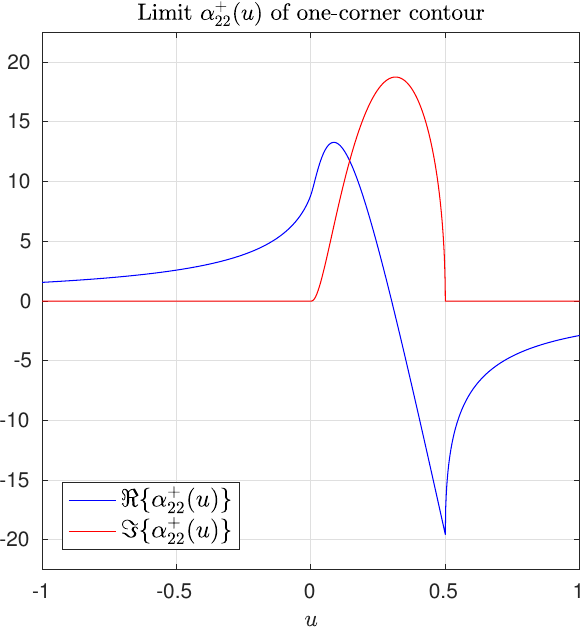}
\caption{\sf Elements $\alpha^+_{11}(u)$ and $\alpha^+_{22}(u)$ of 
  the limit polarizability tensor for the object enclosed by $\Gamma$
  of~(\ref{eq:gamma}) and with $\theta=\pi/2$. The program {\tt
    demo17.m} is used.}
\label{fig:pola}
\end{figure}

We solve~(\ref{eq:inteq8}) and compute the normalized polarizability
\begin{equation}
\alpha(w)=
\frac{1}{|V|}\int_{\Gamma}\rho(r)(e\cdot r)\,{\rm d}\ell\,,
\label{eq:q2}
\end{equation}
where $|V|$ is the area enclosed by $\Gamma$. We are particularly
interested in $\alpha^+(u)$, that is, the limit of $\alpha(w)$ as
$v\to 0^+$. The programs used are extensions of {\tt demo8b.m}. The
construction of the initializer ${\bf R}_*$ is accelerated using
Newton's method and homotopy, compare Section~\ref{sec:HelmTrans}.

The program {\tt demo17.m} computes $\alpha^+(u)$ for $\Gamma$ as
in~(\ref{eq:gamma}) and with $\theta=\pi/2$. The applied electric
field is either $e=(1,0)$, giving the element $\alpha_{11}^+(u)$ of
the limit polarizability tensor, or $e=(0,1)$, giving
$\alpha_{22}^+(u)$. Figure~\ref{fig:pola} shows results. By varying
$\theta$ in {\tt demo17.m}, one can see that a continuous non-zero
$\Im\{\alpha^+(u)\}$ is only possible in the interval
$-|1-\theta/\pi|<u<|1-\theta/\pi|$.

\begin{figure}[t]
\centering 
\includegraphics[height=66mm]{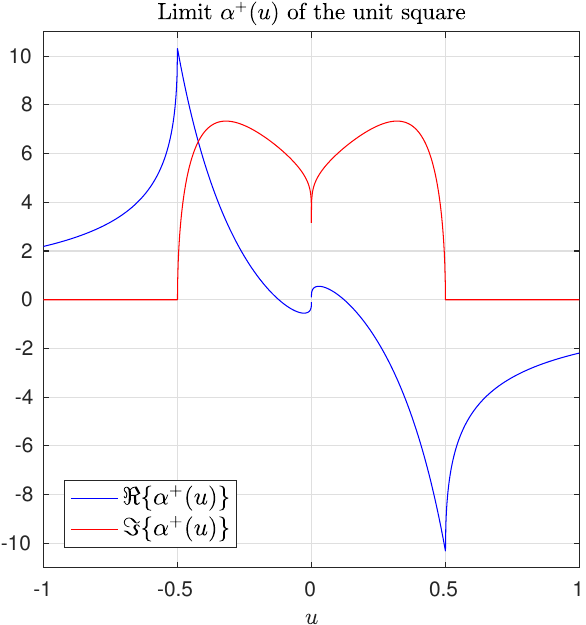}
\includegraphics[height=66mm]{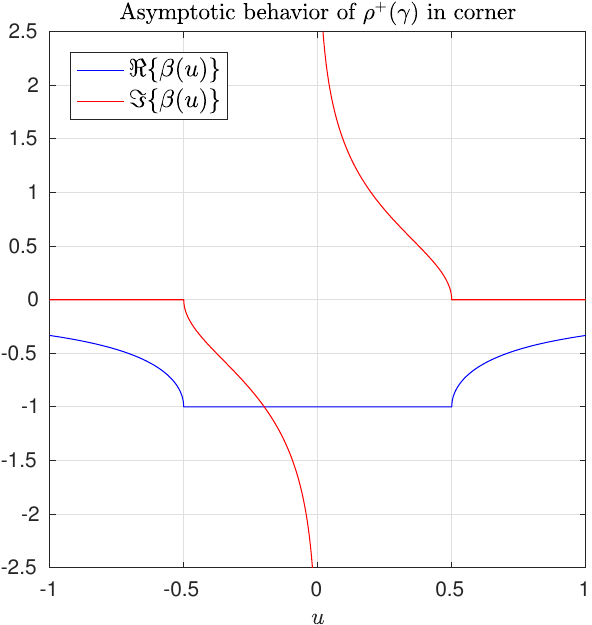}
\caption{\sf Left: the limit polarizability $\alpha^+(u)$ of the unit 
  square. Right: the leading exponent $\beta(u)$ of~(\ref{eq:asymp})
  for the asymptotics of $\rho^+(r)$ close to a corner vertex. The
  program {\tt demo17b.m} is used.}
\label{fig:polasq}
\end{figure}

The program {\tt demo17b.m} computes $\alpha^+(u)$ for $\Gamma$ being
the unit square, compare~\cite[Figure 5(a)]{HelsP12} where a similar
{\sc Matlab} program is used. In addition, {\tt demo17b.m} also
computes the singularity exponent $\beta$ in the leading asymptotic
behavior of $\rho^+(r)$ in the square corners,
\begin{equation}
\rho^+(\gamma)\propto\gamma^\beta\,,
\label{eq:asymp}
\end{equation}
where $\gamma$ is the arc length distance to the nearest corner
vertex, see Section~\ref{sec:asymp}. Figure~\ref{fig:polasq} shows
results. We emphasize that for $v=0$ and $-0.5<u<0.5$, there is no
solution $\rho(r)$ to~(\ref{eq:inteq8}). There is, however, a solution
$\rho(r)\in H^{-1/2}(\Gamma)$ for $v$ arbitrarily close to $0$ and it
is the polarizability $\alpha(w)$ corresponding to this limit
solution, with $v\to 0^+$, that is depicted in the left image of
Figure~\ref{fig:polasq}. Furthermore, the general polarizability
$\alpha(w)$ is simply related to the limit polarizability
$\Im\{\alpha^+(u)\}$ via
\begin{equation}
\alpha(w)=\frac{1}{\pi}\int_{-1}^1\frac{\Im\{\alpha^+(s)\}\,{\rm d}s}{s-w}\,,
\end{equation}
see~\cite[Section 3]{HelsP12}. One can say that
$\Im\{\alpha^+(u)\}/\pi$ is the derivative of a spectral measure
associated with $\alpha(w)$.

\section{Some computations on the cube}
\label{sec:cube}

The applicability of RCIP acceleration to Nyström discretization of
Fredholm second kind integral equations is not restricted to planar
problems. It extends also to 3D. Rotationally symmetric surfaces that
are smooth aside from isolated sharp edges or conical points are
particularly simple to deal with~\cite{HelsKarl16,HelsP17}. Surfaces
that contain a mix of contiguous edges and corners require that RCIP
is applied in a two-step manner: First it is used to find
multiplicative weight corrections that capture the singular behavior
of $\rho(r)$ in directions perpendicular to the edges using the
techniques of Section~\ref{sec:useful}. With these corrections
incorporated into the standard quadrature, $\rho(r)$ can be resolved
on the coarse grid except for in the corners. There intense refinement
has to take place and RCIP is used a second time.

\begin{figure}[t]
\centering 
\includegraphics[height=66mm]{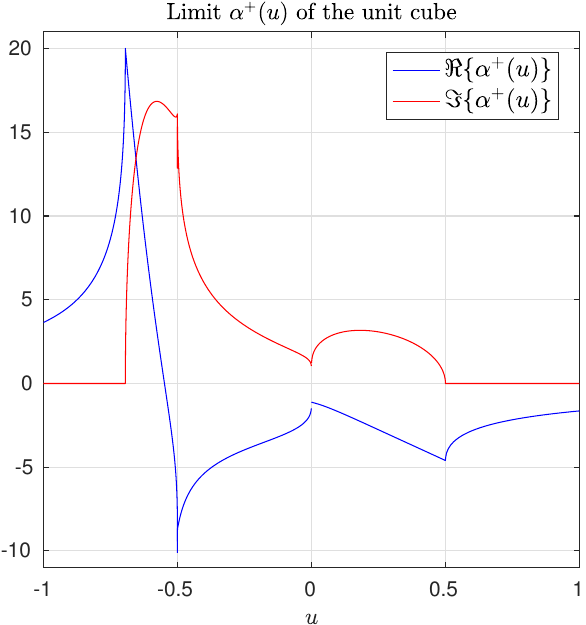}
\includegraphics[height=66mm]{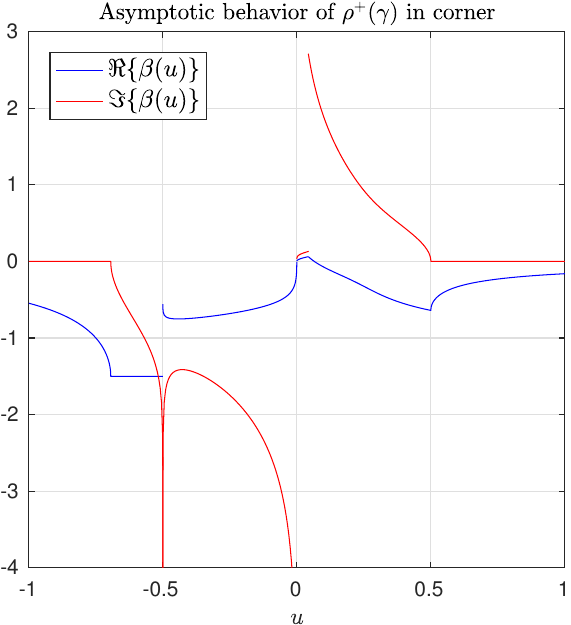}
\caption{\sf Left: the limit polarizability $\alpha^+(u)$ of the unit 
  cube. $\Im\{\alpha^+(u)\}$ has support for $u\in[-0.69452520,0.5]$.
  Right: the leading exponent $\beta(u)$ of~(\ref{eq:asymp}) in the
  direction toward a cube corner vertex. The program {\tt demo18.m} is
  used.}
\label{fig:polacu}
\end{figure}

Reference~\cite{HelsP12} details the two-step procedure for the
solution of~(\ref{eq:inteq8}) on the surface of the unit cube. The
action of the integral operator $K$ in 3D is
\begin{equation}
K\rho(r)=-\int_\Gamma
\frac{\nu\cdot(r-r')\rho(r')\,{\rm d}\ell'}
{2\pi\lvert r-r'\rvert^3}\,.
\end{equation}
The left image of Figure~\ref{fig:polacu}, taken from~\cite{HelsP12},
shows the limit polarizability $\alpha^+(u)$ of the cube computed in
this way.

The right image of Figure~\ref{fig:polacu}, produced by the program
{\tt demo18.m}, shows the vertex singularity exponent $\beta(u)$ of a
cube corner. The function $\beta(u)$ is interesting since for some of
its arguments there exist a number of benchmarks. For example, the
quantity $1+\beta(-1)$ is a so-called ``Fichera-type eigenvalue'' for
which the values $0.45417371$~\cite{Unibw06} and
$0.454173734$~\cite{McKird17} have been reported. Our estimate,
produced by an upgraded version of {\tt demo18.m}, is
$0.45417373430(14)$. The two digits within parenthesis are
extrapolated.

\begin{figure}
\centering 
\noindent\makebox[\textwidth]{
\begin{minipage}{1.1\textwidth}
\includegraphics[height=65mm]{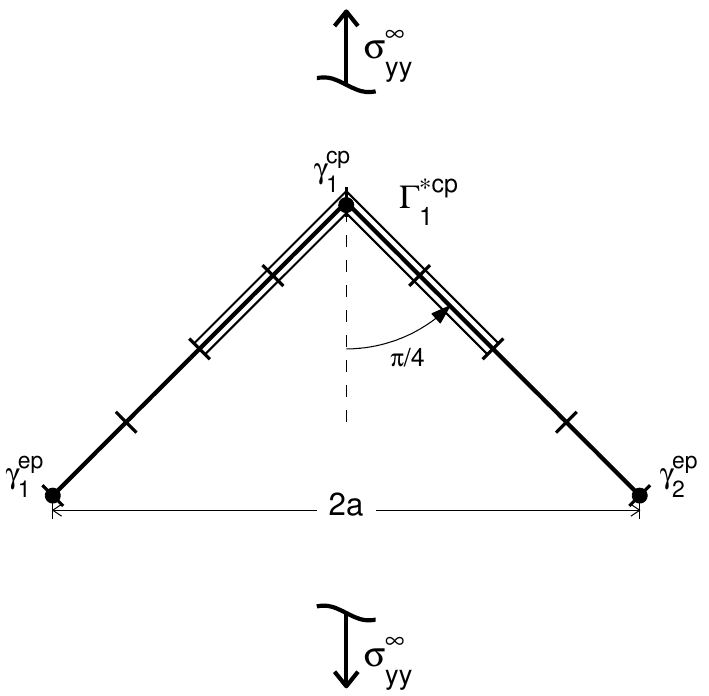}
\includegraphics[height=68mm]{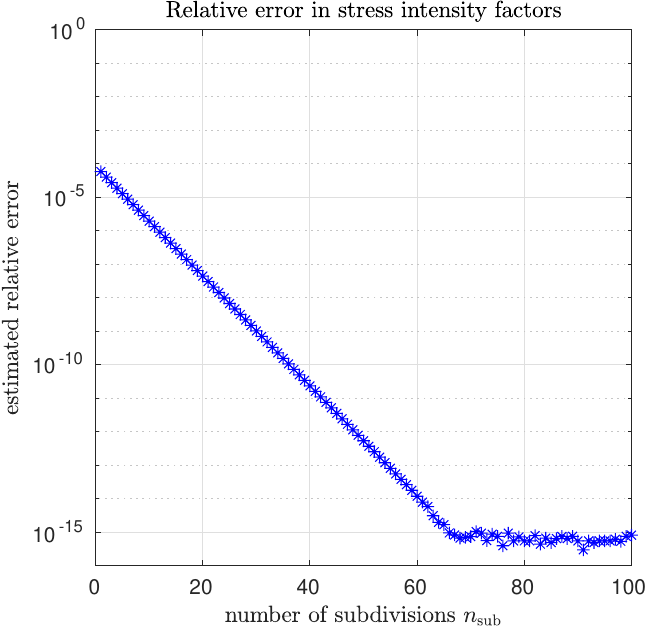}\\

\vspace{5mm}
\includegraphics[height=72mm]{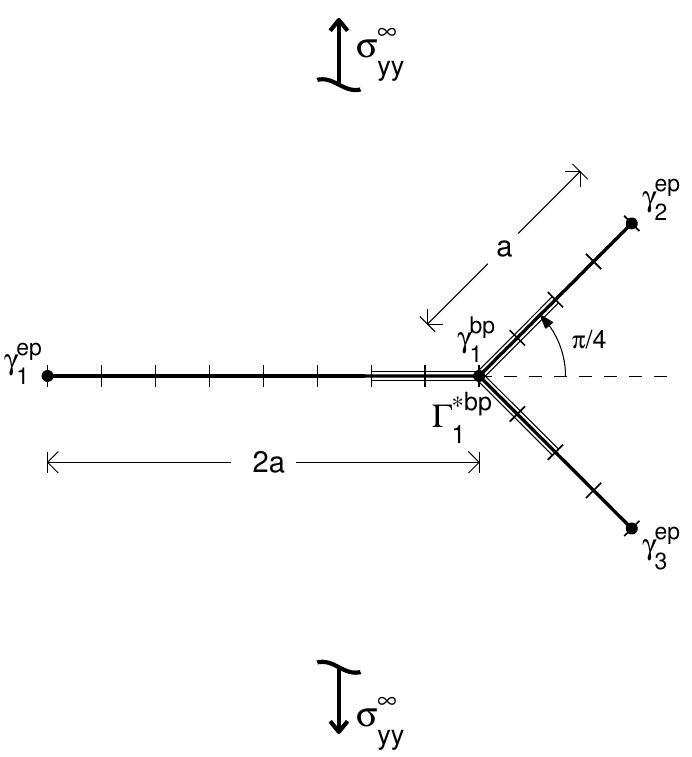}
\includegraphics[height=68mm]{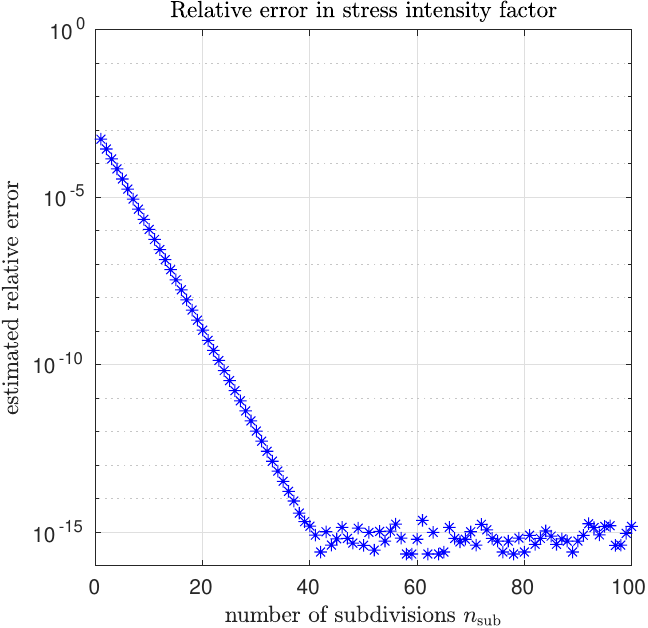}
\end{minipage}}
\caption{\sf Top left: the V-shaped crack under uniaxial load. Top right: 
  convergence of stress intensity factors with $n_{\rm sub}$ using the
  program {\tt demo20.m}. Bottom row: the same for the symmetrically
  branched crack and {\tt demo20c.m}. The boundary subsets
  $\Gamma^\star$ are indicated with extra double lines.}
\label{fig:vbshape}
\end{figure}

\section{Branched cracks in an elastic plane}
\label{sec:Vshape}

As mentioned in Section~\ref{sec:back}, the RCIP method grew out of
work in computational fracture mechanics. A particular goal was to
find an efficient way to compute the so called normalized stress
intensity factors at the tips $\gamma_1^{\rm ep}$ and $\gamma_2^{\rm
  ep}$ of a V-shaped crack in an elastic plane. This biharmonic
boundary value problem can be modeled as a Fredholm second kind
integral equation with composed operators in the
form~(\ref{eq:inteq7}). The stress intensity factors are simple
functionals of the layer density.  See~\cite{Hels11SISC} for details.

One purpose of the present section is to disseminate a {\sc Matlab}
program, {\tt demo20.m}, that
reproduces~\cite[Figure~7.1(a)]{Hels11SISC}. This figure shows how the
stress intensity factors converge with the number of levels $n_{\rm
  sub}$ in the recursion for the compressed inverse ${\bf R}$. The
treatment of composed integral operators in~\cite{Hels11SISC} has
proven to be unnecessarily complicated and is not used in {\tt
  demo20.m}. Instead, {\tt demo20.m} relies on the simplified
expansion technique for~(\ref{eq:inteq7}) that was described in
Section~\ref{sec:compose}.

Introducing the fundamental function~\cite[Section~107]{Musk53}
\begin{equation}
Z(\tau)=(\tau-\gamma_1^{\rm ep})^{-1/2}(\tau-\gamma_2^{\rm ep})^{-1/2}\,,
\label{eq:fundamental}
\end{equation}
the actions of the operators corresponding to $K$ and $M$
of~(\ref{eq:inteq7}) can, for the class of crack problem studied
in~\cite[Section ~6]{Hels11SISC}, be expressed as
\begin{equation}
K\rho(r)=\frac{1}{\pi{\rm i}}\int_{\Gamma}
\frac{\rho(\tau)\,{\rm d}\tau}{(\tau-z)Z(\tau)}
\end{equation}
and
\begin{multline}
M\rho(r)=-\frac{1}{2\pi {\rm i}}\left[\int_{\Gamma}
\frac{\rho(\tau)Z(\tau)\,{\rm d}\tau}{\tau-z}+\frac{\bar{n}_z}{n_z}
\int_{\Gamma}\frac{\rho(\tau)Z(\tau)\,{\rm d}\tau} 
{\bar{\tau}-\bar{z}}\right.\\
\left.+\int_{\Gamma}\frac{\overline{\rho(\tau)}\overline{Z(\tau)}
\,{\rm d}\bar{\tau}}{\bar{\tau}-\bar{z}}
+\frac{\bar{n}_z}{n_z}\int_{\Gamma}
\frac{(\tau-z)\overline{\rho(\tau)}\overline{Z(\tau)}
\,{\rm d}\bar{\tau}}{(\bar{\tau}-\bar{z})^2}\right]\,,
\quad z\in\Gamma\,.
\label{eq:M3}
\end{multline}
The right-hand side in~(\ref{eq:inteq7}) is
\begin{equation}
g(r)=-0.5+0.5\frac{\bar{n}_z}{n_z}\,.
\end{equation}
Figure~\ref{fig:vbshape}, top row, shows the V-shaped crack and output
from {\tt demo20.m}.

The bottom row of Figure~\ref{fig:vbshape} shows convergence of the
stress intensity factor at the tip $\gamma_2^{\rm ep}$ of the
symmetrically branched crack. For this crack, the fundamental function
assumes the more complicated form
\begin{equation}
Z(\tau)=(\tau-\gamma_1^{\rm bp})^{ 1/2}
        \prod_{i=1}^3(\tau-\gamma_i^{\rm ep})^{-1/2}\,,
\end{equation}
but the simplified expansion technique for~(\ref{eq:inteq7}),
described in Section~\ref{sec:compose}, still applies. The results in
the bottom row of Figure~\ref{fig:vbshape} are obtained with {\tt
  demo20c.m} and agree with those in~\cite[Table~7.2]{Hels11SISC}. The
program {\tt demo20b.m} produces the same results (not shown), but
uses the more involved original treatment of composed integral
operators from~\cite{Hels11SISC}.

\section{RCIP in a Method-of-Moments context}

The RCIP method is developed to accelerate and stabilize Nyström
solvers in the presence of boundary singularities. While Nyström
schemes are efficient, they are not the most common in, for example,
computational electromagnetics, where Method-of-Moments (MoM) solvers
dominate. A fair question to ask is therefore: does RCIP apply also to
the MoM? The answer is ``Yes''. Roughly speaking, as we now show, the
MoM amounts to a similarity transformation of the linear system
resulting from Nyström discretization. RCIP still applies, unaffected
by this.

Let us return to~(\ref{eq:inteq3a}) and (\ref{eq:inteq3b}) which,
after the change of variables
\begin{equation}
\boldsymbol{\rho}={\bf L}{\bf c}\,,
\end{equation}
can be written in MoM-form as
\begin{align}
\left({\bf I}_{\rm coa}
+\lambda{\bf L}_{\rm coa}^{-1}{\bf K}_{\rm coa}{\bf L}_{\rm coa}\right)
{\bf c}_{\rm coa}&=\lambda{\bf L}_{\rm coa}^{-1}{\bf g}_{\rm coa}\,,
\label{eq:inteqMoM1}\\
\left({\bf I}_{\rm fin}
+\lambda{\bf L}_{\rm fin}^{-1}{\bf K}_{\rm fin}{\bf L}_{\rm fin}\right)
{\bf c}_{\rm fin}&=\lambda{\bf L}_{\rm fin}^{-1}{\bf g}_{\rm fin}\,.
\label{eq:inteqMoM2}
\end{align}
Here ${\bf c}$ is a column vector with $n_{\rm p}$ entries
(interpreted as coefficients) and ${\bf L}$ is a block diagonal matrix
with identical blocks ${\bf L}_j$, $j=1,\ldots,n_{\rm pan}$ and
$j=1,\ldots,n_{\rm pan}+2n_{\rm sub}$ for~(\ref{eq:inteqMoM1})
and~(\ref{eq:inteqMoM2}), respectively. The ${\bf L}_j$ is of size
$16\times 16$ and has the 16 first standard Legendre polynomials
$P_n(x)$, $n=0,\ldots,15$, evaluated at the nodes {\tt T16}, see
Section~\ref{sec:basepro}, as columns (interpreted as basis
functions). Note that the inverse ${\bf L}^{-1}$ is simple to set up
accurately since
\begin{equation}
{\bf L}_j^{-1}={\bf N}{\bf L}_j^T{\bf W}\,,
\end{equation}
where ${\bf N}$ is a diagonal matrix of normalization constants 
\begin{displaymath}
(2n+1)/2\,, \qquad n=0,\ldots,15\,,
\end{displaymath}
and ${\bf W}$ is a diagonal matrix containing the quadrature weights
{\tt W16}.

Now, repeating the derivation steps of Section~\ref{sec:comp}, we
arrive at~(\ref{eq:inteq4}) which assumes the form
\begin{equation}
\left({\bf I}_{\rm coa}+\lambda\overline{\bf K}_{\rm coa}^\circ
                               \overline{\bf R}\right)
\tilde{\bf c}_{\rm coa}=\lambda{\bf L}_{\rm coa}^{-1}{\bf g}_{\rm coa}\,,
\label{eq:inteqMoM3}
\end{equation}
where
\begin{equation}
\overline{\bf K}_{\rm coa}^\circ=
{\bf L}_{\rm coa}^{-1}{\bf K}_{\rm coa}^\circ{\bf L}_{\rm coa}\,,\qquad
\overline{\bf R}={\bf L}_{\rm coa}^{-1}{\bf R}{\bf L}_{\rm coa}\,,
\end{equation}
and $\tilde{\bf c}_{\rm coa}$ is related to the weight-corrected
density $\hat{\boldsymbol{\rho}}_{\rm coa}$ of~(\ref{eq:wcd}) via
\begin{equation}
\hat{\boldsymbol{\rho}}_{\rm coa}=
{\bf L}_{\rm coa}\overline{\bf R}\tilde{\bf c}_{\rm coa}\,.
\end{equation}

The recursion~(\ref{eq:recur},\ref{eq:rstart}) assumes the form
\begin{align}
&\overline{\bf R}_i={\bf P}^T_{LW\rm{bc}}
\left(
\mathbb{F}\{\overline{\bf R}_{i-1}^{-1}\}+{\bf I}_{\rm b}^\circ
           +\lambda\overline{\bf K}_{i{\rm b}}^\circ
\right)^{-1}{\bf P}_{L\rm{bc}}\,,\quad i=1,\ldots,n_{\rm sub}\,,
\label{eq:recurMoM}\\
&\mathbb{F}\{\overline{\bf R}_0^{-1}\}=
{\bf I}_{\rm b}^\star+\lambda\overline{\bf K}^\star_{1{\rm b}}\,.
\label{eq:rstartMoM}
\end{align}
Here
\begin{equation}
{\bf P}_{L\rm{bc}}={\bf L}_{\rm bb}^{-1}{\bf L}_{\rm bc}\,,
\qquad
{\bf P}^T_{LW\rm{bc}}=
{\bf L}_{\rm bb}^T{\bf W}_{\rm b}{\bf L}_{\rm bc}{\bf N}_{\rm c}\,,
\end{equation}
where ${\bf L}_{\rm bb}$ is ${\bf L}$ on a type {\sf b} mesh, the
matrix ${\bf L}_{\rm bc}$ maps coefficients on a type {\sf c} mesh to
values of Legendre polynomials on a grid on a type {\sf b} mesh, the
diagonal matrix ${\bf W}_{\rm b}$ can be constructed as
\begin{verbatim}
  Wb=diag([W16;0.5*W16;0.5*W16;0.5*W16;0.5*W16;W16]);
\end{verbatim}
and ${\bf N}_{\rm c}$ is ${\bf N}$ on a type {\sf c} mesh.

The program {\tt demo21.m} implements an RCIP-accelerated MoM solver
for the first numerical example in Section~\ref{sec:HelmField}. Simply
put: {\tt demo21.m} is a conversion of {\tt demo11.b} to the MoM
framework, as outlined above. In particular, the Nyström matrix ${\bf
  K}$ has been replaced with the MoM matrix $\overline{\bf K}$. The
results produced by {\tt demo11.b} and {\tt demo21.m} are almost
identical, although {\tt demo11.b} of course has a faster setup phase
than {\tt demo21.m}. Less work is required to obtain the entries of
${\bf K}$ than those of $\overline{\bf K}$.

\bigskip\bigskip\bigskip

\subsection*{Acknowledgements}

The idea to make this tutorial came up during a discussion with Alex
Barnett and Adrianna Gillman of 5/20/12 at the FACM'12 conference at
NJIT. Input and feedback from Alex and Adrianna, and from Shidong
Jiang, Mary-Catherine Kropinski, Mikyoung Lim, David McA. McKirdy,
Rikard Ojala, and Karl-Mikael Perfekt has been of great value. The
work was supported by the Swedish Research Council under contracts
621-2011-5516, 621-2014-5159, and 2021-03720.

\clearpage\newpage
\centerline{*** {\LARGE{\bf Appendicies}} ***}

\bigskip

\renewcommand{\theequation}{A.\arabic{equation}}
\setcounter{equation}{0}
\section*{Appendix A. Proof that ${\bf P}_W^T{\bf P}={\bf I}_{\rm coa}$}

Let ${\bf f}_{\rm coa}$ and ${\bf g}_{\rm coa}$ be two column vectors,
corresponding to the discretization of two panelwise polynomials with
panelwise degree $15$ on the coarse mesh of $\Gamma$. Then
\begin{equation}
{\bf f}_{\rm coa}^T{\bf W}_{\rm coa}{\bf g}_{\rm coa}=
\left({\bf P}{\bf f}_{\rm coa}\right)^T
{\bf W}_{\rm fin}\left({\bf P}{\bf g}_{\rm coa}\right)=
{\bf f}_{\rm coa}^T{\bf P}^T
{\bf W}_{\rm fin}{\bf P}{\bf g}_{\rm coa}\,,
\label{eq:A1}
\end{equation}
because composite $16$-point Gauss--Legendre quadrature has panelwise
polynomial degree $31$. The diagonal matrix ${\bf W}_{\rm coa}$ has
size $16n_{\rm pan}\times 16n_{\rm pan}$.

Since there are $16n_{\rm pan}$ linearly independent choices of ${\bf
  f}_{\rm coa}$ and of ${\bf g}_{\rm coa}$ it follows
from~(\ref{eq:A1}) that
\begin{equation}
{\bf W}_{\rm coa}={\bf P}^T{\bf W}_{\rm fin}{\bf P}\,,
\end{equation}
which, using~(\ref{eq:PW}), can be rewritten
\begin{equation}
{\bf I}_{\rm coa}={\bf W}_{\rm coa}^{-1}{\bf P}^T{\bf W}_{\rm fin}{\bf
  P}={\bf P}_W^T{\bf P}\,.
\end{equation}

\newpage

\renewcommand{\theequation}{B.\arabic{equation}}
\setcounter{equation}{0}
\section*{Appendix B. Derivation of the compressed equation}
  
The compression of~(\ref{eq:inteq3b}), leading up
to~(\ref{eq:inteq4}), was originally described in~\cite[Section
6.4]{Hels08b}. Here we give a summary.
  
The starting point is~(\ref{eq:inteq3}) which, using the operator
split analogous to~(\ref{eq:split1},\ref{eq:split2})
\begin{equation}
K=K^\star+K^\circ
\label{eq:Ko}
\end{equation}
and the variable substitution
\begin{equation}
\rho(r)=\left(I+\lambda K^\star\right)^{-1}\tilde{\rho}(r)\,,
\label{eq:subst2}
\end{equation}
gives the right preconditioned equation
\begin{equation}
\tilde{\rho}(r)+\lambda K^\circ(I+\lambda K^\star)^{-1}\tilde{\rho}(r)
=\lambda g(r)\,,\quad r\in \Gamma\,.
\label{eq:inteq4a}
\end{equation}

Now, let us take a close look at~(\ref{eq:inteq4a}). We observe that
$K^\circ(I+\lambda K^\star)^{-1}$ is an operator whose action on any
function gives a function that is smooth on the innermost two panels
of the coarse mesh on $\Gamma^\star$. This is so since $K^\circ$ is
constructed so that its action on any function gives a function that
is smooth on the innermost two panels of the coarse mesh on
$\Gamma^\star$. Furthermore, the right-hand side $\lambda g(r)$ of
(\ref{eq:inteq4a}) is assumed to be panelwise smooth on the coarse
mesh. Using an argument of contradiction we see that $\tilde{\rho}(r)$
has to be panelwise smooth on the innermost two panels of the coarse
mesh on $\Gamma^\star$.

Having concluded that $\tilde{\rho}(r)$ is panelwise smooth on the two
coarse panels that are closest to the corner we can write
\begin{equation}
\tilde{\boldsymbol{\rho}}_{\rm fin}={\bf P}\tilde{\boldsymbol{\rho}}_{\rm coa}.
\label{eq:b1}
\end{equation}
We also have
\begin{equation}
{\bf g}_{\rm fin}={\bf P}{\bf g}_{\rm coa}\,,
\label{eq:b2}
\end{equation}
the discrete version of~(\ref{eq:subst2}) on the fine grid
\begin{equation}
\boldsymbol{\rho}_{\rm fin}=
\left({\bf I}_{\rm fin}+\lambda {\bf K}_{\rm fin}^\star\right)^{-1}
\tilde{\boldsymbol{\rho}}_{\rm fin}\,,
\label{eq:b3}
\end{equation}
and the relations~(\ref{eq:split2}) and~(\ref{eq:decomp}) which we now
repeat:
\begin{align}
{\bf K}_{\rm fin}&={\bf K}_{\rm fin}^\star
                         +{\bf K}_{\rm fin}^\circ\,,
\label{eq:b4}\\
{\bf K}_{\rm fin}^\circ&={\bf P}{\bf K}_{\rm coa}^\circ{\bf P}_W^T\,.
\label{eq:b5}
\end{align}

Substitution of~(\ref{eq:b1},\ref{eq:b2},\ref{eq:b3},\ref{eq:b4},\ref{eq:b5})
into~(\ref{eq:inteq3b}), which we now repeat:
\begin{equation}
\left({\bf I}_{\rm fin}+\lambda{\bf K}_{\rm fin}\right)
\boldsymbol{\rho}_{\rm fin}=\lambda{\bf g}_{\rm fin}\,,
\label{eq:b6}
\end{equation}
gives
\begin{equation}
{\bf P}\tilde{\boldsymbol{\rho}}_{\rm coa}+
\lambda{\bf P}{\bf K}_{\rm coa}^\circ{\bf P}_W^T
\left({\bf I}_{\rm fin}+\lambda {\bf K}_{\rm fin}^\star\right)^{-1}
{\bf P}\tilde{\boldsymbol{\rho}}_{\rm coa}=
{\bf P}{\bf g}_{\rm coa}\,.
\label{eq:b7}
\end{equation}
Applying ${\bf P}_W^T$ (or ${\bf Q}$) to the left in~(\ref{eq:b7}) and
using the identities~(\ref{eq:PWTP}) (or~(\ref{eq:QP})) gives the
compressed equation~(\ref{eq:inteq4}).

\newpage

\renewcommand{\theequation}{C.\arabic{equation}}
\setcounter{equation}{0}
\section*{Appendix C. Integration of $\rho$ against smooth $f$}

This appendix is about computing integrals
\begin{equation}
\int_\Gamma f(r)\rho(r)\,{\rm d}\ell\,,
\end{equation}
where $\rho(r)$ is the solution to~(\ref{eq:inteq3}). It is assumed
that $f(r(s))|\dot{r}(s)|$ is a piecewise smooth function of the
boundary parameter $s$. The aim is to derive the relation
\begin{equation}
\int_\Gamma f(r)\rho(r)\,{\rm d}\ell
\approx\sum_j \zeta_{{\rm fin}_j}\rho_{{\rm fin}_j}w_{{\rm fin}_j}
    =  \sum_j \zeta_{{\rm coa}_j}\hat{\rho}_{{\rm coa}_j}w_{{\rm coa}_j}\,,
\end{equation}
where
\begin{equation}
\zeta_j=f(r(s_j))|\dot{r}(s_j)|
\end{equation}
and the discrete weight-corrected density
$\hat{\boldsymbol{\rho}}_{\rm coa}$ is defined in~(\ref{eq:wcd}).

The derivation
uses~(\ref{eq:PW},\ref{eq:R},\ref{eq:wcd},\ref{eq:b1},\ref{eq:b3}) and
the diagonal matrices ${\bf W}_{\rm coa}$ and ${\bf W}_{\rm fin}$
defined in Section~\ref{sec:comp} and goes as follows:
\begin{equation}
\begin{split}
\int_\Gamma f(r)\rho(r)\,{\rm d}\ell
&=\int_0^1 f(r(s))\rho(r(s))|\dot{r}(s)|\,{\rm d}s\\
&\approx\sum_j \zeta_{{\rm fin}_j}\rho_{{\rm fin}_j}w_{{\rm fin}_j}\\
&=\boldsymbol{\zeta}_{\rm fin}^T{\bf W}_{\rm fin}\boldsymbol{\rho}_{\rm fin}\\
&=\boldsymbol{\zeta}_{\rm fin}^T{\bf W}_{\rm fin}
\left({\bf I}_{\rm fin}+\lambda {\bf K}_{\rm fin}^\star\right)^{-1}
\tilde{\boldsymbol{\rho}}_{\rm fin}\\
&=\boldsymbol{\zeta}_{\rm coa}^T{\bf P}^T{\bf W}_{\rm fin}
\left({\bf I}_{\rm fin}+\lambda {\bf K}_{\rm fin}^\star\right)^{-1}
{\bf P}\tilde{\boldsymbol{\rho}}_{\rm coa}\\
&=\boldsymbol{\zeta}_{\rm coa}^T{\bf W}_{\rm coa}
{\bf W}_{\rm coa}^{-1}{\bf P}^T{\bf W}_{\rm fin}
\left({\bf I}_{\rm fin}+\lambda {\bf K}_{\rm fin}^\star\right)^{-1}
{\bf P}\tilde{\boldsymbol{\rho}}_{\rm coa}\\
&=\boldsymbol{\zeta}_{\rm coa}^T{\bf W}_{\rm coa}
{\bf P}_W^T\left({\bf I}_{\rm fin}+\lambda {\bf K}_{\rm fin}^\star\right)^{-1}
{\bf P}\tilde{\boldsymbol{\rho}}_{\rm coa}\\
&=\boldsymbol{\zeta}_{\rm coa}^T{\bf W}_{\rm coa}
{\bf R}\tilde{\boldsymbol{\rho}}_{\rm coa}\\
&=\boldsymbol{\zeta}_{\rm coa}^T{\bf W}_{\rm coa}
\hat{\boldsymbol{\rho}}_{\rm coa}\\
&=\sum_j \zeta_{{\rm coa}_j}\hat{\rho}_{{\rm coa}_j}w_{{\rm coa}_j}\,.
\end{split}
\end{equation}

\newpage

\renewcommand{\theequation}{D.\arabic{equation}}
\setcounter{equation}{0}
\section*{Appendix D. Derivation of the recursion}

The recursion~(\ref{eq:recur}) for the rapid construction of the
diagonal blocks of the compressed weighted inverse ${\bf R}$ was
originally derived in~\cite[Section 7]{Hels08b} using different
notation and different meshes than in the present tutorial. The
recursion was derived a second time in~\cite[Section 7]{Hels09IJSS}
using new meshes. Better notation was introduced in~\cite[Section
6]{Hels09JCP}. A third derivation, in a general setting, takes place
in~\cite[Section 5]{Hels11SISC} and it uses the same notation and
meshes as in the present tutorial.

A problem when explaining the derivation of~(\ref{eq:recur}) is that
one needs to introduce intermediate meshes and matrices whose
appearance may cause enervation at a first glance. Particularly so
since these meshes and matrices are not needed in the final
expression~(\ref{eq:recur}). We emphasize that the underlying matrix
property that permits the recursion is the low rank of certain
off-diagonal blocks in discretizations of $K^\circ$ of~(\ref{eq:Ko})
on nested meshes.

\begin{figure}[h]
\centering \includegraphics[height=33mm]{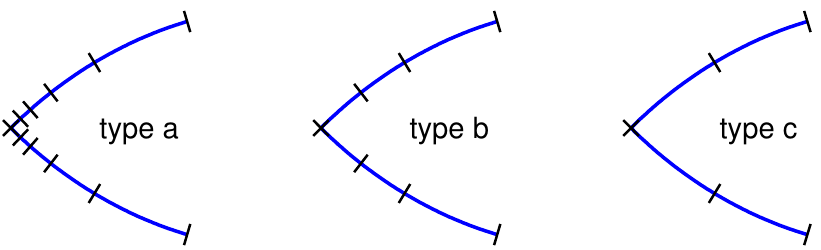}
\caption{\sf Meshes of type {\sf a}, type {\sf b}, and type {\sf c} on
  the boundary subset $\Gamma_i^\star$ for $i=n_{\rm sub}=3$. The type
  {\sf a} mesh has $4+2i$ panels. The type {\sf b} mesh has six
  panels. The type {\sf c} mesh has four panels. The type {\sf a} mesh
  is the restriction of the fine mesh to $\Gamma_i^\star$. For
  $i=n_{\rm sub}$, the type {\sf c} mesh is the restriction of the
  coarse mesh to $\Gamma^\star$. The type {\sf a} mesh and the type
  {\sf b} mesh coincide for $i=1$.}
\label{fig:types}
\end{figure}

The recursion~(\ref{eq:recur}) only uses uses one type of mesh
explicitly -- the type {\sf b} mesh of Figure~\ref{fig:subsets}. On
each $\Gamma_i^\star$ there is a type {\sf b} mesh and a
corresponding discretization of $K^\circ$ denoted ${\bf K}_{i{\rm
    b}}^\circ$. Here we need two new types of meshes denoted type
{\sf a} and type {\sf c}, along with corresponding discrete operators.
For example, ${\bf K}_{i{\rm a}}$ is the discretization of $K$ on a
type {\sf a} mesh on $\Gamma_i^\star$. The three types of meshes are
depicted in Figure~\ref{fig:types}. Actually, a straight type {\sf c}
mesh was already introduced in Figure~\ref{fig:Pbc}.

Now we define ${\bf R}_i$ as
\begin{equation}
{\bf R}_i\equiv{\bf P}_{Wi{\rm ac}}^T
\left({\bf I}_{i{\rm a}}+\lambda {\bf K}_{i{\rm a}}\right)^{-1}
{\bf P}_{i{\rm ac}}\,,
\label{eq:Ridef}
\end{equation}
where ${\bf P}_{Wi{\rm ac}}$ and ${\bf P}_{i{\rm ac}}$ are
prolongation operators (in parameter) from a grid on a type {\sf c}
mesh on $\Gamma_i^\star$ to a grid on a type {\sf a} mesh on
$\Gamma_i^\star$. Note that ${\bf R}_i$ for $i={n_{\rm sub}}$,
according to the definition~(\ref{eq:Ridef}), is identical to the full
diagonal $64\times 64$ block of ${\bf R}$ of~(\ref{eq:R}). Note also
that ${\bf R}_1$ comes cheaply. The rest of this appendix is about
finding an expression for ${\bf R}_i$ in terms of ${\bf R}_{i-1}$ that
is cheap to compute.

Let us split ${\bf K}_{i{\rm a}}$ into two parts
\begin{equation}
{\bf K}_{i{\rm a}}={\bf K}_{i{\rm a}}^\star+{\bf K}_{i{\rm a}}^\circ\,,
\end{equation}
where ${\bf K}_{i{\rm a}}^\star=\mathbb{F}\{{\bf K}_{(i-1){\rm
    a}}\}$ and ${\bf K}_{i{\rm a}}^\circ$ is such that
\begin{equation}
{\bf K}_{i{\rm a}}^\circ=
{\bf P}_{i{\rm ab}}{\bf K}_{i{\rm b}}^\circ{\bf P}_{Wi{\rm ab}}^T
\label{eq:decomp2}
\end{equation}
holds to about machine precision, compare~(\ref{eq:decomp}). The
prolongation operators ${\bf P}_{i{\rm ab}}$ and ${\bf P}_{Wi{\rm
    ab}}$ act from a grid on a type {\sf b} mesh to a grid on a type
{\sf a} mesh. It holds that
\begin{align}
{\bf P}_{i{\rm ac}}&={\bf P}_{i{\rm ab}}{\bf P}_{\rm bc}\,,\\
{\bf P}_{Wi{\rm ac}}&={\bf P}_{Wi{\rm ab}}{\bf P}_{W{\rm bc}}\,.
\end{align}
Summing up, we can rewrite~(\ref{eq:Ridef}) as
\begin{equation}
{\bf R}_i={\bf P}_{W{\rm bc}}^T{\bf P}_{Wi{\rm ab}}^T
\left({\bf I}_{i{\rm a}}+\mathbb{F}\{\lambda{\bf K}_{(i-1){\rm a}}\}+
\lambda{\bf P}_{i{\rm ab}}{\bf K}_{i{\rm b}}^\circ{\bf P}_{Wi{\rm ab}}^T
\right)^{-1}
{\bf P}_{i{\rm ab}}{\bf P}_{\rm bc}\,.
\label{eq:sumup}
\end{equation}

The subsequent steps in the derivation of~(\ref{eq:recur}) are to
expand the inverse of the sum of matrices within parentheses
in~(\ref{eq:sumup}) using a Taylor series
\begin{equation}
({\bf A}+{\bf B})^{-1}={\bf A}^{-1}-{\bf A}^{-1}{\bf B}{\bf A}^{-1}
+{\bf A}^{-1}{\bf B}{\bf A}^{-1}{\bf B}{\bf A}^{-1}-\ldots\,,
\end{equation}
where ${\bf A}$ corresponds to the first two terms and ${\bf B}$
corresponds to the last term; multiply the terms in this series with
${\bf P}_{Wi{\rm ab}}^T$ from the left and with ${\bf P}_{i{\rm ab}}$
from the right; and bring the series back in closed form. The result
is
\begin{equation}
{\bf R}_i={\bf P}_{W{\rm bc}}^T
\left[\left({\bf P}_{Wi{\rm ab}}^T
\left({\bf I}_{i{\rm a}}+\mathbb{F}\{\lambda{\bf K}_{(i-1){\rm a}}\}
\right)^{-1}{\bf P}_{i{\rm ab}}\right)^{-1}+
\lambda{\bf K}_{i{\rm b}}^\circ\right]^{-1}
{\bf P}_{\rm bc}\,,
\label{eq:result}
\end{equation}
which, in fact, is~(\ref{eq:recur}) in disguise. To see this, recall
from~(\ref{eq:Ridef}) that
\begin{equation}
{\bf R}_{(i-1)}\equiv{\bf P}_{W(i-1){\rm ac}}^T
\left({\bf I}_{(i-1){\rm a}}+\lambda{\bf K}_{(i-1){\rm a}}\right)^{-1}
{\bf P}_{(i-1){\rm ac}}\,.
\end{equation}
Then
\begin{align}
\mathbb{F}\{{\bf R}_{(i-1)}\}&=
\mathbb{F}\{
{\bf P}_{W(i-1){\rm ac}}^T
\left({\bf I}_{(i-1){\rm a}}+\lambda{\bf K}_{(i-1){\rm a}}\right)^{-1}
{\bf P}_{(i-1){\rm ac}}\}
\nonumber\\
&={\bf P}_{Wi{\rm ab}}^T
\left({\bf I}_{i{\rm a}}+
\mathbb{F}\{\lambda{\bf K}_{(i-1){\rm a}}\}\right)^{-1}
{\bf P}_{i{\rm ab}}-{\bf I}_{\rm b}^\circ\,,
\label{eq:final}
\end{align}
where the second equality uses $\left({\bf P}_{Wi{\rm ab}}^T{\bf
    P}_{i{\rm ab}}\right)^\circ={\bf I}_{\rm b}^\circ$. Substitution
of~(\ref{eq:final}) in~(\ref{eq:result}) gives the recursion in the
familiar form
\begin{equation}
{\bf R}_i={\bf P}^T_{W\rm{bc}}
\left(
\mathbb{F}\{{\bf R}_{i-1}^{-1}\}+{\bf I}_{\rm b}^\circ+\lambda{\bf K}_{i{\rm
    b}}^\circ\right)^{-1}{\bf P}_{\rm{bc}}\,.
\end{equation}

\renewcommand{\theequation}{E.\arabic{equation}}
\setcounter{equation}{0}
\section*{Appendix E. An inner product preserving scheme}

In~\cite{Brem12b}, Bremer describes a scheme that stabilizes the
solution to the discretized system~(\ref{eq:inteq3b}) on the fine
mesh. The scheme can be interpreted as an inner product preserving
discretization. In practice it corresponds to making a similarity
transformation of the system matrix. While inner product preserving
Nyström discretization elegantly solves problems related to
stability (the condition number of the system matrix is improved) it
does not reduce the number of discretization points (unknowns) needed
to achieve a given precision in the solution. Neither does it affect
the spectrum of the system matrix (similarity transformations preserve
eigenvalues) and hence it does not in any substantial way improve the
convergence rate of the GMRES iterative method~\cite[Lecture
35]{Tref97}.

For completeness, we have implemented inner product preserving
Nyström discretization in the program {\tt demo1d.m}. The program
is a continuation of {\tt demo1b.m} where we also have
replaced~(\ref{eq:inteq1a}) with the more stable integral
equation~(\ref{eq:inteq1b}). This should facilitate comparison with
the program {\tt demo3b.m} and the results shown in
Figure~\ref{fig:conv3}.

\begin{figure}[h]
\centering 
\noindent\makebox[\textwidth]{
\begin{minipage}{1.1\textwidth}
\includegraphics[height=66mm]{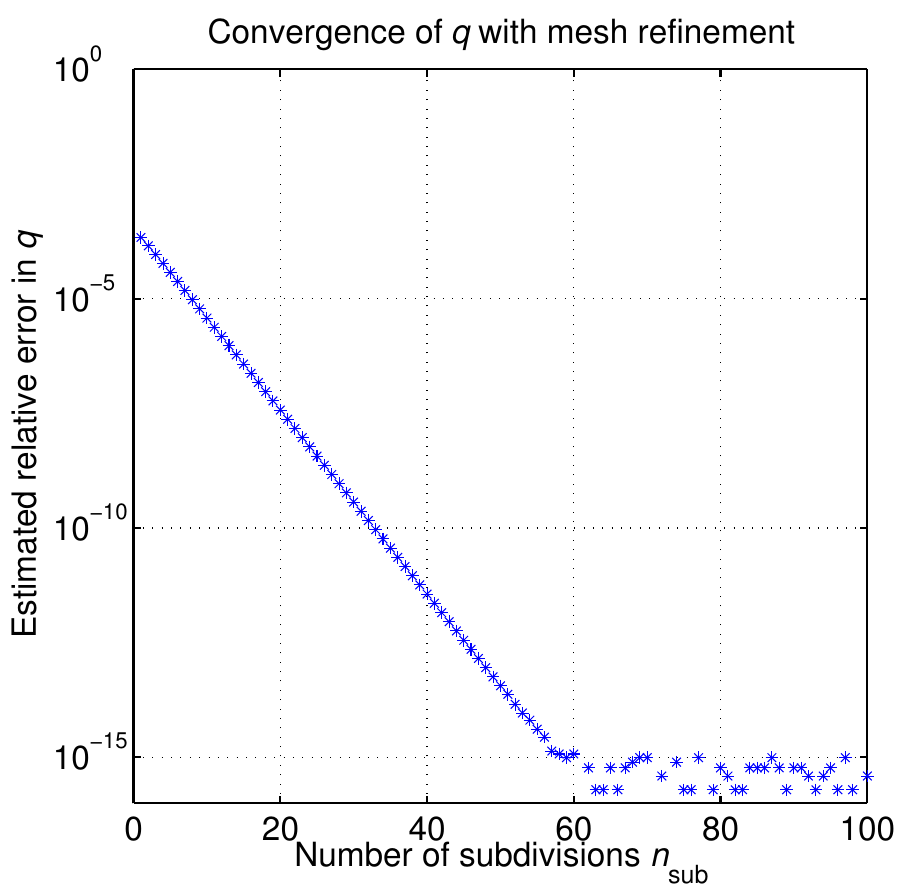}
\includegraphics[height=66mm]{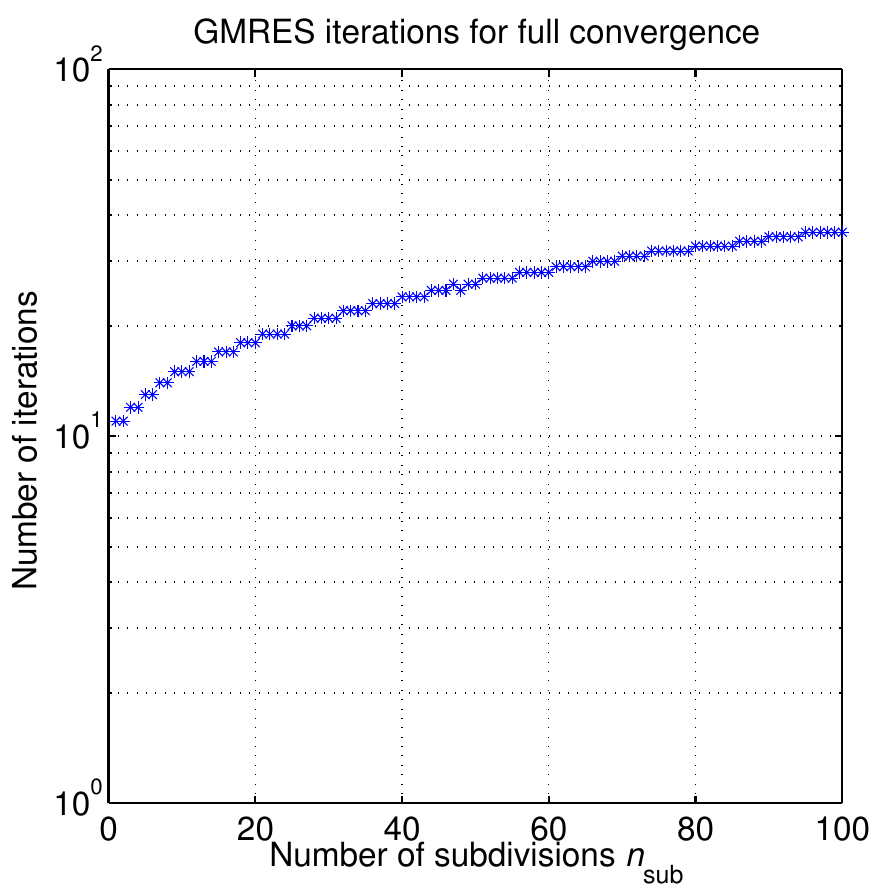}
\end{minipage}}
\caption{\sf Same as Figure~\ref{fig:conv3}, but the program {\tt
    demo1d.m} is used.}
\label{fig:conv1de}
\end{figure}

Figure~\ref{fig:conv1de} shows results produced by {\tt demo1d.m}.
Beyond $n_{\rm sub}=60$ one now achieves essentially full machine
precision in $q$ of~(\ref{eq:q}). Despite this success, inner product
preserving Nyström discretization can perhaps not quite compete
with the RCIP method in this example. The differences in performance
relate to issues of memory and speed. The RCIP method uses a much
smaller linear system ($16n_{\rm pan}$ unknowns) than does inner
product preserving Nyström discretization ($16(n_{\rm pan}+2n_{\rm
  sub})$ unknowns). Besides, the RCIP method converges in only eight
GMRES iterations, irrespective of $n_{\rm sub}$. See
Figure~\ref{fig:conv3}.

\newpage

\begin{small}

\end{small}

\end{document}